\documentclass{elsart}

 \usepackage{graphics}
 \usepackage{graphicx}
 \usepackage{epsfig}
 \usepackage{amssymb}

\def\p{$e^\pm \;$}

\def\bi{\begin{itemize}}
\def\i{\item}
\def\ei{\end{itemize}}
\def\be{\begin{equation}}
\def\ee{\end{equation}}
\def\bea{\begin{eqnarray}}
\def\eea{\end{eqnarray}}

\def\dt{\Delta t}
\def\p{\partial}

\begin{document}

\begin{frontmatter}

\title{Astrophysical Smooth Particle Hydrodynamics}

\author{Stephan Rosswog}

\address{School of Engineering and Science, Jacobs University Bremen, Campus Ring 1, 28759 Bremen, Germany}

\begin{abstract}
The paper presents a detailed review of the smooth particle hydrodynamics (SPH)
method with particular focus on its astrophysical applications. We start by introducing
the basic ideas and concepts and thereby outline all ingredients that are necessary
for a practical implementation of the method in a working SPH code.
Much of SPH's success relies on its excellent conservation properties and therefore
the numerical conservation of physical invariants receives much attention throughout
this review. The self-consistent derivation of the SPH equations from the Lagrangian 
of an ideal fluid is the common theme of the remainder of the text. We derive a modern,
Newtonian SPH formulation from the Lagrangian of an ideal fluid. It
accounts for changes of the local resolution lengths which result in corrective, so-called 
"grad-h-terms". We extend this strategy to special relativity for which
we derive the corresponding grad-h equation set. The variational approach 
is further applied to a general-relativistic fluid evolving in a fixed, curved 
 background space-time. Particular care is taken to explicitely derive all relevant 
equations in a coherent way.
\end{abstract}

\end{frontmatter}

\newpage

{\bf Contents}
\begin{tabbing}
\hspace*{0.5cm}\= 1 \hspace*{0.5cm}\= Introduction \hspace*{10cm} \= \pageref{page_intro}\\
\\
\>                2                \> Basic concepts of Smooth Particle Hydrodynamics \> \pageref{page_basic_concepts}\\
\\
\>                                 \> 2.1 Lagrangian hydrodynamics                    \> \pageref{page_Lag_hydro}\\
\\
\>                                 \> 2.2 The SPH kernel interpolation                \> \pageref{page_kernel}\\
\\
\>                                 \> 2.3 The ``vanilla ice'' SPH equations           \> \pageref{page_vanilla_ice}\\
\\
\>                                 \> 2.4 Conservation properties                     \> \pageref{page_conservation}\\
\\
\>                                 \> 2.5 Alternative SPH discretizations             \> \pageref{page_alternative}\\
\\
\>                                 \> 2.6 Adaptive resolution                         \> \pageref{page_adaptive_res}\\
\\
\>                                 \> 2.7 Artificial dissipation                      \> \pageref{page_AV}\\
\\
\>                                 \> 2.8 Time integration in SPH                     \> \pageref{page_integ}\\
\\
\>                                 \> 2.9 ``Best practice'' suggestions               \> \pageref{page_best_practice}\\
\\
\>                3                \> SPH from a variational principle                \> \pageref{page_var_principle}\\
\\
\>                                 \> 3.1 The Lagrangian and the Euler-Lagrange equations  \> \pageref{page_Lag}\\
\\
\>                                 \> 3.2 The density, its derivatives and ``grad-h''-terms \> \pageref{page_gradh}\\
\\
\>                                 \> 3.3 The SPH equations with  ``grad-h''-terms    \> \pageref{page_SPH_gradh}\\
\\
\>                4                \> Relativistic SPH                                \> \pageref{page_rel_SPH}\\
\\
\>                                 \> 4.1 Special-relativistic SPH                    \> \pageref{page_SR_SPH}\\
\\
\>                                 \> 4.2 General-relativistic SPH on a fixed background metric \> \pageref{page_GR_SPH}\\
\\
\>                5                \> Summary                                         \> \pageref{page_summary}\\
\\
\>                References       \>                                                 \> \pageref{page_references}
\end{tabbing}
\newpage

\section{Introduction}
\label{page_intro}
%
%
Much of what we observe in the physical Universe has been shaped by fluid 
dynamical processes. From the hot gas in galaxy clusters  to the
internal structures of their consituent galaxies down to their stars and 
planets, all has been formed by the interplay between gravity, gas 
dynamics and further physical processes such as the interaction with radiation, 
nuclear burning or magnetic fields. The latter processes often involve intrinsic 
length and time scales that are dramatically different from those of the 
gas dynamical processes, therefore many astrophysical problems are prime 
examples of multi-scale and multi-physics challenges. The complexity of 
the involved physical processes and the lack of symmetry usually prohibit 
analytical treatments and only numerical approaches are feasible.\\
Although fluid dynamics is also crucial for many technical applications,
their requirements usually differ substantially from those of astrophysics and 
this also enters the design of the numerical methods. Typical 
astrophysical requirements include: 
\bi
\i Since fixed boundaries are usually absent, flow geometries are determined by 
the interplay between different physical processes such as gas dynamics and
(self-)gravity which often lead to complicated, dynamically changing flow 
geometries. Thus, a high spatial adaptivity is often required from
astrophysical hydodynamics schemes.
\i Shocks often crucially determine the evolution of cosmic objects. Examples include
in supernova remnants or the Earth's magnetosphere.
\i Physical quantities can vary by many orders of magnitudes between different
regions of the simulation domain. This requires a particularly high robustness 
of the numerical scheme.
\i In many astrophysical problems the numerical conservation of physically conserved 
quantities determines the success and the reliability of a computer simulation. 
Consider, for example, a molecular gas cloud that collapses under the influence of 
its own gravity to form stars. If the simulation for some reason dissipates
angular momentum, a collapsing, self-gravitating portion of gas may form just a 
single stellar object instead of a multiple system of stars and it will thus produce a 
qualitatively wrong result.
\i Many astrophysical questions require dealing with physical processes beyond gas dynamics
and self-gravity. A physically intuitive and flexible formulation of the numerics can
substantially facilitate the implementation of new physics modules into existing codes.   
\ei
No numerical method performs equally well at each of the above requirements, therefore,
the choice of the best-suited numerical approach can often save a tremendous amount of 
effort in obtaining  reliable results. Therefore: {\em horses for 
courses}.\\
In the following, the smooth particle hydrodynamics (SPH) method 
\cite{lucy77,gingold77,hernquist89,benz90a,monaghan92,monaghan05}, a completely 
mesh-free approach to solve the hydrodynamic equations is discussed in
detail. Its conservation properties are a major strength of SPH, therefore 
``hard-wired'' conservation receives much attention 
throughout this text. The derivation of the SPH equations (in the absence of dissipation) 
requires nothing more than a suitable Lagrangian, a density prescription that
depends on the coordinates and the first law of thermodynamics. The resulting 
equations conserve the physically conserved quantities even in their discretized 
form, provided that the original Lagrangian possessed the correct symmetries. 
Therefore, derivations from Lagrangians play a central role in our discussion 
of the subject.\\
%
%
This review has emerged from a lecture series on ``Computational relativistic astrophysics''
as part of a Doctoral Training Programme on the ``Physics of Compact Stars'' that was 
held in summer 2007 at the European Centre for Theoretical Studies in Nuclear Physics and 
Related Areas (ECT$^\ast$) in Trento, Italy. In the pedagogical spirit of this lecture series 
the text is kept in ``lecture'' rather than ``paper'' style, i.e. even rather trivial 
steps are written down explicitely. The goal is to pave a broad and smooth avenue to 
a deep understanding of the smooth particle hydrodynamics method rather than just to provide a 
bumpy trail. Due to this pedagogical scope, the focus of this review needs to be clear-cut, 
but rather narrow: it only discusses the numerical solution of the inviscid hydrodynamics
equations, in the Newtonian, special-relativistic and general-relativistic 
(fixed metric) case. \\
For practical astrophysical simulations often sophisticated additional 
physics modules are required and their implementation may pose additional numerical challenges 
which are beyond the scope of this review. In the following we provide a brief list of 
additional physics that has been implemented into SPH and point to references that are
intended as starting points for further reading.
%
%
\bi
\item{\em Gravity:}\\
Self-gravity is for many astrophysical problems a key ingredient. A straight forward calculation 
of pairwise gravitational forces between $N$ particles requires a prohibitively large $O(N^2)$ 
number of operations and therefore such an approach is not feasible for realistic problems. 
The most natural gravitational force evaluation for a purely meshfree method like SPH is the 
use of a tree, either in the form of an oct-tree 
\cite{barnes86,hernquist89,warren95,dave97,carraro98a,springel01a,springel05a} or a binary tree 
\cite{benz90b,wadsley04,nelson08} both of which require  only  $O(N \log(N))$ operations.
Other possibilities include particle-mesh methods \cite{hockney88,klypin97,dubinski04} in which
the particles are mapped onto a mesh. The latter is used to efficiently solve Poisson's equation 
and to subsequently calculate the accelerations at the particle positions. Dehnen 
\cite{dehnen00,dehnen02} has used ideas from the Fast Multipole Method (FMM) \cite{greengard87} 
for a fast tree code which scales proportional to $O(N)$. Several hybrid approaches which
combine elements of different methods exist \cite{theuns94,theuns98,bode00,knebe01,bode03}.\\

\item{\em Equations of state:}\\
For some problems polytropic equations of state (EOS), $p= K \rho^\Gamma$, with $K$ being a function
of entropy, $\rho$ the matter density and $\Gamma$ the adiabatic exponent, are sufficiently accurate.
Other problems require more sophisticated approaches. This is particularly true in cases where the matter 
compressibility depends on the density regime, as, for example, in neutron stars where the effective 
polytropic exponent above nuclear matter density is $\Gamma \approx 2.5$, but drops to values close to
4/3 at lower densities. Different nuclear EOSs have been used 
\cite{herant94,fryer99b,rosswog99,rosswog02a,fryer06} in the context of neutron stars, at lower densities 
and in cases where the matter composition is important for other processes further physical EOSs 
have been applied \cite{guerrero04,lee04,lee05b,yoon07a,rosswog08b}. \\

\item{\em Solid state mechanics}\\
In the context of planetary impacts additional concepts such as material stresses, fracture 
physics or plasticity criteria need to be implemented into SPH. This branch of development
was initiated by the work of \cite{libersky90} and further continued by 
\cite{libersky93a,benz94,benz95,asphaug96,randles96,hicks97a,vignjevic00,schaefer07}. More recently,
also porosity models were included in SPH \cite{sirono04,jutzi08}.\\

\item{\em Physical viscosity:}\\
For some applications the solution of the non-dissipative hydrodynamics equations is inadequate and 
physical (as opposed to artificial)
viscosity needs to be modelled. Examples include the angular momentum transport in accretion disks, see e.g. 
\cite{frank02}, or the dissipation of sound waves as a mechanism to heat intra-cluster media \cite{fabian03}.
Various authors have implemented the viscous stress tensor into SPH \cite{flebbe94,takeda94,watkins96,morris97,lee02,schaefer04,sijacki06}.
This requires particular care in the implementation of higher order derivatives which -in a straight forward 
kernel approximation- can be sensitive to particle disorder.\\

\item{\em Thermal conduction:}\\
Thermal conduction has, for example, been proposed as a possible heating 
mechanism for offsetting central cooling losses in rich clusters of galaxies.
Its implementation poses some challenges due to second derivatives that enter
the conductive term of the energy equation. Usually,  a particluar 
discretization of the second derivative due to \cite{brookshaw85}
is used in modelling thermal conduction  \cite{cleary99,jubelgas04}. \\

\item{\em Nuclear burning:}\\
Often the energy release due to nuclear reactions is too slow to influence  
gas flows substantially on a dynamical time scale.  Under special
circumstances, however, explosive nuclear burning can be triggered and in such cases
the nuclear energy release can become the main driver of the dynamical gas evolution. 
In such cases the coupling between the hydrodynamics and the nuclear reactions is challenging 
due the huge difference in their intrinsic time scales. The first implementation of 
a nuclear reaction network into SPH goes back to \cite{benz89}. The 14 isotope 
$\alpha$-chain network of this first study has also been used with updated nuclear reaction rates in studies 
of white dwarf coalescences \cite{guerrero04}. More recently, further small networks \cite{hix98,timmes99}
have been coupled with SPH \cite{rosswog08b} to study white dwarf coalescences \cite{yoon07a,dan09}, 
collisions between them \cite{rosswog09c,raskin09} and tidal disruptions of white dwarfs by black 
holes \cite{rosswog08a,rosswog09a}.\\

\item{\em Chemistry:} \\
Chemical abundances carry the imprints of the past evolutionary processes that shaped
today's galaxies. Their chemical enrichment is inseparably interweaved with the formation of 
structure in galaxies since --being closely related to supernovae-- it drives local energy 
feedback and  gas expansion, but on the other hand it also triggers --via metallicity-dependent
cooling-- the contraction and collapse of local gas structures. A large variety of chemical evolution
approaches has been implemented into SPH, for a starting point one may want to consult 
\cite{steinmetz94,steinmetz95,raiteri96,carraro98b,berczik99,kawata03,kobayashi04b,tornatore04,scannapieco05,martinez_serrano08,greif09,wiersma09}. \\

\item{\em Radiation: photons and neutrinos}\\
Closely related to chemical species is the interaction between photons and the ambient gas. 
For some problems relatively simple cooling and heating recipes can be applied, e.g. 
\cite{hernquist89,katz96,springel05a}. Optically thick diffusion has in fact already been implemented
in one of the very first SPH-papers \cite{lucy77}. Later, more sophisticated approaches to calculate
the required second derivatives were suggested \cite{brookshaw85}. For a recent implicit 
diffusion approach, see \cite{viau06}. Flux-limited diffusion schemes have been suggested by 
\cite{whitehouse05,whitehouse06,mayer07}, Monte-Carlo-type approaches have been followed by 
\cite{oxley03,stamatellos05,semelin07,croft08} and ray-tracing has been applied by 
\cite{kessel00,alvarez06,altay08}. Several further methods to treat radiation
in SPH do exist \cite{dale07,johnson07,susa06,pawlik08,gritschneder09,forgan09}.\\
When matter densities are so large that photons are ``trapped'', neutrinos can become 
the only cooling agents. For white dwarfs and less compact stars matter is transparent to neutrinos 
and they can just be treated as a local drain of thermal energy. For higher densities, say in a neutron 
star or the inner regions of the disk around a hyper-accreting black hole, opacity effects 
may prohibit the free escape of neutrinos. A further example of the astrophysical importance of 
neutrino physics is the delayed explosion mechanism \cite{wilson82,wilson85} 
of core-collapse supernovae where the energy deposition via neutrinos is thought to be vital to re-energize 
the stalled shock after it has crossed the outer layers of the stellar iron core. In this context, 
multi-flavour neutrino cooling and heating has been implemented into SPH \cite{herant94,fryer06} in an 
entirely particle-based, flux-limited diffusion approach. To model compact object mergers, neutrinos have 
been implemented into SPH via a hybrid, particle-mesh approach \cite{rosswog03a} and by using local disk 
scale heights to estimate neutrino opacities \cite{lee04,lee05b}. \\

\item{\em Magnetic fields:}\\
Magnetic fields are prevalent everywere in the cosmos, but they pose a notorious numerical challenge not least
because of the difficulty in fulfilling the $\nabla \cdot \vec{B}=0$-constraint.
The most obvious approach is a straight-forward SPH discretization of the ideal MHD equations 
\cite{gingold77,phillips85,dolag02,price04a,price04b,price04c,price05}. While it performs reasonably well
in some problems, particle disorder and finite $\nabla\cdot\vec{B}$ values can trigger numerical 
instabilities \cite{price04c}.
Several approaches have been developed  that apply regularization 
techiques to overcome the problems related to particle disorder \cite{borve01,borve04,borve05,borve06,dolag08}.
Recently, an approach has been implemented \cite{rosswog07c,rosswog08c} that advects so-called Euler-potentials, 
$\alpha$ and $\beta$, with the SPH particles. The magnetic field can be reconstructed from the values of these
potentials at the particle positions via $\vec{B}= \nabla \alpha \times \nabla \beta$. This approach guarantees the 
fulfillment of the $\nabla \cdot \vec{B}=0$-constraint by construction and has shown excellent results in a 
large number of test cases \cite{rosswog07c}.\\

\item{\em Further ``subgrid'' models:}\\
Often, numerical models of astrophysical processes require the feedback from physical processes
on sub-resolution scales. Such processes are usually implemented as (parametrized) ``subgrid'' models
that transmit the main effects from the unresolved processes to the fluid.
Cosmic rays present one such example. They contribute significantly to the pressure of the 
interstellar medium in our Galaxy and therefore they may be important in regulating star 
formation during the formation and evolution of galaxies. For the implementation of the effects 
of cosmic rays into SPH see \cite{ensslin07,jubelgas08,pfrommer08a,pfrommer08b,sijacki08}.
Another example is the modelling of the feedback of (unresolvable) accretion-driven outflows during 
the merger of galaxies to study the self-regulation of black hole growth \cite{dimatteo05,springel05b}.
In simulations on galactic scales stars --their birth, feedback via outflows and possible supernova explosions--
also need to be modelled as effective sub-scale models. By their very nature such approaches are closely linked
to those of radiative transfer and chemical enrichment. For the various approaches we refer to the literature
\cite{navarro93,mihos94,bate95,gerritsen97,sommer_larsen99,springel00,springel01b,thacker00,scannapieco01,thacker01,kay02,ascasibar02,theuns02,springel03,bate03,bonnell03,marri03,tornatore03,vandenbosch03,clark04,kay04,cox04,dale05,springel05a,hoeft06,cattaneo07}.
\ei
The remainder of this review is organized as follows. In Sec. 2
the most basic form of SPH (``vanilla ice'') is derived by discretizing the equations
of Lagrangian hydrodynamics. This form is somewhat ad hoc, yet it performs well in 
practice and it is at the heart of many SPH codes that are regularly used in the 
astrophysics community. In Sec. 3, the Lagrangian of an ideal fluid is discretized
and subsequently used to derive a modern version of SPH. Since it follows from a variational
principle, this formulation avoids any ambiguity with respect to a symmetrization of the 
equations and it naturally leads to additional terms that result from derivatives of the 
local hydrodynamic resolution length, the so-called ``grad-h-terms'', 
\cite{springel02,monaghan02}. Armed with this gadgetery suitable Lagrangians are used 
to elegantly derive the SPH-equations both for the special- and the general-relativistic case, 
see Sec. 4. The summary and a brief outlook are presented in Sec. 5.

\section{Basic concepts of Smooth Particle Hydrodynamics}
\label{page_basic_concepts}
\label{sec:basic:Lag_hydro}

\subsection{Lagrangian hydrodynamics}
\label{page_Lag_hydro}
In contrast to grid-based (Eulerian) methods, Smooth Particle Hydrodynamics 
(SPH) is purely Lagrangian. In the Eulerian picture, derivatives are calculated
at a fixed point in space while, in the Lagrangian description, they are
evaluated in a coordinate system attached to a moving fluid element.
Thus in the Lagrangian approach we follow individual fish rather 
that staring at the pond and watch the swarm pass by.
The Lagrangian (or substantial) time derivative, $d/dt$, is related to the 
Eulerian time derivative, $\p/\p t$, by
\be
\frac{d}{dt}= \frac{dx^i}{dt} \frac{\p}{\p x^i} + \frac{\p}{\p t}=  \vec{v} \cdot \nabla + \frac{\p}{\p t}.
\ee
Applied to the Eulerian continuity equation,
\be
\frac{\p \rho}{\p t} + \nabla \cdot (\rho \vec{v}) = 0,
\ee
one finds, using $d \rho/dt = \vec{v} \cdot \nabla \rho + \p\rho/\p t$, the 
Lagrangian form
\be
\frac{d \rho}{dt}= - \rho \nabla \cdot \vec{v}.
\label{eq:basic:drho_dt}
\ee
The momentum conservation equation for a non-viscous fluid, the so-called 
Euler equation, reads in Lagrangian form
\be
\frac{d\vec{v}}{dt}= - \frac{\nabla P}{\rho} + \vec{f},
\label{eq:basic:Euler_eq}
\ee
i.e. apart from ``body forces'' such as gravity or magnetic fields embodied in
the quantity $\vec{f}$, the fluid is accelerated by gradients of the pressure $P$. The
energy equation follows directly from the (adiabatic) first law of thermodynamics, 
Eq.~(\ref{eq:first_law_3}), together with Eq.~(\ref{eq:basic:drho_dt}):
\be
\frac{du}{dt}= \frac{P}{\rho^2} \frac{d\rho}{dt}= -\frac{P}{\rho} \nabla \cdot
\vec{v}.
\label{eq:basic:Lag_energy_equation}
\ee

\hspace*{0cm}\fbox{
\parbox{14cm}{
\vspace*{0.5cm}
\centerline{\bf First law of thermodynamics}
\vspace*{0.5cm}
For SPH one needs the first law of thermodynamics, $dU= dQ - P dV,$ in terms
of ``specific'' quantities. Restriction to adiabatic processes
allows to drop the $dQ$-term. On a ``per mass basis'', the energy
$U$ becomes $u$ (``energy per mass'') and the volume, $V$, becomes ``volume
per mass'', i.e. $1/\rho$, with $dV \rightarrow d(1/\rho)= -d \rho/\rho^2$. 
For the case without entropy generation the first law reads
\be
du= \frac{P}{\rho^2} d\rho,\label{eq:first_law_2}
\ee
from which 
\be
\frac{du}{dt}= \frac{P}{\rho^2} \frac{d\rho}{dt} \quad \quad{\rm and} \quad \quad
\left(\frac{\partial u}{\partial \rho}\right)_s= \frac{P}{\rho^2}\label{eq:first_law_3}
\ee
follow.
In a relativistic context it can be advantageous to work with quantities
``per baryon'': 
\be
\left(\frac{\partial u}{\partial n}\right)_s= \frac{P}{n^2},\label{eq:first_law_4}
\ee
where $n$ is the baryon number density in the local rest frame.
}}

The set of equations (\ref{eq:basic:drho_dt}), (\ref{eq:basic:Euler_eq}),
(\ref{eq:basic:Lag_energy_equation}) must be closed by an equation of state 
that relates quantities such as the pressure, $P$, or the speed of sound, 
$c_s$, to macroscopic fluid quantities such as density or temperature. All 
the microphysics, say whether the pressure is produced by collisions in a 
Maxwell-Boltzmann gas or by degenerate electrons due to the Pauli principle, 
is embodied in the equation of state. It can be as simple as a polytrope, 
or, for more complicated cases, say for hot nuclear matter,
it may only be available in tabular form, see the introduction for a 
link to the existing literature.

\subsection{The SPH kernel interpolation}
\label{page_kernel}
In the following, discrete representations of the continuous Lagrangian 
hydrodynamics equations are derived. In SPH, the interpolation points 
(``particles'') are moved with the local fluid velocity\footnote{Variants 
that use local averages of fluid velocities exist, see 
the ``XSPH''-approach of \cite{monaghan89}.}, derivatives 
are calculated via a kernel approximation without the need for finite 
differences. In this way the partial differential equations of Lagrangian 
fluid dynamics are transformed into ordinary differential equations.\\
In principle, there is some freedom in discretizing the fluid equations.
But to ensure that the physically conserved quantities are also conserved 
{\em by construction} in the discretized particle equations
the latter need to possess the correct symmetries in the particle indices.
In the simplest, ``vanilla ice'' version
of SPH, the symmetrization is imposed somewhat {\em ad hoc}, yet, it works 
well in practice and it is commonly used throughout astrophysics, see e.g. 
\cite{benz90a,steinmetz96,klessen00b,bate03,wadsley04,rosswog02a,wetzstein08}.
For different symmetrization possibilities see \cite{hernquist89,thomas92,flebbe94,kunze97}
and Sec. \ref{sec:alternative}.

\subsubsection{Interpolating function values}
At the heart of SPH is a kernel approximation in which a function $f(\vec{r})$
is approximated by
\be
\tilde{f}_h(\vec{r})= \int f(\vec{r'}) W(\vec{r} - \vec{r'},h) \;  d^3r',
\label{eq:basic:int_interpol}
\ee
where $W$ is the so-called smoothing kernel (or window function) and the smoothing
length, $h$, determines the width of this kernel. 
Obviously, one would like to recover the original function in the limit of
an infinitely small smoothing region and therefore the kernel should fulfill
\be
\lim\limits_{h \to 0} \tilde{f}_h(\vec{r}) = f(\vec{r}) \; \; {\rm and} \; \; 
\int W(\vec{r} - \vec{r'},h) \; d^3r'= 1,
\ee 
i.e. apart from being normalized, the kernel should have the 
$\delta$-distribution property in the limit of vanishing smoothing length.\\
To arrive at a discrete approximation, one can write the integral as
\be
\tilde{f}_h(\vec{r})= \int \frac{f(\vec{r'})}{\rho(\vec{r'})} W(\vec{r} - \vec{r'},h) \; \rho(\vec{r'}) d^3r',
\ee
where $\rho$ is the mass density and subsequently one can replace the integral by a 
sum over a set of interpolation points (``particles''), whose masses, $m_b$, result 
from the  $\rho(\vec{r}') \; d^3r'$ term\footnote{For 
simplicity, we are omitting the subscript $h$ in what follows. We also
drop the distinction between the function to be interpolated and the interpolant, 
i.e. we use the same symbol $f$ on both sides of the equal signs. Throughout the article
subscripts such as $A_b$ refer to the values at a particle position, i.e. $A_b \equiv A(\vec{r}_b)$.}
\be
f(\vec{r})= \sum_b \frac{m_b}{\rho_b} f_b W(\vec{r} - \vec{r}_b,h).
\label{eq:basic:sum_interpol}
\ee  
The summation interpolant Eq.~(\ref{eq:basic:sum_interpol}) of the density
then reads
\be
\rho(\vec{r})= \sum_b m_b W(\vec{r} - \vec{r}_b,h).
\label{eq:basic:sum_rho}
\ee
This density estimate by summing up kernel-weighted masses in the
neighborhood of the point $\vec{r}$ plays a central role in the derivation of
the SPH equations from a Lagrangian, see
Sec.~\ref{sec:SPH_from_Lagrangian}. Note that $h$ has not been specified yet,
this will be addressed at a later point.  In practice, the density can also be
calculated via integration of a discretized version of 
Eq.~(\ref{eq:basic:drho_dt})
via Eq.~(\ref{eq:basic:divv}), see below. In most applications
the difference between integrated and summed density estimates is completely
negligible. The summation form, however, is certainly 
numerically more robust, integration can, in extremely underresolved
regions, produce negative density values. The summation form does not assume the
density to be a differentiable function while 
Eqs.~(\ref{eq:basic:drho_dt})/(\ref{eq:basic:divv}) do so \cite{price08a}. The 
standard practice in SPH is to keep the particle masses fixed so that the
mass conservation is perfect and there is no need to solve the continuity
equation.

\subsubsection{Approximating derivatives}
To calculate derivatives, one takes the analytical expression of the summation approximation,
Eq.~(\ref{eq:basic:sum_interpol}), 
\be
\nabla f(\vec{r})= \sum_b \frac{m_b}{\rho_b} f_b \nabla W(\vec{r} -
\vec{r}_b,h),
\label{eq:basic:nabla_sum}
\ee
i.e. the exact derivative of the approximated function is used.
The kernel function is known analytically, therefore there is no need for 
finite difference approximations.\\
Processes such as diffusion or thermal conduction require second derivatives.
One could now proceed in a straight forward manner and take another derivative 
of Eq.~(\ref{eq:basic:nabla_sum}). Yet, this estimate is rather sensitive 
to particle disorder and therefore not recommended for practical use. A better
prescription is \cite{brookshaw85}
\be
(\nabla^2 f)_a= 2 \sum_b \frac{m_b}{\rho_b} (f_a - f_b) \frac{w_{ab}}{r_{ab}},
\label{eq:basic:nabla2_sum}
\ee
where $w_{ab}$ is related to the kernel by $\nabla W_{ab}= \hat{e}_{ab} \; w_{ab}$,
$\hat{e}_{ab}$ is the unit vector from particle $b$ to particle $a$, 
$\hat{e}_{ab}=\vec{r}_{ab}/r_{ab}$ and $\vec{r}_{ab}= \vec{r}_a - \vec{r}_b$. 
For more information on higher order derivatives we refer to the existing literature 
\cite{brookshaw85,cleary99,espanol03,monaghan05}.

\subsubsection{The kernel function}
To restrict the number of contributing particles in the sum of Eq.~(\ref{eq:basic:sum_interpol})
to a local subset, the kernel should have compact support, otherwise the summation would extend 
over all $n$ particles and produce a numerically intractable, completely inefficient $n^2$-method. 
Although for geometries such as flattened disks non-radial kernels seem like a natural choice
\cite{fulbright95,shapiro96,owen98} they have the disadvantage that it is difficult to ensure 
exact angular momentum conservation, see below. Therefore, usually radial kernels with
$W(\vec{r} - \vec{r}',h)= W(|\vec{r} - \vec{r}'|,h)$, are used in SPH.\\
The accuracy of the kernel interpolation is in practice rather difficult to 
quantify unless strongly simplifying assumptions about the particle 
distribution are made, which are usually not met in reality.
One can expand $f(\vec{r'})$ in the integral approximant, Eq.~(\ref{eq:basic:int_interpol}),
into a Taylor series around $\vec{r}$
\be
f(\vec{r'})= \sum_{k=0}^{\infty} \frac{(-1)^k h^k f^{(k)}(\vec{r})}{k!} 
\left(\frac{\vec{r}-\vec{r}'}{h}\right)^k.
\ee
By explicitly writing down the integrals over the different terms in the Taylor
expansion on the RHS and comparing to the LHS, one can determine how well the
approximation agrees with the original function. By requiring that error
terms vanish, one can construct kernels of the desired order.  
In practice, however, such kernels may have unwanted properties such as
negative values in certain regions and this may, in extreme cases, lead to unphysical (negative!) 
density estimates from Eq.~(\ref{eq:basic:sum_rho}). Although substantial work has 
been invested into constructing more accurate kernels \cite{monaghan85b,fulk96,price04c,cabezon08}, 
none has been found to overall perform substantially better in practice than the ``standard'' 
cubic spline SPH kernel \cite{monaghan92} which is used in almost all SPH simulations. 
Particular higher-order kernels, however, may have advantages in the context of resolving
Kelvin-Helmholtz instabilities near large density gradients \cite{read09}.
In 3D, the cubic spline kernel reads 
\be
W(q) = \frac{1}{\pi h^3}\left\{\begin{array}{cl}
        1-\frac{3}{2}q^2 + \frac{3}{4} q^3 \quad \quad{\rm for} \; 0 \le q \le 1\\  
        \frac{1}{4} (2-q)^3  \quad \quad \quad \quad{\rm for} \; 1 < q \le 2\\ 
        0 \quad \quad \quad \quad \quad \quad {\rm for} \; q > 2\\
           \end{array}\right. ,\label{eq:basic:standard_kernel}
\ee
where $q= | \vec{r} - \vec{r}' |/h$. This kernel is {\em radial}, i.e. it
depends only on the absolute value of $\vec{r} - \vec{r}'$. With this
kernel the integral approximation, Eq.~(\ref{eq:basic:int_interpol}), is
\be
\tilde{f}_h(\vec{r})= f(\vec{r}) + C \; h^2 + O(h^4),
\label{eq:basic:integral_approximation}
\ee
where the quantity $C$ contains the second derivatives of the function
$f$. Therefore, constant and linear functions are reproduced exactly by 
the integral representation Eq.~(\ref{eq:basic:int_interpol}).\\
In practice, two approximations were applied: i) the integral
interpolation, Eq.~(\ref{eq:basic:int_interpol}) and ii) the discretization,
Eq.~(\ref{eq:basic:sum_interpol}). The accuracy of the discretized equations depends
on the distribution of the interpolation points (``particles''). In early work,
error estimates were based on a purely statistical description assuming a
random distribution of particles. In a simulation, however, the particle distribution is not 
random, but depends on both the kernel used and the dynamics of the considered
system. For this reason, numerical experiments turned out to be far more accurate than 
predicted by these early estimates \cite{monaghan05}.\\
In the following box we collect some expressions for the derivatives of radial kernels
that are frequently used throughout this text.

\subsection{The ``vanilla ice'' SPH equations}
\label{sec:basic:vanilla_ice_SPH}
\label{page_vanilla_ice}
\subsubsection{Momentum equation}
One could now proceed according to ``discretize and hope for the best''
and this was historically the first approach. A straightforward
discretization of Eq.~(\ref{eq:basic:Euler_eq}) (no body forces)
yields
\be
\frac{d\vec{v}_a}{dt}= -\frac{1}{\rho_a} \sum_b \frac{m_b}{\rho_b} P_b
\nabla_a W_{ab}.
\ee
This form solves the Euler equation to the order of the method, but it does
not conserve momentum. To see this, consider the force that particle $b$ exerts on 
particle a,
\be
\vec{F}_{ba}= \left( m_a \frac{d\vec{v}_a}{dt} \right)_b = -
\frac{m_a}{\rho_a}\frac{m_b}{\rho_b} P_b \nabla_a W_{ab},
\label{eq:basic:F_ab}
\ee
and similarly from $a$ on $b$ 
\be
\vec{F}_{ab}= \left( m_b \frac{d\vec{v}_b}{dt} \right)_a = -
\frac{m_b}{\rho_b}\frac{m_a}{\rho_a} P_a \nabla_b W_{ba} = 
\frac{m_a}{\rho_a}\frac{m_b}{\rho_b} P_a \nabla_a W_{ab},
\ee
where Eq.~(\ref{eq:k4}) was used. Since in general $P_a \neq P_b$, this
momentum equation does not fulfill Newton's third law (``actio = reactio'')
by construction and therefore total momentum is not conserved.\\

\hspace*{-2cm}\fbox{
\parbox{17cm}{
\vspace*{0.5cm}
\centerline{\bf Derivatives of radial kernels}
\vspace*{0.5cm}

Throughout the text, we use the notation $\vec{r}_{bk}= \vec{r}_{b} - \vec{r}_{k}$, $r_{bk}=
|\vec{r}_{bk}|$ and $\vec{v}_{bk}= \vec{v}_{b} - \vec{v}_{k}$. 
Derivatives resulting from the smoothing lengths are ignored for the 
moment, they
will receive particular attention in Sec. \ref{sec:SPH_from_Lagrangian}. 
By straight-forward component-wise differentiation one finds
\be
\frac{\partial}{\partial \vec{r}_a} |\vec{r}_b-\vec{r}_k|
= \frac{(\vec{r}_b-\vec{r}_k)
  (\delta_{ba}-\delta_{ka})}{|\vec{r}_b-\vec{r}_k|}
= \hat{e}_{bk} (\delta_{ba}-\delta_{ka}),\label{eq:k1}
\ee
where $\hat{e}_{bk}$ is the unit vector from particle $k$ to particle $b$,
\be
\frac{\partial}{\partial \vec{r}_a} \frac{1}{|\vec{r}_b-\vec{r}_k|}=
- \frac{\hat{e}_{bk} (\delta_{ba}-\delta_{ka})}{|\vec{r}_b-\vec{r}_k|^2}
\label{eq:k2}
\ee
and $\delta_{ij}$ is the usual Kronecker symbol.
Another frequently needed expression is
\bea
\frac{d r_{ab}}{dt}=& &  \frac{\partial r_{ab}}{\partial x_a} \frac{d x_a}{dt} 
                     + \frac{\partial r_{ab}}{\partial y_a} \frac{d y_a}{dt} 
                     + \frac{\partial r_{ab}}{\partial z_a} \frac{d z_a}{dt} 
                     +  \frac{\partial r_{ab}}{\partial x_b} \frac{d x_b}{dt} 
                     + \frac{\partial r_{ab}}{\partial y_b} \frac{d y_b}{dt} 
                     + \frac{\partial r_{ab}}{\partial z_b} \frac{d z_b}{dt} 
                     \nonumber \\
                   =& & \nabla_a r_{ab} \cdot \vec{v}_a + \nabla_b r_{ab}
                   \cdot \vec{v}_b  = \nabla_a r_{ab} \cdot \vec{v}_a - 
                   \nabla_a r_{ab}
                   \cdot \vec{v}_b
                   = \nabla_a r_{ab} \cdot \vec{v}_{ab} 
                   = \hat{e}_{ab} \cdot \vec{v}_{ab},
\label{eq:basic:drab_dt} 
\eea
where 
$\partial r_{ab}/\partial x_b= - \partial r_{ab}/\partial x_a$ etc. was used.
For kernels that only depend on the magnitude of the separation,
$W(\vec{r}_b-\vec{r}_k)= W(|\vec{r}_b-\vec{r}_k|)\equiv W_{bk}$, the derivative with respect
to the coordinate of an arbitrary particle $a$  is
\be
\nabla_a W_{bk}=\frac{\partial}{\partial \vec{r}_a} W_{bk}= \frac{\partial  W_{bk}}{\partial
  r_{bk}}  \frac{\partial r_{bk}} {\partial \vec{r}_a}= 
\frac{\partial  W_{bk}}{\partial r_{bk}} \hat{e}_{bk}
(\delta_{ba}-\delta_{ka})= \nabla_b W_{kb} (\delta_{ba}-\delta_{ka}) \label{eq:k3},
\ee
where Eq.~(\ref{eq:k1}) was used. This yields in particular the important property
\be
\nabla_a W_{ab}=\frac{\partial}{\partial \vec{r}_a}W_{ab}
= \frac{\partial W_{ab}}{\partial r_{ab}} \frac{\partial r_{ab}} {\partial
  \vec{r}_a} 
 = \frac{\partial W_{ab}}{\partial r_{ab}} \hat{e}_{ab}
= - \frac{\partial W_{ab}}{\partial r_{ab}} \frac{\partial r_{ab}} {\partial
  \vec{r}_b} 
= - \frac{\partial}{\partial \vec{r}_b}W_{ab}= - \nabla_b W_{ab}.\label{eq:k4}
\ee
The time derivative of the kernel is given by
\be
\frac{d W_{ab}}{dt}= \frac{\partial W_{ab}}{\partial r_{ab}} \frac{d r_{ab}}
{dt} 
= \frac{\partial W_{ab}}{\partial r_{ab}}
\frac{(\vec{r}_a-\vec{r}_b)\cdot (\vec{v}_a-\vec{v}_b)}{r_{ab}}
= \frac{\partial W_{ab}}{\partial r_{ab}} \hat{e}_{ab} \cdot \vec{v}_{ab}
= \vec{v}_{ab} \cdot  \nabla_a W_{ab}. \label{eq:k5}
\ee
}}

But a slightly more sophisticated approach yields built-in conservation. 
If one starts from
\be
\nabla\left(\frac{P}{\rho}\right)
= \frac{\nabla P}{\rho} - P \frac{\nabla \rho}{\rho^2},
\label{eq:basic:nabla_p_rho}
\ee
solves for $\nabla P/\rho$ and applies the gradient formula 
(\ref{eq:basic:nabla_sum}), the momentum equation reads
\bea
\frac{d\vec{v}_a}{dt}&=& - \frac{P_a}{\rho_a^2} \sum_b m_b \nabla_a W_{ab} -
\sum_b \frac{m_b}{\rho_b} \frac{P_b}{\rho_b}  \nabla_a W_{ab}\nonumber\\
&=& - \sum_b m_b \left(\frac{P_a}{\rho_a^2} + \frac{P_b}{\rho_b^2} \right)
\nabla_a W_{ab}. 
\label{eq:basic:momentum_equation}
\eea
Because the pressure part of the equation is now manifestly symmetric in $a$
and $b$ and $\nabla_a W_{ab}=- \nabla_b W_{ba}$, the forces are now equal and
opposite and therefore total momentum and angular momentum (see below) are
conserved by construction. It is worth pointing out some issues:
\bi
\item So far, any complications resulting from variable
  smoothing lengths have been ignored. Conservation is guaranteed
  if  a) constant smoothing lengths are used (which is a bad idea in practice!) 
  or b) ``$\nabla_a W_{ab}$'' is symmetric in the smoothing lengths, which can be 
  achieved, for example, by using $h_{ab}= (h_a+h_b)/2$ \cite{benz90a} or by replacing 
  the gradient via $[\nabla_a W(r_{ab},h_a) + \nabla_a W(r_{ab},h_b)]/2$ \cite{hernquist89}.
\item The conservation relies on the force being a term
  symmetric in $a$ and $b$ times $\hat{e}_{ab}$ (keep in mind that $\nabla_a
  W_{ab}\propto \hat{e}_{ab}$, see Eq.~(\ref{eq:k4})).
\item SPH's success largely relies on its excellent conservation properties
  which is guaranteed by the correct symmetry in the particle indices.
\ei

\subsubsection{Energy equation}
A suitable energy equation can be constructed from 
Eq.~(\ref{eq:basic:Lag_energy_equation}) in a straight forward manner:
\be
\frac{du_a}{dt}= \frac{P_a}{\rho_a^2} \frac{d\rho_a}{dt} 
               = \frac{P_a}{\rho_a^2} \frac{d}{dt} \left(\sum_b m_b W_{ab}\right)
               = \frac{P_a}{\rho_a^2} \sum_b m_b \vec{v}_{ab} \cdot  \nabla_a W_{ab},
\label{eq:basic:du_dt_a}
\ee
where Eq.~(\ref{eq:k5}) was used.
Together with an equation of state, the equations
(\ref{eq:basic:sum_rho}), (\ref{eq:basic:momentum_equation}) and
(\ref{eq:basic:du_dt_a}) form a complete set of SPH equations.\\
For later use we also note that the velocity divergence can be conveniently
expressed by using Eqs.~(\ref{eq:basic:drho_dt}) and (\ref{eq:k5}) as
\be
(\nabla \cdot \vec{v})_a= - \frac{1}{\rho_a} \frac{d\rho_a}{dt}
= - \frac{1}{\rho_a} \sum_b m_b \vec{v}_{ab}\cdot  \nabla_a W_{ab}.
\label{eq:basic:divv}
\ee
One can derive an alternative energy equation, and this will allow a 
smoother transition to the relativistic equations, by using the 
specific ``thermokinetic'' energy $\hat{e}_a= u_a + \frac{1}{2}v_a^2$ 
instead of the specific thermal energy $u_a$. The corresponding 
continuous evolution equation, 
\be
\frac{d\hat{e}}{dt}= -\frac{1}{\rho}\nabla \cdot (P \vec{v}),
\ee
can  be written as
\be
\frac{d\hat{e}}{dt}= - \frac{P}{\rho^2}\nabla \cdot (\rho \vec{v}) - \vec{v} \cdot
\nabla\left( \frac{P}{\rho}\right).
\ee
Applying Eq.~(\ref{eq:basic:nabla_sum}) yields
\bea
\frac{d\hat{e}_a}{dt}&=& - \frac{P_a}{\rho_a^2} \sum_b \frac{m_b}{\rho_b} 
\rho_b \vec{v}_b \cdot \nabla_a W_{ab} - \vec{v}_a \cdot \sum_b \frac{m_b}{\rho_b}
\frac{P_b}{\rho_b} \nabla_a W_{ab} \nonumber\\
&=& - \sum_b m_b \left( \frac{P_a \vec{v}_b}{\rho_a^2} +
  \frac{P_b\vec{v}_a}{\rho_b^2} \right)\cdot  \nabla_a W_{ab}.
\label{eq:basic:energy_equation_e}
\eea
As we will see later, see Eq.~(\ref{eq:ener_eq_no_diss}), this is 
similar to the relativistic form of the energy equation.

\subsection{Conservation properties}
\label{page_conservation}
We can easily check the numerical conservation of the physically conserved 
quantities. The change of the total particle momentum is
\bea
\sum_a m_a \frac{d\vec{v}_a}{dt} &=& \sum_a \sum_b \vec{F}_{ba} 
= \frac{1}{2} \left(\sum_a \sum_b \vec{F}_{ba} + \sum_a \sum_b \vec{F}_{ba}
\right)\nonumber\\
&=& \frac{1}{2} \left(\sum_{a,b} \vec{F}_{ab} + \sum_{b,a} \vec{F}_{ba}
\right)
= \frac{1}{2} \left(\sum_{a,b} (\vec{F}_{ab} + \vec{F}_{ba})
\right) = 0
\eea
where the (dummy) summation indices were relabeled after the third equal sign
and the notation from Eq.~(\ref{eq:basic:F_ab})  and $\vec{F}_{ab} = - \vec{F}_{ba}$
were used. \\
The  proof for angular momentum is  analogous. The
torque on particle $a$ is 
\be
\vec{M}_a= \vec{r}_a \times \vec{F}_a= \vec{r}_a \times \left(m_a
\frac{d\vec{v}_a}{dt} \right)= \vec{r}_a \times \sum_b \vec{F}_{ba}
\ee
and thus 
\bea
\frac{d\vec{L}}{dt}&=& \sum_a \vec{M}_a= \sum_{a,b} \vec{r}_a \times
\vec{F}_{ba}= \frac{1}{2}\left(\sum_{a,b}\vec{r}_a \times \vec{F}_{ba} +
  \sum_{a,b}\vec{r}_a \times \vec{F}_{ba}\right)\nonumber\\
&=& \frac{1}{2}\left(\sum_{a,b}\vec{r}_a \times \vec{F}_{ba} +
  \sum_{b,a}\vec{r}_b \times \vec{F}_{ab}\right)
= \frac{1}{2}\left(\sum_{a,b}(\vec{r}_a-\vec{r}_b) \times \vec{F}_{ba}\right)
= 0.
\eea
Again, the dummy indices were relabeled and $\vec{F}_{ab} = - \vec{F}_{ba}$
was used. The expression finally
vanishes, because the forces between particles act along the line
joining them: $\vec{F}_{ab} \propto \nabla_a W_{ab} \propto \hat{e}_{ab}
\propto (\vec{r}_a-\vec{r}_b)$, see Eq.~({\ref{eq:k4}).\\
The total energy changes according to
\bea
\hspace*{-0.5cm}\frac{dE}{dt}=\frac{d}{dt} \sum_a \left(m_a u_a + \frac{1}{2} m_a v^2_a
\right) &=& \sum_a m_a \frac{d\hat{\epsilon}_a}{dt} =
\sum_a 
m_a  \left( \frac{du_a}{dt} + \vec{v}_a \cdot \frac{d\vec{v}_a}{dt} \right).
\eea
Inserting Eqs.~(\ref{eq:basic:du_dt_a}) and
(\ref{eq:basic:momentum_equation}) we have
\bea
\frac{dE}{dt}= & & \sum_a m_a \left[ \frac{P_a}{\rho_a^2} \sum_b m_b
  \vec{v}_{ab}\cdot \nabla_a W_{ab} - \vec{v}_a \cdot \sum_b m_b 
\left(\frac{P_a}{\rho_a^2} + \frac{P_b}{\rho_b^2} \right)\nabla_a W_{ab}\right]\nonumber\\
= & & \sum_{a,b} m_a m_b \frac{P_a}{\rho_a^2} \vec{v}_a \cdot\nabla_a W_{ab} 
- \sum_{a,b} m_a m_b \frac{P_a}{\rho_a^2} \vec{v}_b\cdot \nabla_a W_{ab}\nonumber\\
&& - \sum_{a,b} m_a m_b \frac{P_a}{\rho_a^2} \vec{v}_a \cdot \nabla_a W_{ab}
    - \sum_{a,b} m_a m_b \frac{P_b}{\rho_b^2} \vec{v}_a \cdot\nabla_a
    W_{ab}\nonumber\\ 
= & & - \sum_{a,b} m_a m_b \left(\frac{P_a \vec{v}_b}{\rho_a^2} + \frac{P_b
    \vec{v}_a}{\rho_b^2} \right)\cdot \nabla_a W_{ab}.
\label{eq:basic:dE_tot_dt}
\eea
From this equation we could have directly read off the evolution equation for
the specific thermokinetic energy, Eq.~(\ref{eq:basic:energy_equation_e}). To
show the conservation of the total energy we can apply the same
procedure as in the case of the momentum equation since we have again a double
sum over a quantity symmetric in the indices $a, b$ multiplied by a quantity
which is anti-symmetric in the indices $a, b$. Therefore, the double sum
vanishes and total energy is conserved.

\subsection{Alternative SPH discretizations}
\label{page_alternative}
\label{sec:alternative}
Like other numerical schemes, SPH offers some freedom in discretizing 
the continuum equations. At this point it is worth remembering that 
we had enforced momentum conservation by using a particular form of the 
gradient on the RHS of the momentum equation, see Eq.~(\ref{eq:basic:nabla_p_rho}).
Following the same reasoning one could as well have used \cite{monaghan92}
\be
\frac{\nabla P}{\rho}= \frac{P}{\rho^\lambda} \nabla\left(\frac{1}{\rho^{1-\lambda}} \right)
+ \frac{1}{\rho^{2-\lambda}} \nabla \left(\frac{P}{\rho^{\lambda-1}} \right)
\ee
which yields
\be
\frac{d\vec{v}_a}{dt}= - \sum_b m_b \left( 
\frac{P_a}{\rho_a^{\lambda} \rho_b^{2-\lambda}}+
 \frac{P_b}{\rho_a^{2-\lambda} \rho_b^\lambda} 
\right) \nabla_a W_{ab}
\ee
for the momentum equation. This form is symmetric in the particle indices for any value of
$\lambda$ and therefore also conserves momentum by construction. This equation is consistent
with the energy equation
\be
\frac{du_a}{dt}= \frac{P_a}{\rho_a^\lambda} \sum_b m_b \frac{\vec{v}_{ab}}{\rho_b^{2-\lambda}} \cdot \nabla_a W_{ab}
\ee
and the continuity equation
\be
\frac{d\rho_a}{dt}= \rho_a^{2-\lambda} \sum_b m_b \frac{\vec{v}_{ab}}{\rho_b^{2-\lambda}}\cdot \nabla_a W_{ab},
\ee
which is the SPH-discrete form of
\be
\frac{d\rho_a}{dt}= \rho^{2-\lambda} \left[\vec{v}\cdot\nabla(\rho^{\lambda-1}) - \nabla\cdot(\vec{v} \rho^{\lambda-1}) \right].
\ee
The consistency of the above SPH equation set can be shown by using a generalized variational principle 
\cite{price04c} that does not assume that the density is given as a function of the coordinates, but instead assumes a given 
form of the continuity equation.\\
If $\lambda=1$, the continuity equation only depends on the particle volumes $m_b/\rho_b$ rather than
the particle masses. This can reduce numerical noise in regions of large density contrasts where particles of very 
different masses interact \cite{ritchie01}. Marri and White \cite{marri03} found improved results in their galaxy formation
calculations by using (among other modifications) a value of $\lambda=3/2$.\\
The SPH equations can be generalized even further after noting that the continuity equation can be written as \cite{price04c}
\be
\frac{d\rho}{dt}= \Psi \left\{\vec{v}\cdot\nabla\left(\frac{\rho}{\Psi}\right) - 
\nabla\cdot\left(\frac{\rho \vec{v}}{\Psi} \right)\right\},
\ee
with $\Psi$ being a scalar variable defined on the particle field. This leads to SPH equations of the form
\bea
\frac{d\rho_a}{dt}&=& \Psi_a \sum_b m_b \frac{\vec{v}_{ab}}{\Psi_b} \cdot \nabla_a W_{ab}\\
\frac{d\vec{v}_a}{dt}&=& - \sum_b m_b \left( \frac{P_a}{\rho_a^2}\frac{\Psi_a}{\Psi_b} + \frac{P_b}{\rho_b^2}\frac{\Psi_b}{\Psi_a} \right) \nabla_a W_{ab}\\
\frac{du_a}{dt}&=& \frac{P_a}{\rho_a^2} \sum_b m_b \frac{\Psi_a}{\Psi_b} \vec{v}_{ab} \cdot \nabla_a W_{ab}.
\eea
The ``vanilla ice'' equation set, Eqs.~(\ref{eq:basic:momentum_equation}) and (\ref{eq:basic:du_dt_a}), is recovered
for $\Psi=1$, choosing $\Psi= \rho/\sqrt{P}$ one recovers the momentum equation of Hernquist and Katz \cite{hernquist89}.\\
The above equation set has been used in recent work \cite{read09} to address SPH's weakness of resolving Kelvin-Helmholtz 
instabilities across large density jumps. The authors choose a different auxiliary function for each of the above equations, 
$\Psi^{\rho}$, $\Psi^{v}$, $\Psi^{u}$, and perform an error and stability analysis leading them 
to suggest $\Psi^{\rho} = \Psi^{u}= 1/A^{\Gamma}$ and $\Psi^{v}= \rho$. Here $A$ is the entropy-dependent part of the polytropic
equation of state, $P= A(s) \rho^{\Gamma}$ and $\Gamma$ the polytropic exponent. This prescription leads to a sharper density
transition that is more consistent with the entropy transitions between both parts of the fluid. Together with a 
higher-order kernel and large neighbor numbers they find convincing results in Kelvin-Helmholtz instability simulations.

\subsection{Adaptive resolution}
\label{page_adaptive_res}
Whenever densities and length scales vary by large amounts, the smoothing length
$h$ should be adapted in space and time. Several ways to adjust the smoothing lengths 
have been used over the years
\cite{gingold77,gingold78,gingold82,hernquist89,benz90a,steinmetz93}. One 
of them seeks to keep the number of neighbors of each particle (approximately) 
constant \cite{hernquist89}, another to integrate an additional equation that
makes use of the continuity equation \cite{benz90a}. One can use the ansatz
\be
\frac{h(t)}{h_0}= \left(\frac{\rho_0}{\rho(t)}\right)^{1/3},
\ee 
where the index $0$ labels the quantities at the beginning of the
simulation. Taking Lagrangian time derivatives of both sides 
together with Eq.~(\ref{eq:basic:drho_dt}) yields
\be
\frac{d h_a}{dt}= \frac{1}{3} h_a (\nabla \cdot \vec{v})_a.
\ee
Another convenient way is to evolve $h$ according to
\be
h_a= \eta \left(\frac{m_a}{\rho_a}\right)^{1/3},
\label{eq:basic:h_adaption}
\ee
where $\eta$ should be chosen in the range between 1.2 and 1.5 
\cite{gingold82,price04c,monaghan05}.\\
Note, that the above SPH equations were derived under the assumption
that the smoothing lengths are constant. Varying the smoothing lengths 
while using the above equation set is strictly speaking inconsistent.
A consistent formulation that involves extra-terms will be discussed in 
detail in Sec.~\ref{advanved:sec:grad_h}. The importance of these corrective,
or ``grad-h'' terms is problem- and resolution-dependent \cite{springel02,rosswog07c}.

\subsection{Artificial dissipation}
\label{subsec:artificial_dissi}
\label{page_AV}
\subsubsection{Purpose and general reasoning}
In gas dynamics, even perfectly smooth initial conditions can steepen into discontinuous solutions, 
or ``shocks'', see e.g. \cite{landau59,whitham74,shu92}, which are nearly omnipresent in astrophysics. 
On length scales comparable to the gas mean free path these solutions are smooth (as a result of the
{\em physical viscosity} that is always present to some extent), but on the macroscopic scales of a 
simulation (which is usually orders of magnitude larger), the very steep gradients appear as discontinuous.\\
The strategies to deal numerically with shocks are (broadly speaking) two-fold: i) one can either
make use of the analytical solution of a Riemann problem between two adjacent computational entities
(either cells or particles) or ii) broaden the discontinuity to a numerically resolvable length 
across which gradients can be calculated. The latter is instantiated by adding extra, ``artificial''
dissipation to the flow. This can be achieved by choosing particular numerical 
discretization schemes that in fact solve continuum equations that are only similar to the original ones,
but contain additional, higher-order deviatoric terms, such as, for example, the Lax scheme \cite{press92}. 
Alternatively, one may explicitely add pressure-like terms to the fluid equations. The addition of such 
``ad-hoc'' terms is one of the oldest techniques \cite{vonneumann50} in the relatively young field of
computational physics. Most astrophysical SPH implementations follow the latter strategy, e.g. 
\cite{hernquist89,wadsley04,springel05a,rosswog07c}, but variants that use 
Riemann-solvers do also exist \cite{inutsuka02,cha03}.\\
Strictly speaking, real discontinuities with their infinite derivatives are no proper
solutions of the ideal hydrodynamics equations Eqs.~(\ref{eq:basic:drho_dt})-(\ref{eq:basic:Lag_energy_equation}).
Artificial viscosity (AV) aims at spreading such discontinuities over a numerically resolvable length.
In the words of von Neumann and Richtmyer (from their seminal paper \cite{vonneumann50}), the ``idea
is to introduce (artificial) dissipative terms into the equations so as to give the shocks a thickness
comparable to (but preferentially larger than) the spacing ... [of the grid points]. Then the differential
equations (more accurately, the corresponding difference equations) may be used for the entire calculation,
just as though there were no shocks at all.''
Although guidance by physical viscosity can sometimes be helpful, artificial viscosity is not meant
to mimic physical viscosity, it is instead an {\em ad hoc} method to produce on a resolvable scale
what is the result of unresolvable, small-scale effects. In this sense one can
think of it as a kind of sub-grid model.\\
AV should have a number of desirable properties \cite{caramana98}
and must not introduce unphysical artifacts. It should always be dissipative, i.e. transfer kinetic energy into
internal one and never vice versa, which is not a completely trivial task in 3D. AV should be absent in rigid
and (shockless) differential rotation and in uniform compression. In other words, it should be ``intelligent'' in 
the sense that it distinguishes uniform compression from a shock. This can be done either via a tensor 
formulation of artificial viscosity, e.g. \cite{owen04}, or via ``limiters'' that detect such motion and subsequently 
suppress the action of AV. Generally, AV should go smoothly to zero as the compression vanishes and it should be 
absent for expansion. Moreover, to not destroy one of SPH's most salient strength, it must be implemented consistently 
into momentum and energy equation, so that the conservation of energy, momentum and angular momentum is still guaranteed.

\subsubsection{Bulk and von-Neumann-Richtmyer viscosity}
Guided by the requirements that i) no real discontinuities occur, ii) the thickness of the ``shock layer''
is everywhere of the order of the resolvable length scale $l$, iii) no noticeable effects occur away
from shock layers and iv) the Rankine-Hugoniot relations \cite{landau59} hold when considering length scales
that are large in comparison to the thickness of the shock layers, von Neumann and Richtmyer suggested an 
artificial pressure of the form
\be
q_{\rm NR}= c_2 \rho l^2 (\nabla \cdot \vec{v})^2,
\ee
where $c_2$ is a dimensionless parameter of order unity, to be added to the hydrodynamic pressure. Their approach
yielded reasonably good results in strong shocks, yet it still allowed oscillations in the post-shock region to occur.
To avoid them, usually an additional term \cite{landshoff55} is introduced that vanishes less rapidly  and has the 
form of a bulk viscosity \cite{landau59}
\be
q_{\rm b}= - c_1 \rho c_{\rm s} l (\nabla \cdot \vec{v}).
\ee
Here $c_{\rm s}$ is the sound velocity and $c_1$ is a parameter of order unity. Often, the two contributions are combined
into an artificial viscous pressure,
\be
q_{\rm visc} = - c_1 \rho c_{\rm s} l (\nabla \cdot \vec{v}) + c_2 \rho l^2 (\nabla \cdot \vec{v})^2.
\label{eq:basic:qvisc}
\ee
The term $l \; (\nabla \cdot \vec{v})$ is thereby an estimate for the velocity jump between adjacent cells or 
particles.

\subsubsection{"Standard" SPH viscosity: reasoning and limitations}
A straight-forward approach to transform these ideas into an SPH artificial viscosity prescription would be to insert
Eq.~(\ref{eq:basic:divv}) into (\ref{eq:basic:qvisc}), to use particle properties for the densities and sound 
velocities and identify the resolution length scale $l$ with the smoothing length. Similar approaches have indeed
been pursued \cite{hernquist89}, but we will focus here on an approach due to \cite{monaghan83} that is probably the
most wide-spread form of SPH artificial viscosity and therefore often referred to as the ``standard viscosity''.\\
Instead of just increasing the hydrodynamic pressures, $P_a$, by viscous contributions, $q_{\rm visc}$, one augments 
the pressure terms in the momentum equation (\ref{eq:basic:momentum_equation}) by an artificial contribution $\Pi_{ab}$:
\be
\left(\frac{P_a}{\rho_a^2} + \frac{P_b}{\rho_b^2} \right) \rightarrow 
\left(\frac{P_a}{\rho_a^2} + \frac{P_b}{\rho_b^2} + \Pi_{ab}\right).
\ee
For simplicity, let us consider the bulk viscosity 
contribution first and restrict ourselves to 1D. The bulk contribution to $\Pi_{ab}$ is then of the form 
$- c_1 c_{\rm s} (h/\rho) \; \p v/\p x$ and a Taylor expansion of the velocity field, $v(x_a + (x_b-x_a))$, yields
\be
\left(\frac{\p v}{\p x} \right)_a= \frac{v_b-v_a}{x_b-x_a} + O \left((x_b-x_a)^2 \right).
\ee
Inserting this approximation, replacing particle properties by their averages, $\bar{A}_{ab}= (A_a+A_b)/2$, using the 
notation $x_{ab}= x_a-x_b$ and $v_{ab}=v_a-v_b$ and replacing the possibly diverging denominator, 
\be
\frac{1}{x_{ab}} \rightarrow \frac{x_{ab}}{x_{ab}^2 + \epsilon \bar{h}_{ab}^2},
\ee
the SPH prescription for the bulk viscosity reads
\be
\Pi_{ab, {\rm bulk}} = \left\{\begin{array}{cl}
           - c_1 \frac{\bar{c}_{s,ab}}{\bar{\rho}_{ab}} \mu_{ab}  \; \; {\rm for} \; \;
           x_{ab} \; v_{ab} < 0\\  
           0 \; \quad \quad \quad  {\rm otherwise}\\
           \end{array}\right. , {\rm where} \quad
         \mu_{ab}= \frac{\bar{h}_{ab} \; x_{ab} \; v_{ab}}{x_{ab}^2 + \epsilon \bar{h}_{ab}^2}. 
\label{eq:basic:PI_AV}
\ee
The product $x_{ab} \; v_{ab}$ thereby detects whether particles are approaching ($<0$) and only in this 
case artificial viscosity becomes active. The quantity $\mu_{ab}$ has taken over the role of the 
term $l (\nabla \cdot \vec{v})$ in the original prescription of von Neumann and Richtmyer. Having said 
this, we can now treat the von Neumann-Richtmyer term, $q_{\rm NR}$, in exactly the same way. Adapting to 
the usual SPH notation, $c_1 \rightarrow \alpha$,  $c_2 \rightarrow \beta$, the artificial viscosity term
reads
\be
\Pi_{ab} = \Pi_{ab,{\rm bulk}} + \Pi_{ab,{\rm NR}} = \left\{\begin{array}{cl}
           \frac{- \alpha \bar{c}_{ab} \mu_{ab} + \beta
             \mu_{ab}^2}{\bar{\rho}_{ab}} \quad{\rm for} \; \;
           \vec{r}_{ab} \cdot \vec{v}_{ab} < 0\\  
           0 \quad \quad \quad \quad  {\rm otherwise}\\
           \end{array}\right. ,
\label{eq:basic:PI_AV}
\ee
with 
\be
\mu_{ab}= \frac{\bar{h}_{ab}\vec{r}_{ab} \cdot \vec{v}_{ab}}{r_{ab}^2 + \epsilon \bar{h}_{ab}^2}.
\label{eq:basic:mu_ab}
\ee
Note that the scalar quantities have now been replaced by vector quantities for the use in 3D. 
Numerical experiments suggest the following values for the involved parameters:
$\alpha \approx 1$, $\beta \approx 2$ and $\epsilon \approx 0.01$.\\
Accounting for artificial viscosity, the momentum equation reads
\be
\frac{d\vec{v}_a}{dt} = - \sum_b m_b \left(\frac{P_a}{\rho_a^2} +
  \frac{P_b}{\rho_b^2} + \Pi_{ab} \right)
\nabla_a W_{ab}. 
\label{eq:basic:momentum_equation_AV}
\ee
To have a consistent formulation, the energy equation must be modified according to
\be
\frac{du_a}{dt}= \frac{P_a}{\rho_a^2} \sum_b m_b \vec{v}_{ab} \cdot \nabla_a W_{ab}
+ \frac{1}{2} \sum_b m_b \Pi_{ab} \vec{v}_{ab} \cdot \nabla_a W_{ab}.
\label{eq:basic:energy_equation_u_AV}
\ee
This combination still conserves energy, linear and angular momentum by construction, the
proofs are analogous to the ones outlined above.\\
To illustrate the performance of this artificial viscosity prescription, let us consider a 
standard test case for hydrodynamic schemes, the so-called ``Sod shocktube'' \cite{sod78} 
which is a particular realization of a Riemann problem. A tube filled with gas, such as illustrated in 
Fig.~\ref{fig:basic:shock_tube},
\begin{figure}
\centerline{\includegraphics[width=3in]{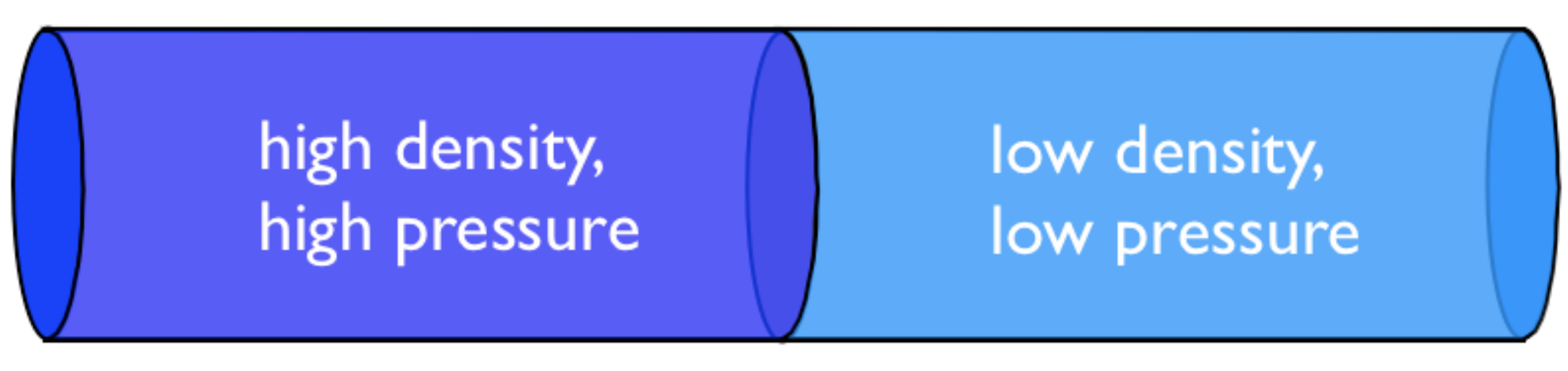}}
\caption{Schematic illustration of the shock tube problem.}
\label{fig:basic:shock_tube}
\end{figure} 
\begin{figure}[t]
\centerline{\includegraphics[width=5in]{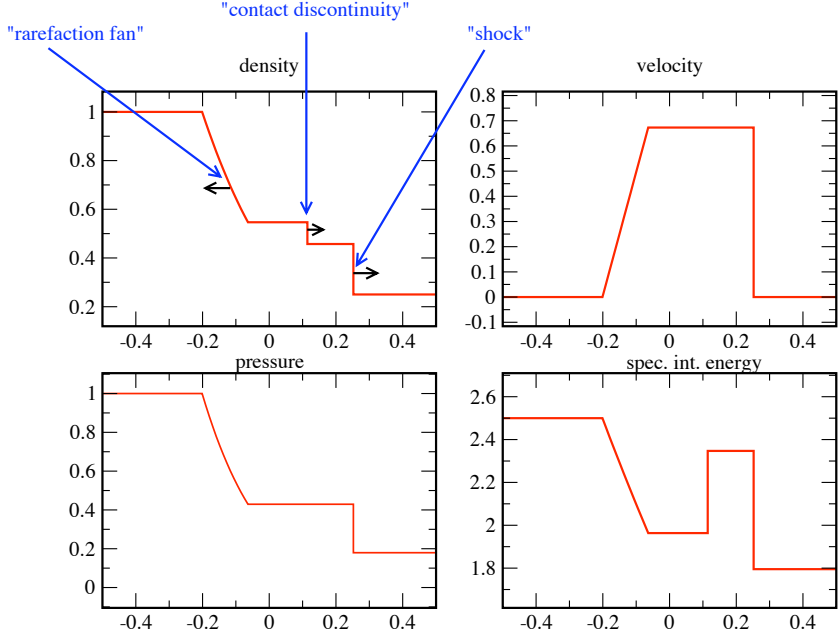}}
\caption{The exact solution of the non-relativistic shock tube problem.}
\label{fig:basic:shock_tube_exact}
\end{figure}
contains a high density ($\rho_1=1$)/high pressure ($p_1=1$) region that is separated by
a wall from a low density ($\rho_2=0.25$)/low pressure ($p_2=0.1795$) region. 
For the Sod test a polytropic equation of state with exponent $\Gamma=1.4$ is used.
Once the wall is removed, the
discontinuity decays into two elementary, non-linear waves that move in
opposite directions: a shock moves into the unperturbed original low density
region, while a rarefaction wave travels into the original high-density 
region. Between the shock and the tail of the rarefaction wave (the rightmost
end of the rarefaction) two new states develop which are separated by a
contact discontinuity across which the pressure is constant. The exact solution
is sketched in Fig.~\ref{fig:basic:shock_tube_exact}.\\
If {\em no artificial viscosity} is applied in such a problem, strong post-shock 
oscillations occur as illustrated in Fig.~\ref{fig:basic:post_shock_osc} for 
the velocities in the Sod test.
If the standard artificial viscosity prescription, Eq.~(\ref{eq:basic:PI_AV}) is used,
the numerical solution agrees very well with the exact one, see Fig.~\ref{fig:basic:non_rel_Sod},
provided that the interesting regions of the flow are adequately resolved\footnote{Consistent
with the finite resolution width of numerical shocks, we set up the initial density distribution
as a Fermi function with a transition width of 
$\Delta x$: $\rho(x)= (\rho_1-\rho_2)/\left(1+\exp(\frac{x-x_S}{\Delta x})\right) + \rho_2$, where $x_S$ is
the initial position of the shock. For the presented test $\Delta x$ was set to 1.5 times the
average particle separation.}.\\
\begin{figure}
\centerline{\includegraphics[width=3in]{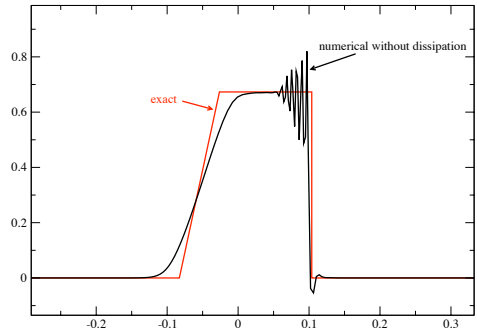}}
\caption{Velocity snapshot of the Sod shock tube problem. Shown is the 
exact solution together with a numerical solution {\em without} artificial
viscosity.}\label{fig:basic:post_shock_osc}
\end{figure}
\begin{figure}[t]
\centerline{\includegraphics[width=5in]{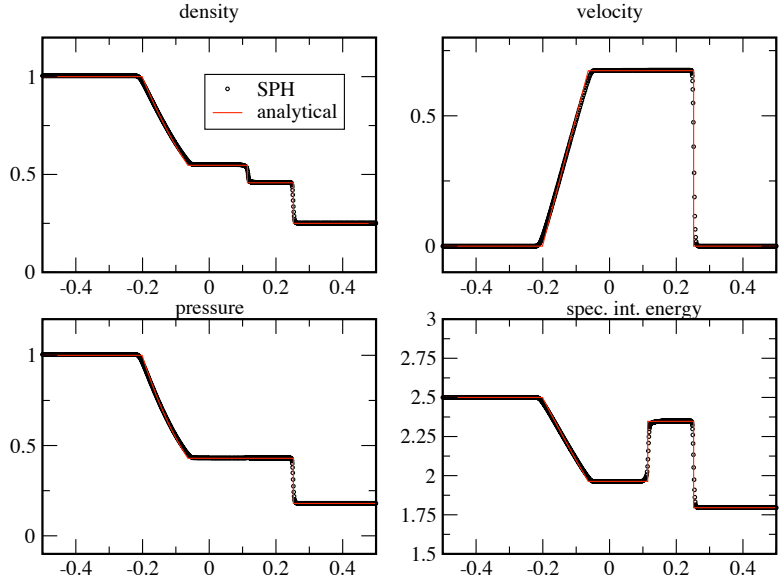}}
\caption{Sod's shock tube: the exact solutions are shown by the red line, the
  numerical solution obtained with SPH are shown as circles. For this
  simulation 1000 particles were used, density was calculated via summation, 
and the smoothing lengths were updated according to $h_a= 1.4 (m_a / \rho_a)$.
For this test a second-order Runge-Kutta integrator was used.}
\label{fig:basic:non_rel_Sod}
\end{figure}
While this artificial viscosity form performs very well in 1D, it suffers from
deficiencies if used in multi-D: i) for fixed parameters $\alpha$ and $\beta$
it may affect the flow even if it is not really needed and ii) under certain conditions
it can introduce spurious shear forces. Consider for example an idealized shear flow
as sketched in Fig.~\ref{fig:basic:shear}. Assume that the velocity decreases vertically
as sketched on the left and no shocks occur. For such a situation no artificial viscosity 
is needed. Nevertheless, as sketched for two example particles (``1'' and ``2'') the scalar
product $\vec{r}_{ab} \cdot \vec{v}_{ab}$ is finite and thus, via $\mu_{ab}$ in
Eq.~(\ref{eq:basic:mu_ab}), introduces a viscous force that is unwanted. Similar 
configurations are encountered for the (astrophysically important) cases of
accretion disks.

\subsubsection{Reducing artificial viscosity where unnecessary}
Several recipes were suggested to cure the deficiencies that the standard
artificial viscosity introduces in multi-D.\\ 
\begin{figure}[t]
\centerline{\includegraphics[width=3in]{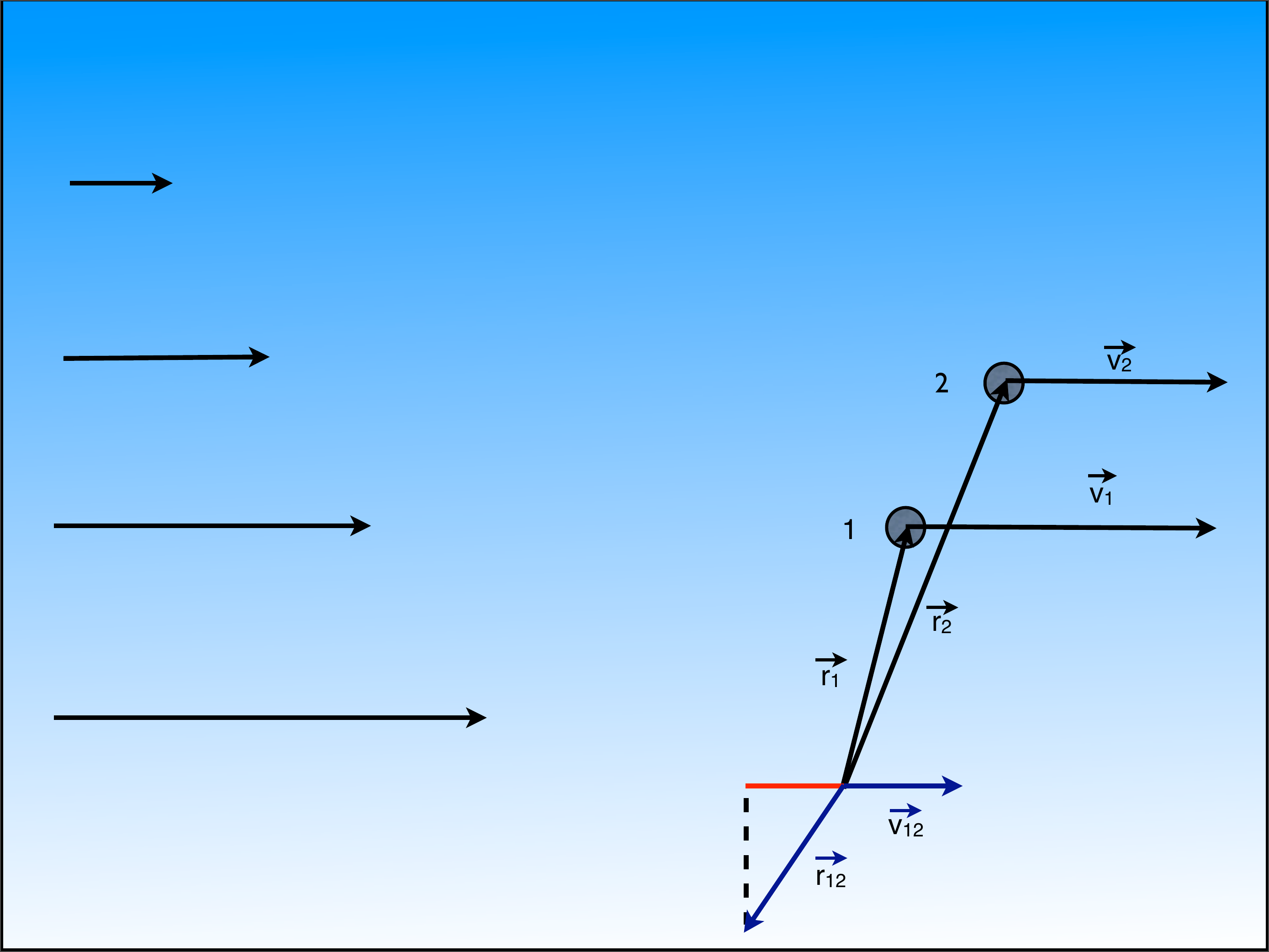}}
\caption{Illustration how the ``standard'' SPH artificial viscosity introduces spurious 
shear forces: in this pure shear flow the difference position vector has a finite projection
on the difference velocity vector (red) and thus introduces unwanted forces.}
\label{fig:basic:shear}
\end{figure}
One such recipe is the so-called {\em "Balsara-switch"} \cite{balsara91}. Its strategy
is to define a ``limiter'' that distinguishes shock from shear motion and suppresses AV
in the latter case. One defines a quantity
\be
f_a= \frac{|\langle \nabla \cdot \vec{v}\rangle_a|}{|\langle \nabla \cdot \vec{v}\rangle_a| 
+ |\langle \nabla \times \vec{v}\rangle_a| + 0.0001 c_{s,a}/h_a}
\ee
and, as before, applies a symmetrized average of the limiters $f$, $\Pi_{ab}'= \Pi_{ab} \; \bar{f}_{ab}$. 
In pure compressional motion ($|\langle \nabla \cdot \vec{v}\rangle| \ne 0$
and $|\langle \nabla \times \vec{v}\rangle|=0$) the limiter reduces to unity and the standard viscosity is recovered,
and in pure shear flow  ($|\langle \nabla \cdot \vec{v}\rangle|=0$ and $|\langle \nabla \times \vec{v}\rangle|\ne 0$) the
action of artificial viscosity is suppressed ($\bar{f}_{ab} \ll 1$). This limiter has been found very useful in many cases
\cite{steinmetz96,navarro97,rosswog00}, but it reaches its limitations if shocks occur in a shearing environment such as in an
accretion disk \cite{owen04}.\\
Morris and Monaghan \cite{morris97} suggested the use of {\em time-dependent artificial viscosity parameters}. The main
idea is to have $\alpha$ and $\beta$ only at non-negligible levels where they are really needed. They suggested to fix
 $\beta$ to $2 \alpha$, to assign each particle its individual parameter, $\alpha_a$, and to evolve it according to an 
additional differential equation,
\be
\frac{d \alpha_{a}}{dt}= - \frac{\alpha_{a}- \alpha_{\rm min}}{\tau_{a}} +
S_{a}, 
\label{eq:basic:alpha_t}
\ee
so that $\alpha_a$ decays exponentially with an e-folding time $\tau_a$ to a minimum value $\alpha_{\rm min}$, 
unless it is triggered to rise by the source term $S_a$. The time scale $\tau_a$ should be chosen so that
the viscosity persists for a few smoothing lengths behind the shock. This can be obtained by
\be
\tau_a= \frac{h_a}{\xi c_{s,a}},
\ee
where $\xi\approx 0.1$\footnote{Note the similarity to the SPH-version of the Courant time step criterion.}.
Morris and Monaghan suggested $S_a= {\rm max}\left[-( \vec{\nabla} \cdot \vec{v})_{a},0\right]$ for the source term.
If one wishes to restrict the growth of $\alpha$ to $\alpha_{\rm max}$, one can use 
 $S_a= {\rm max}\left[-( \vec{\nabla} \cdot \vec{v})_{a} (\alpha_{\rm max}-\alpha_{a}),0\right]$ \cite{rosswog00}. 
Typical values for the numerical parameters are $\alpha_{\rm min}= 0.1$ and $\alpha_{\rm max}=1.5$.
In a Sod test
this prescription leads to non-negligible values of $\alpha_a$ only in the vicinity of the shock front, elsewhere
artificial viscosity is practically absent. This prescription has largely removed unwanted effects from AV in
three-dimensional SPH simulations \cite{rosswog00,dolag05}. Note, however, that in homologous flow, 
$\vec{v} \propto \vec{r}$, $\alpha_a$ can still rise although it is not needed. This is counterbalanced
to some extent by the increase of $c_{s,a}$ and the decrease of $h_a$, so that -via $\tau_a$- the decay term in 
Eq.~(\ref{eq:basic:alpha_t}) becomes more dominant, but there is certainly still room for improvement.

\subsubsection{New forms inspired by Riemann solvers}
Monaghan used the analogy to Riemann solvers to motivate a new form of artificial dissipation
which involves signal velocities and jumps in variables across characteristics \cite{monaghan97}.
The main idea of these ``discontinuity capturing terms'' is that for any conserved scalar 
variable $A$ with $\sum_a m_a dA_a/dt=0$ a dissipative term of the form
\be
\left( \frac{dA_a}{dt}\right)_{\rm diss}= \sum_b m_b \frac{\alpha_{A,b} v_{\rm sig}}{\bar{\rho}_{ab}} 
(A_a-A_b)\hat{e}_{ab} \cdot \nabla_a W_{ab}
\ee
should be added, where the parameter $\alpha_{A,b}$ determines the exact amount of dissipation and $v_{\rm sig}$ 
is the maximum signal velocity between particle $a$ and $b$. Applied to velocity and thermokinetic energy
this yields
\bea
\left( \frac{d \vec{v}_a}{dt} \right)_{\rm diss}&=& \sum_b m_b \frac{\alpha v_{\rm sig}(\vec{v}_a-\vec{v}_b)\cdot 
\hat{e}_{ab}}{\bar{\rho}_{ab}} \nabla_a W_{ab}\label{basic:eq:v_diss}\\
\left( \frac{d \hat{e}_a}{dt} \right)_{\rm diss}&=& \sum_b m_b \frac{e^\ast_a - e^\ast_b}{\bar{\rho}_{ab}} 
\hat{e}_{ab}\cdot \nabla_a W_{ab},\label{basic:eq:e_diss}
\eea
where, following \cite{price08a}, the energy including velocity components along the line of sight between particles $a$ and $b$,
$e^\ast_a= \frac{1}{2} \alpha v_{\rm sig} (\vec{v}_a\cdot\hat{e}_{ab})^2 + \alpha_u v_{\rm sig}^u u_a$, has been used and
different signal velocities and dissipation parameters were explicitely allowed for. If one uses 
$du_a/dt= d\hat{e}_a/dt -\vec{v}_a \cdot d\vec{v}_a/dt$, one finds the dissipative terms of the thermal energy
equation
\be
\left( \frac{du_a}{dt} \right)_{\rm diss}= - \sum_b \frac{m_b}{\bar{\rho}_{ab}} \left[ \alpha v_{\rm sig} \;  
\frac{1}{2} (\vec{v}_{ab}\cdot\hat{e}_{ab})^2 + \alpha_u v^u_{\rm sig} (u_a-u_b) \right] \hat{e}_{ab} \cdot \nabla_a W_{ab}.
\ee
The first term in this equation bears similarities with the ``standard'' artificial viscosity prescription, see
Eq.~(\ref{eq:basic:energy_equation_u_AV}), the second one expresses the exchange of thermal energy between particles
and therefore represents an artificial thermal conductivity which smoothes discontinuities in the specific energy. Such 
artificial conductivity had been suggested earlier to cure the so-called ``wall heating problem''\cite{noh87}. 
Tests have shown that artificial conductivity substantially improves SPH's performance
in simulating Sedov blast waves \cite{rosswog07c}.\\
For non-relativistic hydrodynamics the maximum signal velocity between two particles can be estimated as \cite{monaghan97}
\be
v_{\rm sig}= c_{{\rm s},a} + c_{{\rm s},b} - \vec{v}_{ab}\cdot\hat{e}_{ab},
\ee
where $c_{{\rm s},k}$ is the sound velocity of particle $k$.
Price \cite{price08a} had realized that SPH's difficulty to treat Kelvin-Helmholtz instabilities across contact
discontinuities with large density jumps 
\cite{agertz07} is closely related to a ``blip'' that occurs in the pressure at the contact discontinuities\footnote{It 
is not visible in our Sod shock tube in Fig.~\ref{fig:basic:non_rel_Sod} since we had started from smoothed initial 
conditions.}. He suggested to use artificial conductivity only to eliminate spurious pressure gradients across contact
discontinuities and to this end suggested
\be
v^u_{\rm sig}= \sqrt{\frac{|P_a-P_b|}{\bar{\rho}_{ab}}}.
\ee
Clearly, this quantity has the dimensions of a velocity and vanishes in pressure equilibrium. This approach has substantially 
improved SPH's ability to treat Kelvin-Helmholtz instabilities \cite{price08a}.\\
To avoid conductivity where it is unwanted, one can follow again a strategy  with time-dependent parameters.
For the artificial viscosity one can use Eq.~(\ref{eq:basic:alpha_t}}), and proceed in a similar way for $\alpha_u$.
One can use the second derivative of the thermal energy,
\be 
S_{u,a}= \frac{h_a |\nabla^2 u|_a} {\sqrt{u_a+\epsilon}},
\ee
to control the growth $\alpha_u$. The second derivative can be calculated as in Eq.~(\ref{eq:basic:nabla2_sum}), 
the parameter $\epsilon$ avoids that $S_{u,a}$ diverges as $u_a \rightarrow 0$.\\
According to a recent analysis \cite{read09}, SPH's difficulty to treat Kelvin-Helmholtz instabilities results
from a mismatch in the sharpness of pressure and density across the density jump. This can be either cured by
generating entropy at the boundary and thus smoothing the pressure as in \cite{price08a}, or by obtaining
a sharper density estimate. By a combination of using the freedom in discretization, see Sec.~\ref{sec:alternative},
a particular, higher-order kernel and an entropy-weighted density estimate together with large neighbor
numbers, \cite{read09} also find convincing results Kelvin-Helmholtz instability simulations.

\subsection{Time integration in SPH}
\label{page_integ}
To integrate the ordinary differential equations of SPH one has to find a 
reasonable tradeoff between accuracy and efficiency of an integrator and 
most often the available computer power is better invested in larger 
particle numbers rather than in high-order integration schemes. Since the 
evaluation of derivatives is usually very expensive, and in particular so 
for self-gravitating fluids, one tries to minimize the number of force 
evaluations per time step, which gives preference to low-order integrators. 
If the storage of derivatives (from earlier time steps) is not a concern, 
one can resort to multi-step methods such as Adams-Bashforth-type 
integrators, e.g. \cite{press92,burden01}, giving higher-order time integration
accuracy at moderate costs of force evaluations.\\
After a short collection of commonly used time step criteria, we want to discuss briefly two 
methods that we find particularly useful: the St\"ormer-Verlet/leap frog algorithm,
which appeals by its conservation properties and the class of Fehlberg methods whose
advantage is the appropriate choice of the time step size based on monitoring 
the quality of the numerical solution.\\

\subsubsection{Time stepping}
We will briefly collect 
here commonly used time step criteria, depending on the considered physical system,
further criteria that capture the specific physical time scales have to be added.
A criterion that triggers on accelerations, $\vec{a}_a$, is \cite{monaghan92}
\be
\Delta t_{f,a} \propto \sqrt{h_a/|\vec{a}_a|},
\ee 
and 
\be
\Delta t_{CV,a} \propto \frac{h_a}{v_{{\rm s},a} + 0.6(c_{{\rm s},a}+2 \; {\rm max}_b
  \; \mu_{ab})}
\ee 
is a combination of a Courant-type\footnote{The Courant or Courant-Friedrichs-Levi 
(or CFL for short) criterion ensures that the numerical propagation speed of information
does not exceed the physical one. If a spatial scale of $\Delta x$ can be resolved, the numerical
time step has to be $\Delta t < \Delta x/c_{\rm s}$ to ensure numerical stability\cite{press92}.} 
and viscous time step control, where $\mu_{ab}$ is the quantity defined in
Eq.~(\ref{eq:basic:mu_ab}).
If one wishes to restrict the change of the smoothing length over a time step, one can use
additionally\cite{wetzstein08}
\be
\Delta t_{h,a} \propto \frac{h_a}{\dot{h}_a}.
\ee
In a simulation, the minimum of the different time step criteria (with a suitably chosen prefactor) 
determines the hydrodynamical time step. If further physical processes, say nuclear burning, occur 
on much shorter time scales, one may resort to an
``operator splitting'' approach and integrate different processes with separate integration schemes,
e.g.\cite{rosswog09a}.\\
For inexpensive test problems or  cases where the physical time 
scales are (more or less) the same for each particle, it is easiest to evolve 
all the particles on the same (shortest particle) time step. In many 
astrophysical examples, say cosmological structure formation, the collapse of 
a molecular cloud or the tidal disruption of a star by a black hole, the 
required time steps in different parts of the fluid may span many orders of 
magnitude. In such cases it is beneficial to group particles
into block time steps of $\Delta t_n= 2^n \Delta t_{\rm min}$, where $n$ is an integer 
and $\Delta t_{\rm min}$ is the smallest required time step, and to evolve each group
of particles separately. In this way the number of (expensive) force evaluations 
can be reduced by orders of magnitude. Successful implementations of
such individual time step schemes can be found, for example, in 
\cite{porter85,ewell88,hernquist89,bate95,klessen00a,springel05a,rosswog05a,wetzstein08}.

\subsubsection{St\"ormer-Verlet and leap frog}
The Verlet or leapfrog algorithm\footnote{This algorithm has many 
names: in astronomy it is often called {\em St\"ormer method}, in 
molecular dynamics it is usually named {\em Verlet method} in the 
context of partial differential equations it is usually referred 
to as {\em leap-frog method}.\\ Note that the scheme needs modification if
the acceleration depends on the velocity.} is particularly appealing due to 
its exact time reversibility. The original Verlet algorithm 
\cite{verlet67} is very easy to derive: start from two Taylor expansions
for $\vec{r}(t+\Delta t)$ and $\vec{r}(t-\Delta t)$, add them and solve for
$\vec{r}(t+\Delta t)$ to find the position update prescription of the 
Verlet algorithm
\be
\vec{r}(t+\Delta t)= 2 \vec{r}(t) - \vec{r}(t-\Delta t) + \vec{a}(t) \Delta
t^2 + O(\dt^3). 
\label{eq:ODE:verlet_position}
\ee
It is interesting to note that the position is updated without using the 
velocities. But this comes at a price:  for a position update two positions 
at earlier time steps are needed. The velocity can be reconstructed 
from the positions via centered finite differences:
\be
\vec{v}(t)= \frac{\vec{r}(t+\Delta t)-\vec{r}(t-\Delta t)}{2 \Delta t} + O(\dt^3)
\label{eq:ODE:verlet_velocity}.
\ee
This simple integrator is only second-order accurate, but it is time reversible 
and has excellent conservation properties.
To kick off a simulation at $t= 0$, one needs $\vec{r}(-\Delta t)$. It can be obtained by
solving for $\vec{r}(-\Delta t)$ after inserting $t=0$ into 
Eq.~(\ref{eq:ODE:verlet_velocity}):
\be
\vec{r}(-\Delta t)= \vec{r}(\Delta t) - 2 \Delta t \vec{v}_0,
\ee
where $\vec{v}_0= \vec{v}(t=0)$.
Inserting this into Eq.~(\ref{eq:ODE:verlet_position}) provides the 
position update
\be
\vec{r}(\Delta t)= \vec{r}_0 + \Delta t \; \vec{v}_0 + \frac{1}{2} \vec{a}_0 \;
\Delta t^2, 
\ee
which looks like the first terms of a Taylor expansion around $t=0$.\\
The ``leapfrog form'' is obtained by defining velocities at half-steps, again
via centered differences
\bea
\vec{v}\left(t-\frac{\Delta t}{2}\right)= \frac{\vec{r}(t)-\vec{r}(t-\Delta
  t)}{\Delta t} \; {\rm and} \;
\vec{v}\left(t+\frac{\Delta t}{2}\right)= \frac{\vec{r}(t+\Delta t)-\vec{r}(t)}{\Delta t}
\label{eq:ODE_v+}.
\eea
The last equation can be solved for the leapfrog form of the position equation
\be
\vec{r}(t+\Delta t)= \vec{r}(t) + \Delta t \; \vec{v}\left(t+\frac{\Delta
    t}{2}\right).
\label{eq:ODE:position_leapfrog}
\ee
Starting from Eq.~(\ref{eq:ODE:verlet_position}) and inserting
Eq.~(\ref{eq:ODE_v+}) we find the velocity update of the leapfrog
algorithm
\be
\vec{v}\left(t+\frac{\Delta t}{2}\right)= \vec{v}\left(t-\frac{\Delta
    t}{2}\right) + \vec{a}(t) \Delta t.
\label{eq:ODE:velocity_leapfrog}
\ee
So positions and accelerations are always evaluated
at ``full'' time steps $t^n$, while the velocities are evaluated
in between, either at $t^{n-1/2}$ or $t^{n+1/2}$, so the velocities are always
``leaping'' over the positions. If velocities at $t$ are
needed, they can be calculated according to 
\be
\vec{v}(t)= \frac{\vec{v}\left(t+\frac{\Delta t}{2}\right)+
\vec{v}\left(t-\frac{\Delta t}{2}\right)}{2}.
\ee

\begin{figure}
\centerline{
\includegraphics[width=3.5in]{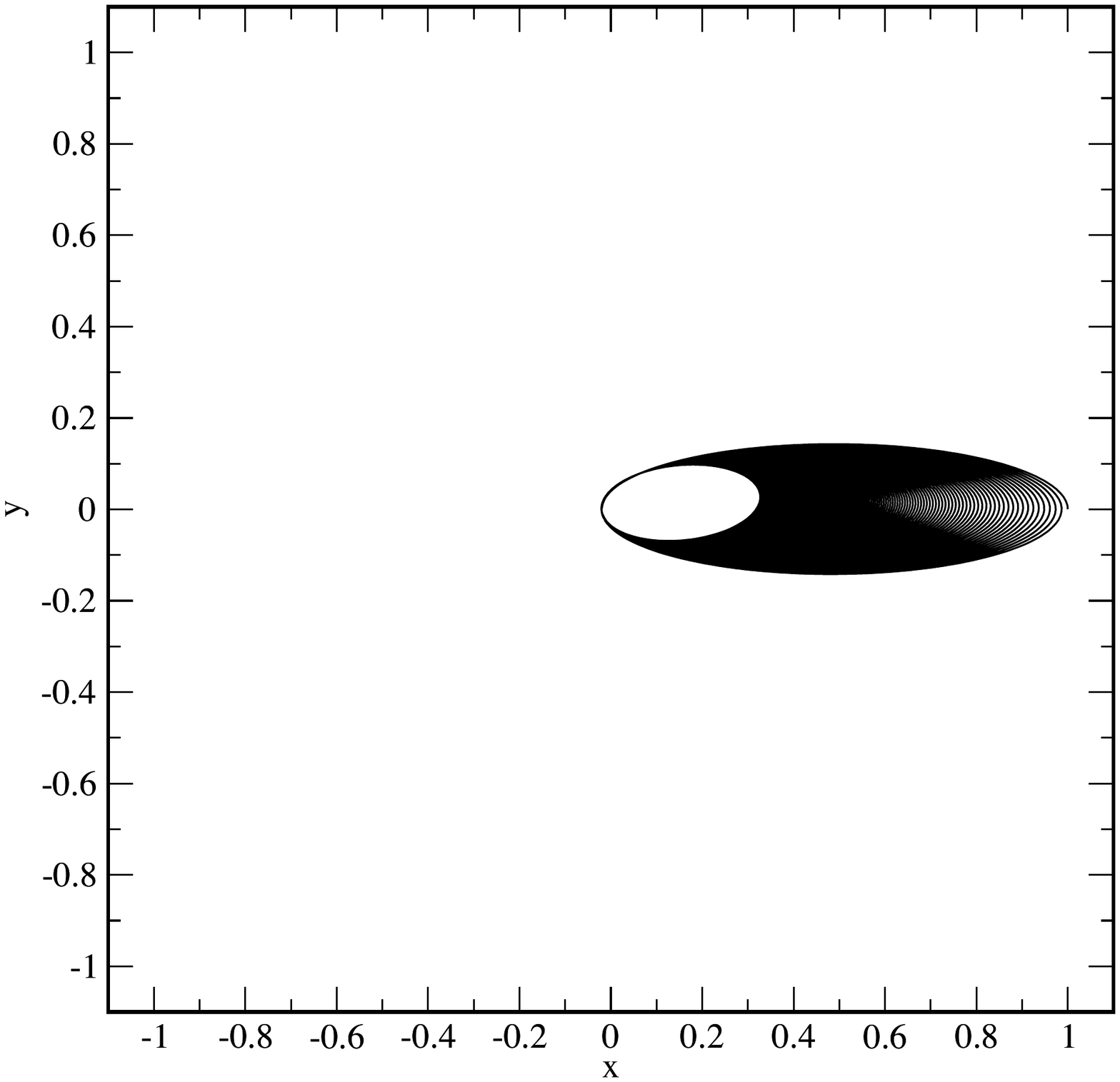} \hspace*{-2.cm}
\includegraphics[width=3.5in]{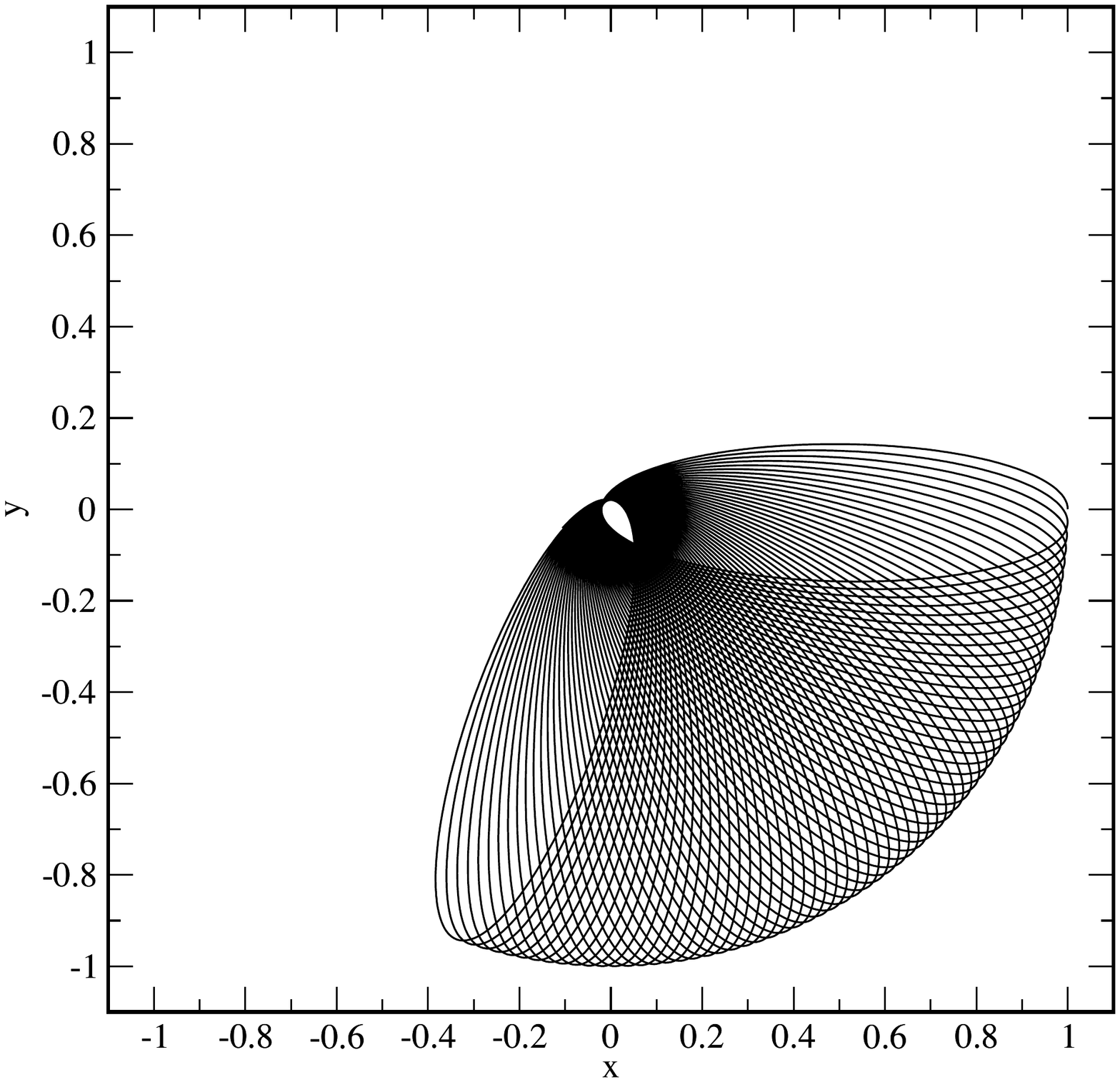}
}
\caption{Integration of the test particle orbit up to $t=150$ for an
eccentric orbit ($\vec{v}_0=(0,0.2)$, $\dt=10^{-3}$). Left:
fourth-order Runge-Kutta. Right: velocity St\"ormer-Verlet.}
\label{fig:ODE:long_term_integ_RK4_vel_V}
\end{figure}

The ``St\"ormer-Verlet form'' is particularly useful, since all quantities 
are evaluated at the same point of time. Start from Eq.~(\ref{eq:ODE:verlet_velocity}) 
and solve for 
\be
\vec{r}(t-\Delta t) = \vec{r}(t+\Delta t) - 2 \Delta t \; \vec{v}(t).
\ee
Insert this into the Verlet position update,
Eq.~(\ref{eq:ODE:verlet_position}), to find the position update of the
  velocity St\"ormer-Verlet algorithm
\bea
\vec{r}(t+\Delta t)
&=& \vec{r}(t) + \vec{v}(t) \; \Delta t + \frac{1}{2} \vec{a}(t) \Delta t^2.
\label{eq:ODE:pos_Stoermer_verlet}
\eea
To find the velocity update for the algorithm start from
Eq.~(\ref{eq:ODE:verlet_velocity}) and insert the Verlet position update,
Eq.~(\ref{eq:ODE:verlet_position}),
\bea
\vec{v}(t)
&=&\frac{\vec{r}(t) - \vec{r}(t-\Delta t)}{\Delta t} + \frac{\vec{a}(t)}{2}
\Delta t. 
\eea
Completely analogous one finds for the velocity at $t+\Delta t$
\be
\vec{v}(t+\Delta t)= \frac{\vec{r}(t+\Delta t) - \vec{r}(t)}{\Delta t} +
\frac{\vec{a}(t+\Delta t)}{2} \Delta t.
\ee
Adding the last two equations and using Eq.~(\ref{eq:ODE:verlet_velocity})
gives 
\be
\vec{v}(t+\Delta t)= \vec{v}(t) + \Delta t\; \frac{\vec{a}(t)+\vec{a}(t+\Delta
  t)}{2}. 
\ee
Together with Eq.~(\ref{eq:ODE:pos_Stoermer_verlet}) this update forms the St\"ormer-Verlet algorithm.
It is a simple example of a ``geometric integrator''
which preserves qualitative features of the exact flow of the ODE by 
construction\cite{leimkuhler04,hairer06}. In particular, it conserves angular
momentum exactly\footnote{For an elegant proof based on Newton's {\em Principia}
see \cite{hairer06}.}.\\
The integration of an elliptical Kepler orbit represents a simple, yet significant
numerical experiment. As long as no deviations from purely Newtonian point mass
gravity are considered, bound orbits are closed ellipses, e.g. \cite{landau76,goldstein02},
any ``non-closure'' is a numerical artifact due to finite integration accuracy.\\
To define the problem, we use units in which a test body (``planet'') 
is accelerated by 
\be
\vec{a}= - \frac{\vec{r}}{r^3},
\ee 
and, since angular momentum is conserved in a central force field, one
only needs to consider two spatial dimensions. We choose $\vec{r}_0= (1,0)$
and $\vec{v}_0= (0,0.2)$ as initial conditions. We perform a simulation up 
to $t= 150$ with a slightly too large time step, $\Delta t=10^{-3}$. For 
comparison we also use a fourth-order Runge-Kutta scheme with the same time step.
The results are displayed in Fig.~\ref{fig:ODE:long_term_integ_RK4_vel_V}. 
\begin{figure}
\centerline{\includegraphics[width=4in]{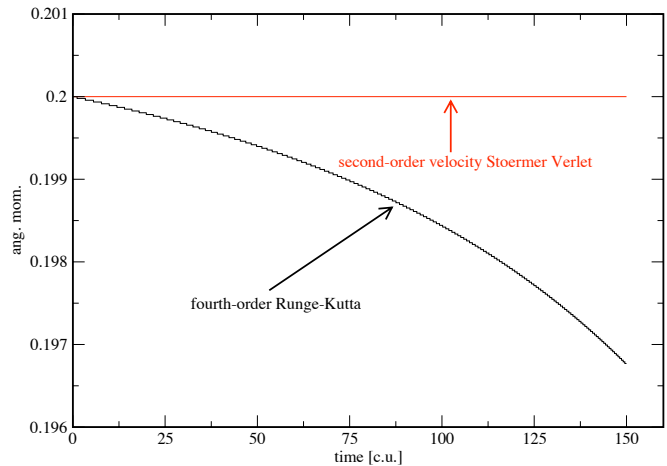}}
\caption{Angular momentum evolution of the long-term comparison. While the
  velocity St\"ormer-Verlet method conserves angular momentum to machine
  precision, the fourth-order Runge-Kutta method loses about 2 percent. 
As a result of this purely numerical dissipation the planet spirals inward.}
\label{fig:ODE:long_term_integ_RK4_vel_V_ang_mom}
\end{figure}
Due to the inappropriate time step the numerical inaccuracies in the velocity 
St\"ormer-Verlet case lead to a perihelion shift, but otherwise the behavior 
is physical: the test particle orbits the center of gravity on an 
eccentric orbit. In the Runge-Kutta case a qualitatively different 
and unphysical effect occurs: energy and angular momentum are dissipated for 
purely numerical reasons. The evolution of angular momentum is shown in
Fig.~\ref{fig:ODE:long_term_integ_RK4_vel_V_ang_mom}. The ``steps'' in the
Runge-Kutta case occur at perihelion since here the time steps are much too
large.  Therefore, the test particle drifts into a more circular, lower energy
orbit. This illustrates that {\em exact conservation can be more
  important that high order!}

\subsubsection{Time stepping with quality control: Fehlberg methods}
The size of the time step $\dt$ is crucial for the accuracy (and stability!)
of the numerical solution of an ODE. It has to be a small fraction of the
shortest physical time scale of the problem at hand. For a complex multi-physics
system this may require to determine the time scales of every involved physical 
process. In the worst case, these may not be known and, on top, the meaning of 
``a small fraction'' may subject to trial and error.\\
Ideally, a method should decide  itself which step size to take. This is
the philosophy behind {\em adaptive step size control} or {\em Fehlberg methods}
\cite{fehlberg64,fehlberg66,fehlberg69,press92}. The main idea is to use two estimates of 
different order for the solution and use them to estimate the error growth rate.
Based on the comparison of the estimated and the tolerable growth rate, the 
chosen time step may be accepted or rejected. Independent of whether the time 
step was accepted or not, the growth rate can provide an educated guess for the 
size of the next time step.\\
The procedure is the following:
\bi
\i Choose a trial time step $\Delta t$.
\i Use two different methods to obtain results at the end of the time step, $a_1$ and $a_2$. 
To be reasonably efficient, the two 
methods should use nearly the same evaluations of the RHS of the ODE.
\i Use  $a_1$ and $a_2$ to calculate an approximate local truncation error
$E$. 
\i Compare the error growth rate $\epsilon \approx E/\Delta t$ with some 
 acceptable error growth rate, $\epsilon^{\rm acc}$. 
If the rate is small enough, $\epsilon < \epsilon^{\rm acc}$, accept
the time step, $t^{n+1}= t^n + \Delta t$ and $y^{n+1} = a_2$, where $a_2$ is
the more accurate method. Otherwise, re-take the step with a reduced step
size determined from $\epsilon$ and the previously tried step size.
\ei 

Obviously, the above strategy can be applied to a large variety of 
integrator combinations. An often used example is the combination of 
fourth- and fifth-order Runge-Kutta schemes \cite{fehlberg69}. As a low-order
example, we apply the outlined strategy to a second- and a third-order
method. Let $\Phi(t)$ be the exact solution of the differential equation 
and assume that an approximation to the solution at  $t^n$ is known: 
$\Phi(t^n)\simeq y^n$. Take three RHS evaluations of the ODE:
\bea
f_1 &=& f(t^n,y^n)\\
f_2 &=& f(t^n + \Delta t,y^n + \Delta t \; f_1)\\
f_3 &=& f\left(t^n+\frac{\Delta t}{2},y^n + \frac{\Delta t}{4} [f_1+f_2]\right)
\eea
to produce a second-order (this is just the modified Euler method)
\be
a_1= y^n + \Delta t \frac{f_1+f_2}{2}
\ee
and a third-order method
\be
a_2= y^n + \frac{\dt}{6} [f_1+f_2+4f_3].
\ee
Thus, we have
\be
\Phi(t^n+\Delta t)= a_1 + K \Delta t^3 + O(\Delta t^4)
= a_2 + O(\Delta t^4).
\ee
Therefore, the leading error is 
\be
E= |a_2-a_1| = K \Delta t^3 + O(\dt^4)
\ee
and the error grows at a rate
\be
\epsilon\approx \frac{E}{\Delta t} \approx K \Delta t^2.
\ee
The step can either be accepted ($\epsilon<\epsilon^{\rm acc}$), 
$t^{n+1}= t^n + \Delta t$, $y^{n+1}= a_2$, or
rejected ($\epsilon > \epsilon^{\rm acc}$) and re-taken. In both cases one
can use the error measure to estimate the new time step. Since $\epsilon \propto \dt^2$,
one finds
\be
\frac{\epsilon^{\rm new}}{\epsilon^{\rm tried}}= \left(\frac{\Delta t^{\rm new}}{\Delta t^{\rm tried}}\right)^2,
\ee
and the suggestion for the new time step is
\be
\Delta t^{\rm new}= s \; \left(
\frac{\epsilon^{\rm acc}}{\epsilon^{\rm tried}} \right)^{1/2} \Delta t^{\rm tried},
\ee
where $s < 1$ is a ``safety factor'' for a conservative choice of the next 
time step. \\
For further reading on ODE-integration the excellent text books of 
\cite{hockney88,leimkuhler04,hairer06,griebel07} are recommended. 

\subsection{``Best practice'' suggestions}
\label{page_best_practice}
We want to collect here a couple of suggestions born form practical experience
that should help to carry out reliable simulations and to avoid numerical artifacts.
This is somewhat ``soft'', yet hopefully still useful knowledge.
\begin{itemize}
\item {\bf Neighbor numbers and resolution}\\
We recommend a large neighbor number, typically 100 or more. This somewhat
increases the smoothing lengths and thus deteriorates the integral approximation,
Eq.~(\ref{eq:basic:integral_approximation}). On the other hand, it can substantially
reduce numerical noise, i.e. fluctuations around the true solution. To compensate,
it is advisable to run simulations at the highest affordable particle number,
numerical resolution {\em does} matter. Substructures can only be considered
resolved if they are substantially larger than the local smoothing lengths.
If just an idea about the mass distribution is required, one may get away with
a small particle number, say a few thousand particles, but reliable thermodynamic 
properties usually require {\em much} larger particle numbers. Contrary to some prejudices, SPH {\em can}
resolve shocks properly, but since shocks are spread across a few smoothing lengths
an appropriate numerical resolution is required.
\item {\bf Particle masses}\\
If at all possible, equal-mass particles should be used. Light particles that 
interact with much more massive ones easily  become``nervous'' (``ping pong and 
cannon ball effect''), i.e. they can exhibit large  fluctuations. In some
problems different particle numbers are unavoidable, if, for example, a highly
centrally condensed star is considered, the Courant time step criterion may
restrict the central time steps to prohibitively small values. In such cases one
should make sure that interacting particles only differ in masses by factors of 
very few.
\item {\bf Initial conditions}\\
Accurate initial conditions are absolutely crucial for reliable simulations. 
The SPH particles should start out from their true numerical equilibrium positions. 
This can be obtained by applying an artificial damping term $\propto \vec{v}_a$ 
in the momentum equation that drags the particles to the desired initial conditions.
Once equilibrium is found the particles should not move considerably during a 
few dynamical time scales after the damping has been switched off.  
\item {\bf Time integration}\\
Ideally, an integration scheme should monitor the achieved accuracy for
the tried time step and {\em reject} it if the error becomes
unacceptably large. Just estimating a time step that seems reasonable 
and than hope for the best, may produce noisy, or in the worst case, spurious 
results. The rejection of time steps in individual time step schemes may, 
however, pose bookkeeping challenges. A simple workaround has been proposed 
by \cite{saitoh09}. For each and every simulation the conservation of
energy, momentum and angular momentum should be monitored. Reducing the time step 
size and increasing the force accuracy, say, if a tree is used for gravity, should improve
the conservation properties. A correct code should ensure conservation
to better than 1\% over several thousand time steps.
\item {\bf Artificial dissipation}\\
Artificial dissipation should only be applied where really necessary. Schemes
with time-dependent parameters are highly recommended. Modern versions of
artificial dissipation terms such as Eqs.~(\ref{basic:eq:v_diss}) and
(\ref{basic:eq:e_diss}) should be used. Recent years have seen a large leap forward
in eliminating unwanted effects from artificial dissipation, yet, there is certainly 
room for further improvement.
\item {\bf Displaying results}\\
Generally, display continuous quantities and avoid particle plots to present
physical results. The particles are just auxiliary constructs, ``moving
interpolation points'' that represent a continuum. Just showing particle
positions projected on a plane may mimic a ``resolution'' that does not exist, 
a contour plot is a more honest presentation. To check that the numerics is well 
behaved, it may, however, be useful to inspect the particle distribution. A 
noisy particle distribution should always raise doubts about the quality of a
simulation. 
\end{itemize}

%
%
\hspace*{0cm}\fbox{
\parbox{14cm}{
\vspace*{0.5cm}
\centerline{\bf Summary of ``vanilla ice'' SPH}

\vspace*{0.5cm}
We summarize here the most basic form of the SPH
equations. As the particle masses are kept fix, there is no need to solve
the continuity equation. Densities can be obtained via summation from
\be
 \rho_a= \sum_b m_b W_{ab}.
\label{eq:basic:summary_sum_rho}
\ee
Alternatively, the continuity equation can be integrated, see
Eq.~(\ref{eq:basic:drho_dt}). The evolution equation for the specific internal
energy can be written as
\be
\frac{du_a}{dt}= \sum_b m_b \left(\frac{P_a}{\rho_a^2} + \frac{1}{2} \Pi_{ab}  \right)\vec{v}_{ab} \cdot \nabla_a W_{ab}
\label{eq:basic:summary_energy_equation_u_AV}
\ee
and the momentum equation as
\be
\frac{d\vec{v}_a}{dt}= - \sum_b m_b \left(\frac{P_a}{\rho_a^2} +
  \frac{P_b}{\rho_b^2} + \Pi_{ab} \right)
\nabla_a W_{ab}. 
\label{eq:basic:summary_momentum_equation_AV}
\ee
Particles are advanced in time according to this equation.
$\Pi_{ab}$ is the ``standard'' artificial viscosity term given 
in Eq.~(\ref{eq:basic:PI_AV}), more modern prescriptions are given in
Eq.~(\ref{basic:eq:v_diss}) and Eq.~(\ref{basic:eq:e_diss}). Alternative
forms of energy may be integrated forward in time, see, for example,
Eq.~(\ref{eq:basic:energy_equation_e}).
}}

\section{SPH from a variational principle}
\label{sec:SPH_from_Lagrangian}
\label{page_var_principle}
In the previous section we had seen how the ``vanilla ice''
SPH equations can be derived directly from the Lagrangian form of the
inviscid hydrodynamic equations. Although this form ensures conservation
of energy, linear and angular momentum and works well in practice, the
corresponding symmetries were enforced somewhat ad hoc. SPH can also
be derived using nothing more than a suitable fluid Lagrangian, the 
first law of thermodynamics and a prescription on how to obtain a 
density estimate via summation \cite{gingold82,monaghan01,springel02}.

\subsection{The Lagrangian and the Euler-Lagrange equations}
\label{page_Lag}
The Lagrangian of a perfect fluid is given by \cite{eckart60}
\be
L= \int \rho \left(\frac{v^2}{2} - u(\rho,s) \right) dV,
\ee
where $\rho$ is the mass density, $v$ the fluid velocity, $u$ the specific
energy and $s$ the specific entropy. In SPH-discretization it reads
\be
L_{\rm SPH,h}= \sum_b m_b \left(\frac{v_b^2}{2} - u(\rho_b,s_b) \right).
\label{eq:L_SPH_discr}
\ee
The discretized equations for the fluid are then derived by applying
the Euler-Lagrange equations
\be
\frac{d}{dt} \left(\frac{\partial L}{\partial \vec{v}_a} \right) -
\frac{\partial L}{\partial \vec{r}_a} = 0,\label{eq:EL_DGL}
\ee
where $\vec{r}_a$ and $\vec{v}_a$ refer to the position and velocity of
particle $a$. The term in brackets yields the canonical particle momentum 
(like usual, we keep the particle mass fixed in time)
\be
\frac{\partial L}{\partial \vec{v}_a}= \frac{\partial}{\partial
    \vec{v}_a} \left[ \sum_b m_b \left(\frac{v_b^2}{2} - u(\rho_b,s_b)
    \right) \right]= m_a \vec{v}_a \label{eq:dL_dv}
\ee
and therefore, the first term in Eq.~(\ref{eq:EL_DGL}) yields the change of
particle momentum $m_a \frac{d\vec{v}_a}{dt}$.\\
The second term in the Lagrangian
acts like a potential, and if self-gravity is considered,
$ - \sum_b m_b u_b$ has to be replaced by $- \sum_b  m_b (u_b + \Phi_b)$, where
$\Phi_b$ is the gravitational potential \cite{price07a}. In the following, we will only
consider $\Phi_b=0$. The second term in Eq.~(\ref{eq:EL_DGL})
becomes
\bea
\frac{\partial L}{\partial \vec{r}_a}= \frac{\partial}{\partial \vec{r}_a}
\left[\sum_b m_b \left(\frac{v_b^2}{2} - u(\rho_b,s_b)
    \right) \right]
= - \sum_b m_b \left.\frac{\partial u_b}{\partial \rho_b}\right\vert_s 
\cdot \frac{\partial \rho_b}{\partial \vec{r}_a}. 
\label{eq:dL_dx}
\eea
We can make use of the first law of thermodynamics, Eq.~(\ref{eq:first_law_3}),
to find
\be
m_a \frac{d\vec{v}_a}{dt}= - \sum_b m_b \frac{P_b}{\rho_b^2} \frac{\partial \rho_b}{\partial \vec{r}_a}.\label{eq:mi_dot_vi}
\ee

\subsection{The density, its derivatives and ``grad-h''-terms}
\label{page_gradh}
So far, we had ignored all extra terms that result from variable smoothing
lengths \cite{nelson94,springel02,monaghan02}, this is what we address now. 
From now on, we use the smoothing length of the particle itself in the density summation, 
\be
\rho_a= \sum_b m_b W(r_{ab},h_a).\label{eq:Advanced_rho}
\ee
and use Eq.~(\ref{eq:basic:h_adaption}) to adapt the smoothing length.
Thus, $\rho_a$ depends on $h_a$ and vice versa, which requires an iteration to reach
consistency.\\ 
If we take the changes of $h$ into account, the Lagrangian time derivative of
the density is given by 

\bea
\frac{d\rho_a}{dt}&=& \frac{d}{dt} \left(\sum_b m_b W_{ab}(h_a)\right)
                  = \sum_b m_b \left\{ \frac{\partial W_{ab}(h_a)}{\partial r_{ab}}
                       \frac{d r_{ab}}{dt} 
                      + \frac{\partial W_{ab}(h_a)}{\partial h_{a}} 
                         \frac{d h_{a}}{dt}\right\}\nonumber \\
                  &=& \sum_b m_b \frac{\partial W_{ab}(h_a)}{\partial r_{ab}}
                      \hat{e}_{ab} \cdot \vec{v}_{ab}
                      +
                      \sum_b m_b \frac{\partial W_{ab}(h_a)}{\partial h_a}
                       \cdot \frac{\partial h_a}{\partial \rho_a} 
                      \frac{d \rho_a}{dt}\nonumber\\
                  &=& \sum_b m_b \vec{v}_{ab} \cdot \nabla_a W_{ab}(h_a)
                      + \frac{\partial h_a}{\partial \rho_a} 
                      \frac{d \rho_a}{dt} \sum_b m_b 
                      \frac{\partial W_{ab}(h_a)}{\partial h_a},
\eea
where we have used Eqs.~(\ref{eq:basic:drab_dt}) and (\ref{eq:k4}).
If we collect the $d\rho_a/dt$-terms  and introduce

\be
\Omega_a\equiv \left(1 - \frac{\partial h_a}{\partial \rho_a}  \sum_b m_b
\frac{\partial W_{ab}(h_a)}{\partial h_a}  \right)\label{eq:omega_a}
\ee
the time derivative of the density reads
\be
\frac{d\rho_a}{dt}= \frac{1}{\Omega_a} \sum_b m_b \vec{v}_{ab} \cdot \nabla_a
W_{ab}(h_a) \label{eq:advanced:drho_dt_omega}.
\ee
This is the generalization of the SPH expression for the 
density change that follows from Eq.~(\ref{eq:basic:divv}). 
In a similar way, the spatial derivatives can be calculated
\bea
\frac{\p\rho_b}{\p\vec{r}_a}
&=&  \sum_k m_k \left\{ \nabla_a W_{bk}(h_b)
   + \frac{\partial W_{bk}(h_b)}{\partial h_b} \frac{\p h_b}{\p \rho_b} 
      \frac{\p \rho_b}{\p\vec{r}_a} \right\}\nonumber \\
&=& \frac{1}{\Omega_b} \sum_k m_k \nabla_a W_{bk}(h_b).\label{eq:drho_dx}
\eea
To summarize: if derivatives of the smoothing length $h$ are accounted for,
the ``standard'' SPH expressions for the density derivatives have to be corrected 
by factors $1/\Omega$, see Eq.~(\ref{eq:omega_a}), (\ref{eq:advanced:drho_dt_omega}) and (\ref{eq:drho_dx}). 

\subsection{The SPH equations with ``grad-h''-terms}
\label{advanved:sec:grad_h}
\label{page_SPH_gradh}
We keep again the masses fixed in time, so there is no need to solve the
continuity equation. By inserting  Eq.~(\ref{eq:advanced:drho_dt_omega}) into
Eq.~(\ref{eq:basic:du_dt_a}) we find the energy equation
\be
\frac{d u_a}{dt}= \frac{1}{\Omega_a}\frac{P_a}{\rho_a^2}
\sum_b m_b \vec{v}_{ab} \cdot \nabla_a W_{ab}(h_a) \label{eq:energy_equation}.
\ee
For the momentum equation we need to insert the density gradient, 
Eq.~(\ref{eq:drho_dx}), into Eq.~(\ref{eq:mi_dot_vi})
\bea
m_a \frac{d\vec{v}_a}{dt}= - \sum_b m_b \frac{P_b}{\rho_b^2} \nabla_a \rho_b
= - \sum_b m_b \frac{P_b}{\rho_b^2} \left(\frac{1}{\Omega_b} \sum_k m_k \nabla_a W_{bk}(h_b)\right).
\eea
Using Eq.~(\ref{eq:k3}), the above equation becomes
\bea
\hspace*{-0.cm}m_a \frac{d\vec{v}_a}{dt} \hspace*{-0.cm}&=& \hspace*{-0.cm}- \sum_b m_b \frac{P_b}{\rho_b^2} \frac{1}{\Omega_b} 
\sum_k m_k \nabla_b W_{kb}(h_b) \; (\delta_{ba}-\delta_{ka})\nonumber\\
&=&\hspace*{-0.cm}- m_a \frac{P_a}{\rho_a^2} \frac{1}{\Omega_a} \sum_k m_k \nabla_a W_{ka}(h_a)
+ \sum_b m_b \frac{P_b}{\rho_b^2}\frac{1}{\Omega_b} m_a \nabla_b
W_{ab}(h_b)\nonumber\\
&=&\hspace*{-0.cm}- m_a \frac{P_a}{\rho_a^2} \frac{1}{\Omega_a} \sum_b m_b
\nabla_a W_{ba}(h_a) - m_a \sum_b m_b \frac{P_b}{\rho_b^2}\frac{1}{\Omega_b} \nabla_a
W_{ab}(h_b)\nonumber\\
&=&\hspace*{-0.cm}- m_a \sum_b m_b 
\left(
\frac{P_a}{\Omega_a \rho_a^2} \nabla_a W_{ab}(h_a) + 
\frac{P_b}{\Omega_b \rho_b^2} \nabla_a W_{ab}(h_b)
\right).
\eea
Thus the final momentum equation reads
\be
\frac{d\vec{v}_a}{dt}= - \sum_b m_b 
\left(
\frac{P_a}{\Omega_a \rho_a^2} \nabla_a W_{ab}(h_a) + 
\frac{P_b}{\Omega_b \rho_b^2} \nabla_a W_{ab}(h_b)
\right).
\label{eq:SPH_from_Lag_dvdt}
\ee
Note that now all arbitrariness (such as the particular form of 
Eq.~(\ref{eq:basic:nabla_p_rho})) has been eliminated from the derivation.
The ``grad-h'' terms increase the accuracy of SPH and the conservation 
properties in the presence of varying smoothing lengths. How important
they are in practice depends on both the problem and the numerical 
resolution \cite{springel02,rosswog07c}.

\hspace*{0cm}\fbox{
\parbox{14cm}{
\vspace*{0.5cm}
\centerline{\bf SPH equations from a variational principle}

\vspace*{0.5cm}
With 
\be
\rho_a= \sum_b m_b W(r_{ab},h_a) \quad {\rm and} \quad 
h_a= \eta \left(\frac{m_a}{\rho_a}\right)^{1/3}
\ee
the energy equation reads
\be
\frac{d u_a}{dt}= \frac{1}{\Omega_a}\frac{P_a}{\rho_a^2}
\sum_b m_b \vec{v}_{ab} \cdot \nabla_a W_{ab}(h_a)
\ee
and the momentum equation is
\be
\frac{d\vec{v}_a}{dt}= - \sum_b m_b 
\left(
\frac{P_a}{\Omega_a \rho_a^2} \nabla_a W_{ab}(h_a) + 
\frac{P_b}{\Omega_b \rho_b^2} \nabla_a W_{ab}(h_b)
\right),
\ee
 where
\be
\Omega_a\equiv \left(1 - \frac{\partial h_a}{\partial \rho_a} \sum_b m_b
\frac{\partial}{\partial h_a} W_{ab}(h_a) \right)
\ee
are the so-called ``grad-h-terms''. Additional artificial dissipation 
terms as discussed in Sec.~\ref{sec:basic:Lag_hydro} need to be applied
in order to resolve shocks.
}}

\section{Relativistic SPH}
\label{sec:rel_SPH}
\label{page_rel_SPH}
Several relativistic versions of SPH exist.  
Early formulations due to Mann \cite{mann91,mann93} were able to 
capture the overall solutions, but they were --by today's
standards-- not particularly accurate.\\  
Laguna et al. \cite{laguna93a} developed a 3D, general-relativistic SPH code. 
Due to their
choice of variables their continuity equation contains a gravitational
source term. This term requires to introduce SPH kernel functions for curved
space-times which are in general not isotropic and invariant under
translations. Moreover, they did not use a conservative form of the equations
and their approach requires additional time derivatives of Lorentz factors
which they approximated by first order finite differences. Similar to
the approach of Mann, only mildly relativistic shocks could be treated. 
The approach of \cite{laguna93a} is also the basis of the more recent code
described in \cite{rantsiou07}.\\    
Kheyfets et al. \cite{kheyfets90} suggested an alternative approach 
in which they define the spatial kernel function in a local
rest frame and assume that space is approximately flat in this frame. The
interaction between particles requires the choice of a particular frame whose
calculation requires the solution of four coupled algebraic equations. While
this approach is completely covariant, it requires a large computational
effort.\\
Like in grid-based methods, a conservative formulation is crucial
to handle strong shocks. As we had seen in Sec.\ref{subsec:artificial_dissi}, 
concepts of Riemann solvers were used as a guide to improve SPH's 
artificial viscosity by treating the interaction between two particles as a 
local Riemann problem \cite{monaghan97}. This approach replaces wave
propagation speeds by an appropriately chosen signal velocity.
With this approach (together with evolving the total energy rather than the
thermal energy) Chow and Monaghan \cite{chow97} obtained accurate results even
for ultra-relativistic flows.\\
More recently, Siegler and Riffert \cite{siegler00a} and Siegler
\cite{siegler00b} have suggested a set of equations for both the special- and 
general-relativistic (fixed background metric) case. It is based on a 
formulation of the Lagrangian hydrodynamics equations  
in conservative form. By a particular choice of dynamical variables they avoid
many complications that have plagued earlier relativistic SPH formulations. In
particular, by choosing the relativistic rest mass density for the SPH
formalism the continuity equation has the same form as in non-relativistic SPH,
time derivatives of Lorentz factors do not appear and flat space
kernels can be used. This comes at the price of inverting the dynamically
evolved variables into the physical variables. Siegler \cite{siegler00b} was 
able to accurately simulate some test cases with Lorentz factors up to 
$\gamma= 1000$.\\ 
A more elegant approach is based on the use of the discretized Lagrangian
of a perfect fluid \cite{monaghan01}. Guided by the canonical momentum and
energy, suitable numerical variables can be chosen which lead to evolution
equations that are similar to the Newtonian ones. Due to the consistent
derivation from a Lagrangian the conservation of mass, energy, linear and
angular momentum are guaranteed. This approach can be applied both to the
special- and the general-relativistic case.\\  
In recent years, SPH has also been applied to study flows in time-varying
space-times. The first approaches used Post-Newtonian approximations 
\cite{ayal01,faber00,faber01,faber02b}, more recently the conformal 
flatness approximation \cite{isenberg80,wilson96} has been implemented
 \cite{oechslin02,faber04,faber06}.  \\
In the following subsections we derive the special- and general-relativistic
SPH equations from a variational principle similar to \cite{monaghan01}.

\subsection{Special-relativistic SPH}
\label{page_SR_SPH}
Here, we generalize the approach of Monaghan and Price \cite{monaghan01}
to include the special-relativistic ``grad-h'' terms.
We use the conventions that the metric tensor has the signature (-,+,+,+), 
Latin indices run from 1 to 3, Greek ones from 0 to 3 with the zero
component being time. In addition, the
usual Einstein sum convention and $c=G=1$ are used and the flat spacetime
metric is denoted by $\eta_{\mu\nu}$.\\
With these conventions, the line element reads
\be
ds^2= \eta_{\mu \nu} dx^\mu dx^\nu
\ee
and the proper time is given by $d\tau^2= -ds^2$.
The 4-velocity is defined as 
\be
U^\mu= \frac{dx^\mu}{d\tau} \label{eq:SR:4vel}
\ee
and, due to the above two relations, it is normalized to $-1$:
\be
U_\mu U^\mu= \eta_{\mu \nu} \frac{dx^\nu}{d\tau}
\frac{dx^\mu}{d\tau}= \frac{ds^2}{d\tau^2}= -1.\label{eq:SR:U_norm}
\ee
The velocity components can be written as 
\be
(U^\mu)= (\gamma, \gamma \vec{v}),\label{eq:SR:4vel}
\ee 
where $\gamma$ is the Lorentz factor
\be
\gamma= \frac{1}{\sqrt{1-v^2}}
\ee 
with $\vec{v}$ being the 3-velocity. Due to Eq.~(\ref{eq:SR:4vel})
the 3-velocity can be expressed as
\be
v^i= \frac{U^i}{U^0}. \label{eq:SR:3vel}
\ee
The 4-momentum of a particle with rest mass $m$ is given by
\be
(p^\mu)= (m U^\mu)= (\gamma m, \gamma m \vec{v})= (E,\vec{p}),
\ee
where $E$ is the particle energy and $\vec{p}$ its relativistic momentum. The
last expression is also correct for particles with vanishing rest mass.

\subsubsection{The Lagrangian}
Consistent with our overall strategy, we derive the SPH equations 
from a Lagrangian to ensure exact conservation. We start from the 
Lagrangian of a perfect fluid \cite{fock64}
\be
L_{\rm pf,sr}= - \int T^{\mu\nu} U_\mu U_\nu \; dV\label{eq:fluid_Lag_SRT}.
\ee
\hspace*{0cm}\fbox{
\parbox{14cm}{
\vspace*{0.5cm}
\centerline{\bf The energy momentum tensor of a perfect fluid}

\vspace*{0.5cm}

The energy-momentum tensor, $T^{\mu\nu}$, is a key element of relativistic
hydrodynamics as it contains the sources of the relativistic
gravitational field. Its components $T^{\mu \nu}$ are defined as
``flux   of $\mu$-component of the 4-momentum in $\nu$-direction''.
Thus, the $T^{00}$-component is the {\em energy density}, the $T^{0j}$
component is the {\em energy flux}, the $T^{i0}$ component is the {\em
  momentum density} and the $T^{ij}$ is the {\em momentum flux}. For a perfect
fluid with zero viscosity and heat conduction it is given by
\be
T^{\mu \nu}= (P+e) U^\mu U^\nu + P \eta^{\mu\nu}, \label{eq:SR:en_mom_tensor}
\ee
where $P$ is the fluid pressure, $U^\mu$ the 4-velocity and $e$ the energy
density in the local rest frame. The conservation of energy and momentum can be expressed as
\be
T^{\mu\nu},\nu= 0,
\ee
where the comma denotes the partial derivative (in general relativity it has
to be replaced by the covariant derivative, usually denoted by a semicolon).
$\mu=0$ expresses energy and $\mu=i$ momentum conservation.\\
In the Newtonian limit (with $c=1$) the following relations hold
 $U^0 = \gamma \simeq 1$, $U^i \simeq v^i$, $e \simeq \rho$ and $P/e \ll 1$. 
Thus the components read
\bea
T^{00}&=& (P+e) U^0  U^0 + P \approx \rho\\
T^{0j}&=& (P+e) U^0  U^j \approx \rho v^j\\
T^{ij}&=& (P+e) U^i  U^j + P \delta^{ij}\approx \rho v^i v^j + P \delta^{ij}.
\eea
Therefore, $T^{0\mu},_\mu=0$ reduces to the usual continuity equation
\be
T^{0\mu},_\mu= T^{00},_0 + T^{0j},_j = \frac{\p \rho}{\p t} + \frac{\p (\rho
  v^j)}{\p x^j}=0
\ee
and $T^{i\mu},_\mu=0$ becomes
\be
T^{i\mu},_\mu= T^{i0},_0 + T^{ij},_j = \frac{\p (\rho v^i)}{\p t} + \frac{\p (\rho
  v^i v^j)}{\p x^j} + \frac{\p P}{\p x^j}=0,
\ee
which is the Euler equation.}}

This Lagrangian is, apart from the
flat space volume element, the same as the general-relativistic one, see
Sec.~\ref{chap:GR}. The energy density, $e$\footnote{This quantity is usually denoted by
  $\rho$ in the literature. To avoid confusion with the non-relativistic mass
  density that we have used previously, we choose a different symbol
  here.}, that is needed in the energy-momentum tensor,  see 
Eq.~(\ref{eq:SR:en_mom_tensor}), has contributions from the rest mass and 
the thermal energy:
\be
e= e_{\rm rest} +  e_{\rm therm}= \rho_{\rm rest} c^2+ u \rho_{\rm rest}
= n m_0 c^2 (1 + u/c^2),
\ee
where $m_0$ is the baryon mass\footnote{Note that this quantity depends on the
relative number of neutrons and protons, i.e. on the nuclear composition.} 
and $u$ is again the specific thermal energy and $n$ is the baryon number 
density, the latter two are measured in the local rest frame.
Here, we have kept the $c^2$s for clarity. From now on we will measure all our
energies in units of $m_0 c^2$ (and use $c\equiv 1$). With these conventions the
energy momentum tensor reads
\be
T^{\mu\nu}= (n[1 + u(n,s)] + P)  U^\mu U^\nu + P \eta^{\mu\nu}.
\ee
 Here $s$ is the specific entropy per baryon in the local
fluid rest frame and $P$ the fluid pressure. For a practical
computation we fix a particular frame (``computing frame'') and
transform the quantities into this frame where necessary.\\
Using Eq.~(\ref{eq:SR:U_norm}), we can considerably simplify the above
Lagrangian. The quantity under the integral  becomes
\be
T^{\mu\nu} U_\mu U_\nu = \left\{(n [1 + u] + P)  U^\mu U^\nu\right\} U_\mu U_\nu 
+ P \eta^{\mu\nu} U_\mu U_\nu= n(1+u)
\ee
and the Lagrangian simplifies to
\be
L_{\rm pf,sr}= - \int n(1+u) \; dV. \label{eq:fluid_Lag_SRT_simp}
\ee
The baryon number density in the rest frame of a fluid element, $n$, is related to
the number density in our computing frame, $N$, by a Lorentz factor,
$\gamma$. To see this, assume a volume in the local rest frame
of a fluid element, $V_{\rm lrf}$,  that contains $k$ baryons. In general, 
this fluid
element will move with some velocity with respect to our computing frame. If
we assume that the motion is in, say, $x-$direction, the $x-$component of the volume
element in the computing frame, $V_{\rm cf}$, appears Lorentz-contracted by a 
factor $\gamma$:  $V_{\rm cf}= V_{\rm lrf}/\gamma$. Since this volume element 
contains the same number of baryons, $k$, the density in the computing frame is
\be
N= \frac{k}{V_{\rm cf}}= \frac{k \gamma}{V_{\rm lrf}}=\gamma n. 
\label{eq:N_vs_n}
\ee
As before, see Eq.~(\ref{eq:L_SPH_discr}), we want to discretize the
Lagrangian to derive the SPH equations. We subdivide space in
volumes $\Delta V_b$, so that a volume labeled by index $b$ contains 
$\nu_b= \Delta V_b N_b$ baryons, where $N_b$ is the baryon density at particle $b$ 
(all quantities measured in the computing frame). Or, the other way around,
if a fluid parcel labeled by $b$ contains  $\nu_b$ baryons, it has a volume
\be
\Delta V_b= \frac{\nu_b}{N_b}.
\ee
Similar to Eq.(\ref{eq:basic:sum_interpol}),
we can use this volume element in the SPH-approximation of 
a quantity $f$\footnote{For ease of notation we have again dropped the distinction
between the function to be interpolated and the interpolant.}:
\be
f(\vec{r})= \sum_b f_b \frac{\nu_b}{N_b} W(|\vec{r}-\vec{r}_b|,h).
\ee
With this prescription the baryon number density in the computing
frame at position of particle $a$ reads
\be
N_a= N(\vec{r}_a)= \sum_b \nu_b W(|\vec{r}_a-\vec{r}_b|,h_a),
\label{eq:dens_summ_SR}
\ee
where we have used $h_a$ to conform with the non-relativistic version, 
Eq.~(\ref{eq:Advanced_rho}). The latter can be recovered by the following 
replacements:
$m_b \rightarrow \nu_b$ and  $\rho_b \rightarrow N_b$. If we keep the baryon
number per SPH particle, $\nu_b$, fixed, the total baryon number is conserved
and there is no need to evolve a continuity equation. Analogously to 
Eq.~(\ref{eq:basic:divv}) one could also discretize the continuity equation 
and calculate the density by integration. We adapt the smoothing length similar 
to the Newtonian case, see Eq.~(\ref{eq:basic:h_adaption}), 
\be
h_a= \eta \left(\frac{\nu_a}{N_a}\right)^{1/3},
\ee
which, again, requires an iteration for consistent values of $N_a$ and $h_a$.
We can now discretize the Lagrangian of Eq.~(\ref{eq:fluid_Lag_SRT_simp})
\be
L_{\rm SPH,sr}= - \sum_b \frac{\nu_b}{N_b} n_b [1+ u(n_b,s_b)],
\ee
or, by using Eq.~(\ref{eq:N_vs_n}),
\be
L_{\rm SPH,sr}= - \sum_b \frac{\nu_b}{\gamma_b} [1+ u(n_b,s_b)]
=- \sum_b \nu_b \; \sqrt{1-v_b^2} \; [1+ u_b]. \label{eq:SR:L_SPH}
\ee

\subsubsection{The momentum equation}
The Euler-Lagrange equations, Eq.~(\ref{eq:EL_DGL}), determine the evolution
of special-relativistic momentum:
\be
\vec{p}_a \equiv \frac{\partial L_{\rm SPH,sr}}{\partial\vec{v}_a}
= - \sum_b \nu_b \frac{\partial}{\partial \vec{v}_a} 
\left(\frac{1+ u(n_b,s_b)}{\gamma_b}\right).
\ee
The specific energy $u$ depends, via the density, on the velocity:
\be
\frac{\partial u_b}{\partial \vec{v}_a}= \left(\frac{\partial u_b}{\partial
    n_b}\right)_s \frac{\partial n_b}{\partial \vec{v}_a}
= \frac{P_b}{n_b^2} \frac{\partial n_b}{\partial \vec{v}_a},
\ee 
where we have used Eq.~(\ref{eq:first_law_4}). Via Eq.~(\ref{eq:N_vs_n}) we have
\be
\frac{\partial n_b}{\partial \vec{v}_a}= N_b \frac{\partial}{\partial
  \vec{v}_a} \left(\frac{1}{\gamma_b}\right)= N_b \frac{\partial}{\partial
  \vec{v}_a} (1-v_b^2)^{1/2}= - N_b \gamma_b \vec{v}_b \delta_{ab}.
\ee
Thus, the canonical momentum is
\bea
\vec{p}_a&=& \frac{\partial L_{\rm SPH,sr}}{\partial\vec{v}_a}=  - \sum_b \nu_b [1+u_b]
\frac{\partial}{\partial \vec{v}_a}\left(\frac{1}{\gamma_b}\right)  -  \sum_b
\frac{\nu_b}{\gamma_b} \frac{\partial u_b}{\partial \vec{v}_a}\nonumber\\
&=& \sum_b \nu_b (\vec{v_b} \gamma_b \delta_{ab}) [1+u_b]
+ \sum_b \frac{\nu_b}{\gamma_b} \left(\frac{P_b}{n_b^2}\right)(N_b \gamma_b
\vec{v}_b \delta_{ab})\nonumber\\
&=& \nu_a \gamma_a \vec{v}_a \left(1+u_a+\frac{P_a}{n_a}\right),\label{eq:SR:dL_dva}
\eea
where we have again used Eq.~(\ref{eq:N_vs_n}), the term in brackets after 
the last equal sign is the enthalpy per baryon. Obviously, in relativity the
pressure contributes to the momentum density.
We define the canonical momentum per baryon as
\be
\vec{S}_a\equiv \gamma_a \vec{v}_a \left(1+u_a+\frac{P_a}{n_a}\right) \label{eq_Sa},
\ee
whose time evolution is governed by
$\partial L_{\rm SPH,sr}/\partial\vec{r}_a$, see Eq.~(\ref{eq:EL_DGL}), 
and which requires the gradient of the number density. The latter is obtained 
similar to the Newtonian case, see Eq.~(\ref{eq:drho_dx}),
\bea
\frac{\partial N_b}{\partial \vec{r}_a}&=& \sum_k \nu_k \left\{ 
\frac{\partial W(r_{bk},h_b)}{\partial r_{bk}}
\frac{\partial r_{bk}} {\partial \vec{r}_a} + \frac{\partial W(r_{bk},h_b)}
{\partial h_b} \frac{\partial h_b}{\partial N_b} \frac{\partial N_b}{\partial \vec{r}_a}\right\}\nonumber\\
&=& \frac{1}{\tilde{\Omega}_b} \sum_k \nu_k \nabla_a W_{bk}(h_b) (\delta_{ba} - \delta_{ka})
\label{eq:SR:dNb_dra}
\eea
with
\be
\tilde{\Omega}_b= 1 - \frac{\p h_b}{\p N_b} \sum_k \nu_k \frac{\p W(r_{bk},h_b)}{\p h_b}.
\label{eq:SR:omega_tilde}
\ee
Thus
\bea
\frac{\partial L_{\rm SPH,sr}}{\partial\vec{r}_a}
&=& - \sum_b \frac{\nu_b}{\gamma_b} \frac{\partial u_b}{\partial \vec{r}_a}
= - \sum_b \frac{\nu_b}{\gamma_b} \frac{P_b}{n_b^2} \frac{\partial
  n_b}{\partial \vec{r}_a} = - \sum_b
\frac{\nu_b}{\gamma_b^2} \frac{P_b}{n_b^2} \frac{\partial N_b}{\partial \vec{r}_a}\nonumber\\
&=& - \sum_{b,k} \frac{\nu_b \nu_k}{\gamma_b^2}
\frac{P_b}{\tilde{\Omega}_b n_b^2} \nabla_a W_{bk}(h_b) (\delta_{ba}-\delta_{ka})\nonumber\\
&=& - \nu_a \sum_{b} \nu_b \left\{\frac{P_a}{\tilde{\Omega}_a N_a^2} \nabla_a W_{ab}(h_a) + 
\frac{P_b}{\tilde{\Omega}_b N_b^2} \nabla_a W_{ab}(h_b)
\right\},\label{eq:SR:dL_dra}
\eea
where we have used Eqs.~(\ref{eq:first_law_4}), (\ref{eq:N_vs_n}), (\ref{eq:dens_summ_SR}),
(\ref{eq:SR:dNb_dra}), and (\ref{eq:k4}). With Eq.~(\ref{eq:SR:dL_dra}) and  Eq.~(\ref{eq:SR:dL_dva}),
the special-relativistic SPH momentum equation becomes
\be
\frac{d\vec{S}_a}{dt}= - \sum_{b} \nu_b \left\{\frac{P_a}{\tilde{\Omega}_a N_a^2} \nabla_a W_{ab}(h_a) + 
\frac{P_b}{\tilde{\Omega}_b N_b^2} \nabla_a W_{ab}(h_b)
\right\} \label{eq:SR:mom_eq}.
\ee
This form can be obtained from  the non-relativistic case, Eq.~(\ref{eq:SPH_from_Lag_dvdt}), by
the following replacements: 
$\vec{v}_a \rightarrow \vec{S}_a$,
 $m_k \rightarrow \nu_k$,
$\Omega_k \rightarrow \tilde{\Omega}_k$ and
$\rho_k \rightarrow N_k$.\\
Since the Lagrangian is invariant under infinitesimal translation the total
canonical momentum
\be
\vec{P}= \sum_b \vec{p}_b = \sum_b \nu_b \vec{S}_b= \sum_b \nu_b \gamma_b
\vec{v}_b \; \left(1+u_b+\frac{P_b}{n_b}\right)
\ee
is conserved. Similarly, due to invariance under rotation, the angular
momentum
\be
\vec{L}= \sum_b \vec{r}_b \times \vec{p}_b = \sum_b \nu_b \vec{r}_b \times \vec{S}_b= \sum_b \nu_b \gamma_b
 \left(1+u_b+\frac{P_b}{n_b}\right) \vec{r}_b \times \vec{v}_b
\ee
is conserved by construction.

\subsubsection{The energy equation}
To choose a suitable numerical energy variable
we start from the conserved canonical energy 
(no explicit time dependence in the Lagrangian) 
\be
E \equiv \sum_a \frac{\partial L_{\rm SPH,sr}}{\partial \vec{v}_a} \cdot \vec{v}_a - L_{\rm SPH,sr}.
\ee
With Eqs.~(\ref{eq:SR:L_SPH}) and Eq.~(\ref{eq:SR:dL_dva}) the energy reads
\bea
E &=& \sum_a \nu_a \left(\vec{v}_a \cdot \vec{S}_a +
  \frac{1+u_a}{\gamma_a}\right) = \sum_a \nu_a \hat{\epsilon}_a
\eea
where, we have defined
\be
\hat{\epsilon}_a \equiv
\vec{v}_a \cdot \vec{S}_a + \frac{1+u_a}{\gamma_a}.\label{eq:SR:epsilon_a}
\ee
For later use, we note that by using $v_a^2= 1 - 1/\gamma_a^2$ and 
Eq.~(\ref{eq:N_vs_n}) we can express
$\hat{\epsilon}_a$ as
\bea
\hat{\epsilon}_a= \gamma_a v_a^2 \left(1+u_a+\frac{P_a}{n_a}\right)  +
\frac{1+u_a}{\gamma_a} = \gamma_a \left(1+u_a+\frac{P_a}{n_a}\right) -
  \frac{P_a}{N_a}.
\label{eq:SR:epsilon2}
\eea

To find the evolution equation for $\hat{\epsilon}_a$, we need the time 
derivative of the second term in Eq.~(\ref{eq:SR:epsilon_a}). Ideally, it should not contain a time 
derivative of the Lorentz factor since such terms are known to destabilize 
numerical schemes \cite{norman86}.  
Using 
\be
\frac{d\gamma_a}{dt}= \gamma_a^3 \vec{v}_a  \cdot \frac{d\vec{v}_a}{dt}
\quad \;
{\rm and}
\quad \;
\frac{d n_a}{dt}= \frac{d}{dt}\left(\frac{N_a}{\gamma_a}\right)=
\frac{1}{\gamma_a} \frac{d N_a}{dt} - N_a \gamma_a \vec{v}_a \cdot
\frac{d\vec{v}_a}{dt} 
\label{eq:SR:dgamma_dt}
\ee
we find
\bea
\frac{d}{dt} \left(\frac{1+u_a}{\gamma_a}\right)&=& \frac{1}{\gamma_a^2}
\left(\gamma_a \frac{du_a}{dt} - (1+u_a) \frac{d\gamma_a}{dt}\right)\nonumber\\
&=& \frac{1}{\gamma_a} \frac{P_a}{n_a^2} \frac{d n_a}{dt} -
\frac{1+u_a}{\gamma_a^2} \gamma_a^3 \vec{v}_a  \cdot \frac{d\vec{v}_a}{dt}\nonumber\\
&=& \frac{P_a}{N_a^2} \frac{d N_a}{dt} - \left(1+ u_a+
  \frac{P_a}{n_a}\right)\gamma_a \vec{v}_a \cdot
\frac{d\vec{v}_a}{dt}\nonumber\\
&=& \frac{P_a}{N_a^2} \frac{dN_a}{dt} - \vec{S}_a \cdot \frac{d\vec{v}_a}{dt},\label{eq:SR:h1}
\eea
where we used Eqs.~(\ref{eq:first_law_4}), (\ref{eq:N_vs_n}) and (\ref{eq:SR:dgamma_dt}).
With equation (\ref{eq:SR:h1}) at hand, we can take the derivative of
Eq.~(\ref{eq:SR:epsilon_a})
\bea
\frac{d \hat{\epsilon}_a}{dt} &=& \frac{d}{dt} \left\{\vec{v}_a \cdot
  \vec{S}_a + \frac{1+u_a}{\gamma_a} \right\}\nonumber\\
&=& \vec{S}_a \cdot \frac{d\vec{v}_a}{dt}  + \vec{v}_a
\cdot \frac{d\vec{S}_a}{dt} + \frac{P_a}{N_a^2} \frac{dN_a}{dt} - \vec{S}_a
\cdot \frac{d\vec{v}_a}{dt} \nonumber\\
&=& \vec{v}_a \cdot \frac{d\vec{S}_a}{dt} + \frac{P_a}{N_a^2}
\frac{dN_a}{dt},
\label{eq:SR:deps_dt}
\eea
i.e. apart from the relativistic momentum equation (\ref{eq:SR:mom_eq}),
we need the Lagrangian time derivative of the number density in the
computing frame:
\bea
\frac{dN_a}{dt} &=& \sum_b \nu_b \left\{ \frac{\p W_{ab}(h_a)}{\p r_{ab}} 
\frac{d r_{ab}}{dt} + \frac{\p W_{ab}}{\p h_a} \frac{\p h_a}{\p N_a} \frac{dN_a}{dt}   
 \right\}\nonumber\\
&=& \frac{1}{\tilde{\Omega}_a} \sum_b \nu_b \vec{v}_{ab} \cdot \nabla_a W_{ab}(h_a)
, \label{eq:SR:dN_dt}
\eea
where we have used Eq.~(\ref{eq:SR:omega_tilde}). If we insert Eqs.~(\ref{eq:SR:mom_eq}) 
and (\ref{eq:SR:dN_dt}) into Eq.~(\ref{eq:SR:deps_dt}) the energy evolution equation becomes
\be
\frac{d \hat{\epsilon}_a}{dt} = - \sum_b \nu_b 
\left(
\frac{P_a \vec{v}_b}{\tilde{\Omega}_a N_a^2} \cdot \nabla_a W_{ab}(h_a) + 
\frac{P_b \vec{v}_a}{\tilde{\Omega}_b N_b^2} \cdot \nabla_a W_{ab}(h_b)
\right).
\label{eq:ener_eq_no_diss}
\ee 
This relativistic energy equation looks similar to the non-relativistic
equation for the thermokinetic energy, see
Eq.~(\ref{eq:basic:energy_equation_e}).

\subsubsection{Recovery of the primitive variables}
Due to our choice of the numerical variables neither the momentum nor 
the energy equation contain time derivatives of Lorentz factors. 
Such terms restricted earlier versions of relativistic SPH to moderate 
Lorentz factors only. But this comes at the price that the physical variables
need to be recovered at every time step from the numerical ones, a price that
relativistic Eulerian codes also have to pay.\\
\begin{figure}
\hspace*{-1cm}\includegraphics[width=6.5in]{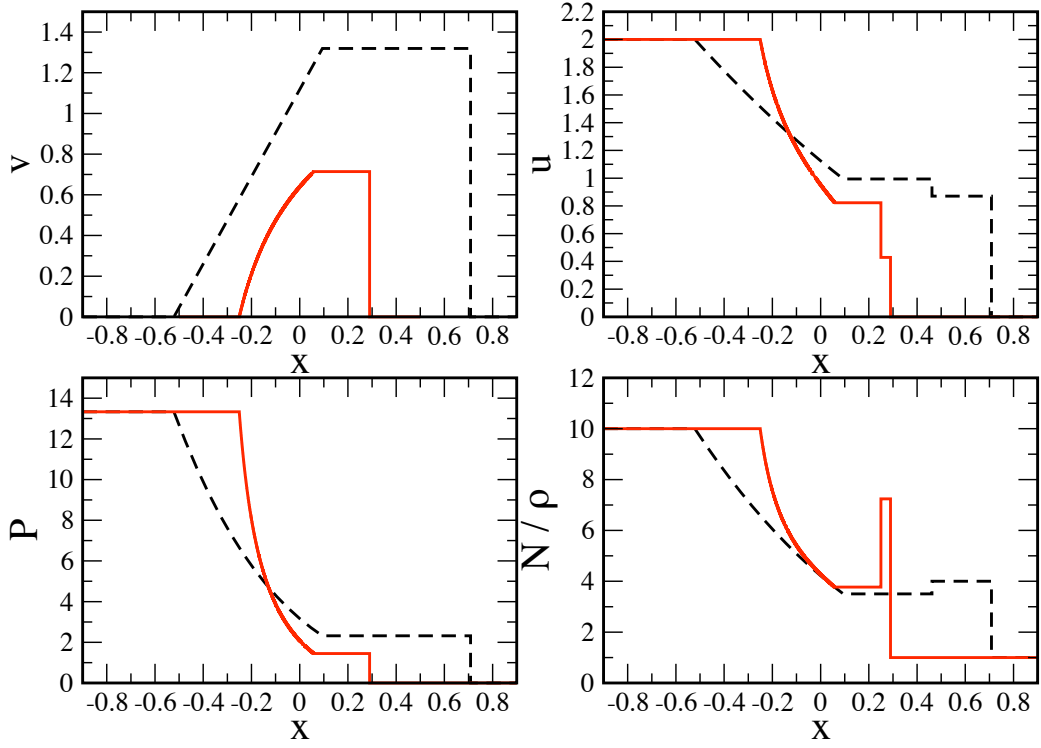}
\caption{Mildly relativistic ($\gamma_{\rm max} \approx 1.4$) shock tube test: 
comparison between the Newtonian (dashed) and the special-relativistic results 
(solid, red). Upper left: velocity in units of $c$, upper right: thermal energy, 
lower left: pressure, lower right: (computing frame) number density $N$ and Newtonian
mass density $\rho$.}
\label{fig:SR:shock_tube_comparison_N_SR}
\end{figure} 
At the end of a time step $\gamma$/$v$, $u$ and $n$ need to be calculated 
from the updated numerical quantities $N$, $\vec{S}$ and $\hat{\epsilon}$. 
One can express all variables in the equation of state
\be
P_a= (\Gamma-1) n_a u_a 
\label{eq:polytrope}
\ee
as a function of the updated numerical variables and the pressure 
$P_a$. The resulting equation is solved numerically for $P_a$ and once
it is found, the other physical variables can be 
recovered \cite{marti96,monaghan97}.
From Eqs.~(\ref{eq_Sa}) and (\ref{eq:SR:epsilon2}) one finds
\be
\vec{v}_a= \frac{\vec{S}_a}{\hat{\epsilon}_a+P_a/N_a}
\label{eq:v_of_S_eps}
\ee
and thus the Lorentz factor is
\be
\gamma_a= \frac{1}{\sqrt{1-S_a^2/(\hat{\epsilon}_a+P_a/N_a)^2}}.
\label{eq:gamma_of_P}
\ee
From Eq.~(\ref{eq:v_of_S_eps}) we have
\be
\left(\hat{\epsilon}_a + \frac{P_a}{\gamma_a n_a}\right) \vec{v}_a
= \vec{S}_a = \gamma_a \vec{v}_a \left(1+u_a+\frac{P_a}{n_a}\right)
\ee
which can be solved for
\be
u_a= \frac{\hat{\epsilon}_a}{\gamma_a} + \frac{P_a}{\gamma_a N_a} (1-\gamma_a^2)-1. 
\label{eq:u_of_eps}
\ee
With aid of Eqs.~(\ref{eq:N_vs_n}) and (\ref{eq:u_of_eps}),
Eq.~(\ref{eq:polytrope}) can be solved, e.g. via a Newton-Raphson scheme,
for the new pressure $P_a$. Once $P_a$ is known, the Lorentz factor can 
be calculated from Eq.~(\ref{eq:gamma_of_P}), the specific energy from 
Eq.~(\ref{eq:u_of_eps}) and the velocity from Eq.~(\ref{eq:v_of_S_eps}). \\

\subsubsection{Numerical tests}
\begin{figure}
\centerline{
\includegraphics[width=3in]{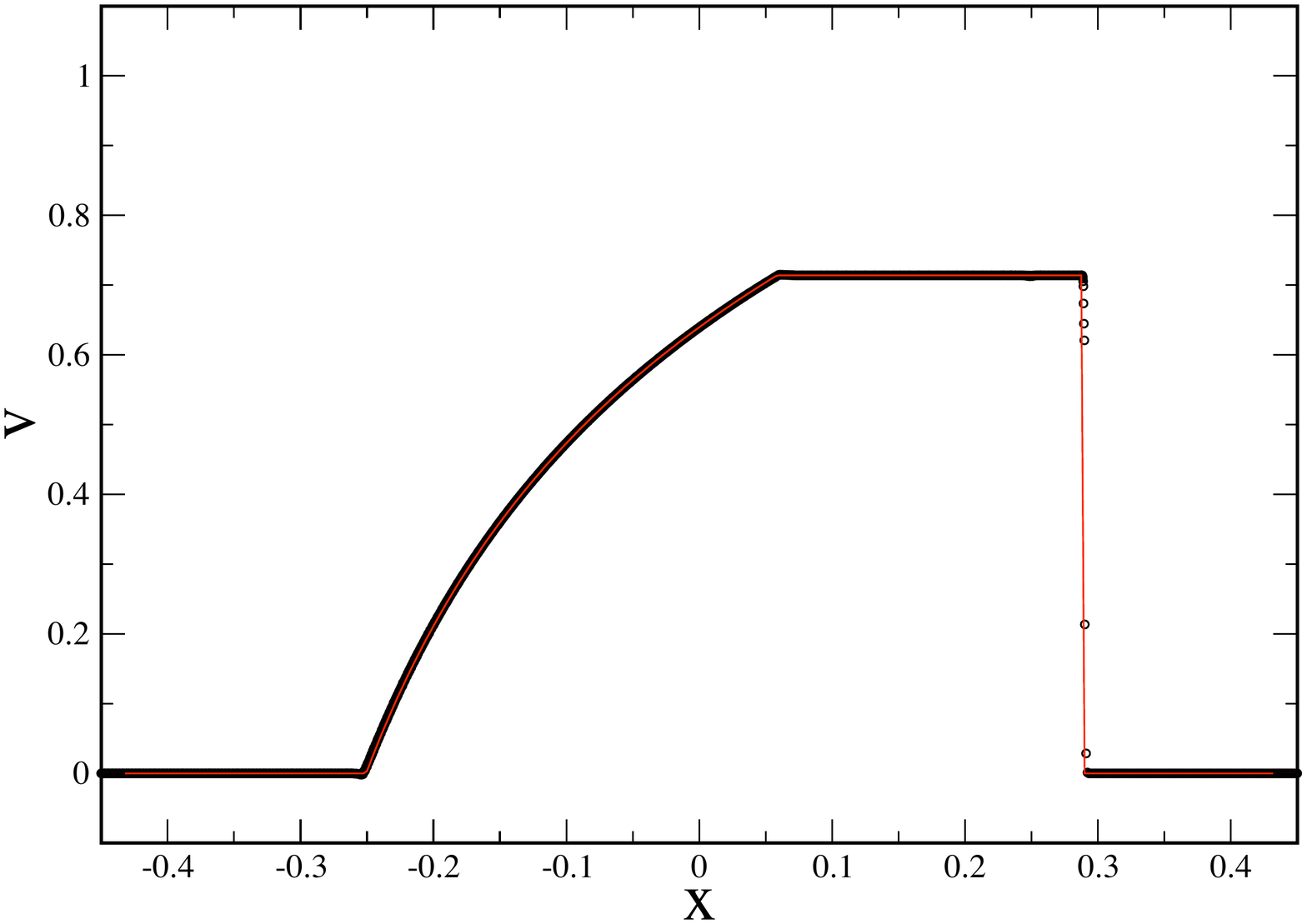}
\hspace*{-0.8cm}
\includegraphics[width=3in]{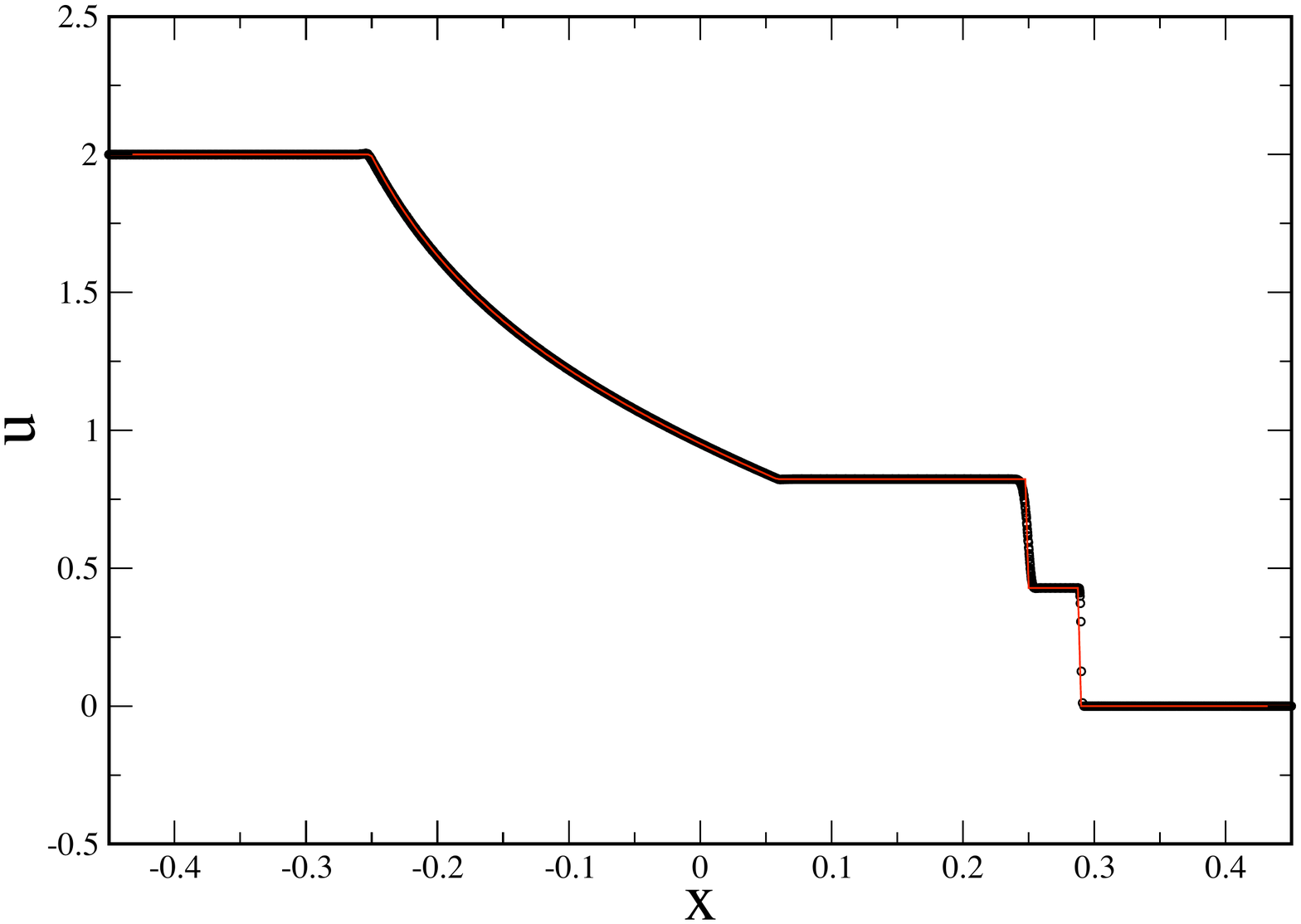}}
\vspace*{-0.45cm}
\centerline{
\includegraphics[width=3in]{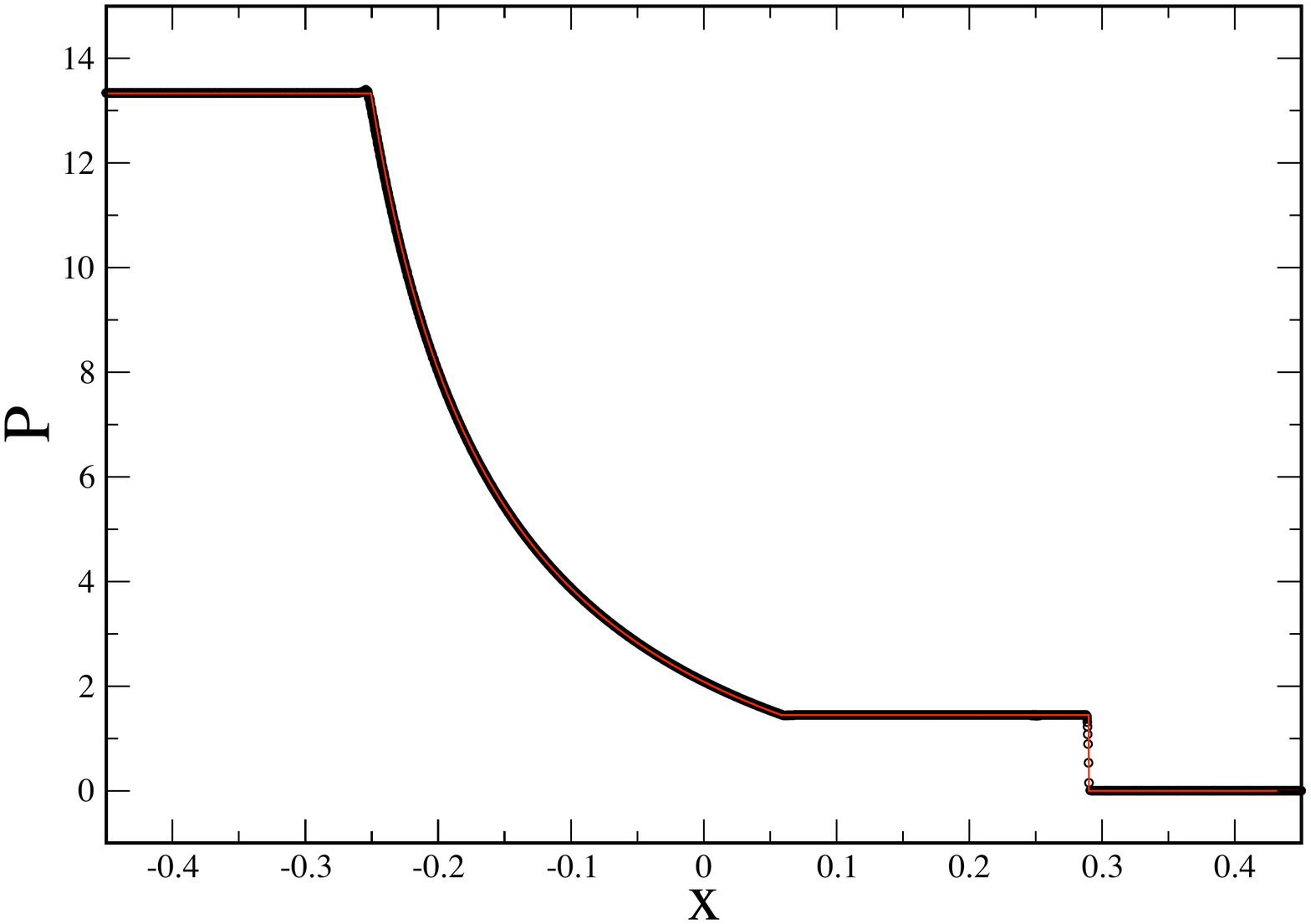}
\hspace*{-0.8cm}
\includegraphics[width=3in]{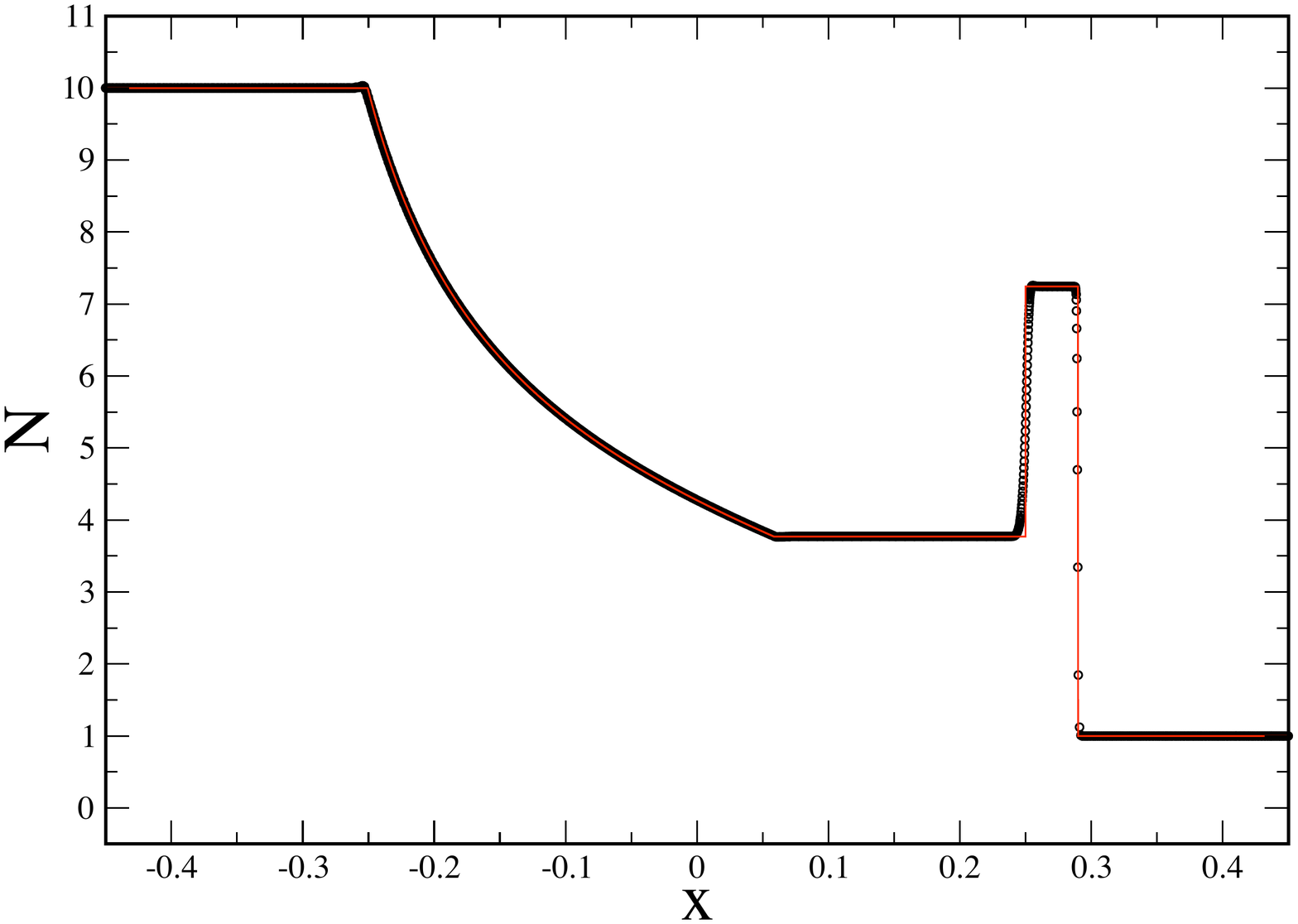}}
\vspace*{-0.cm}
\caption{SPH solution of the relativistic shock tube test of \cite{marti96} 
(shown are 3000 particles; 
upper left: velocity in units of $c$, upper right: thermal energy, lower 
left: pressure, lower right: (computing frame) number density $N$).
The black circles show the SPH solution, the red line marks the exact 
solution \cite{marti03}.}
\label{fig:SR:shock_tube}
\end{figure}

To explore the performance of this SPH formulation we show a numerical 
test that can be considered a special-relativistic generalization of 
Sod's shock tube test, see Sec. \ref{subsec:artificial_dissi}.
This test has become a widespread benchmark for relativistic hydrodynamics codes, 
see for example \cite{marti96,chow97,siegler00a,delZanna02,marti03}. The test is
performed with a polytropic equation of state with exponent $\Gamma= 5/3$, vanishing 
initial velocities everywhere, the left state has a pressure $P_L=40/3$ and a density 
$N_L=10$, while the right state is prepared with $P_R=10^{-6}$ and $N_R=1$. 
Although the resulting velocities are only mildly relativstic ($\gamma_{\rm max} \approx 1.4$), 
the deviations from the purely Newtonian result are already substantial. This is
demonstrated in Fig.~\ref{fig:SR:shock_tube_comparison_N_SR}, where we compare the 
exact solutions for these initial conditions for the Newtonian (dashed) and the
special-relativistic case (solid).\\
Fig.~\ref{fig:SR:shock_tube} shows the SPH result (black circles,
about 3000 particles) at t= 0.35 together with the exact solution (red, solid line). 
The general agreement is excellent, noticeable deviations are only observed 
in the form of a slightly smeared out contact discontinuity at $x\approx 0.25$. 
A striking difference to earlier SPH formulations \cite{laguna93a,siegler00a}
is the absence of any spike in $u$ and $P$ at the contact discontinuity.\\
This SPH formulation has been further explored in a large set of special-relativistic 
benchmark tests \cite{rosswog09d}. As expected, it performs very well 
in pure advection problems. Maybe more remarkable, it also shows convincing results in extremely 
strong, relativistic shocks. For example, it yields accurate solutions in a wall
shock test with a Lorentz factor of $\gamma= 50 \; 000$. The special-relativistic 
``grad-h''-terms generally improve the accuracy, but (at least in the performed set of 
tests) only to a moderate extent. For more details we refer to \cite{rosswog09d}.
\clearpage
%
%
\hspace*{0cm}\fbox{
\parbox{14cm}{
\vspace*{0.5cm}
\centerline{\bf Summary of the special-relativistic SPH equations}

\vspace*{0.5cm}

The momentum equation derived from the special-relativistic Lagrangian reads 
\be
\frac{d\vec{S}_a}{dt}= - \sum_{b} \nu_b \left\{\frac{P_a}{\tilde{\Omega}_a N_a^2} \nabla_a W_{ab}(h_a) + 
\frac{P_b}{\tilde{\Omega}_b N_b^2} \nabla_a W_{ab}(h_b)
\right\}
\label{eq:SR_summary:mom_eq}
\ee
where 
\be
\vec{S}_a= \gamma_a \vec{v}_a
\left(1+u_a+\frac{P_a}{n_a}\right)\label{eq:SR_summary:mom_var} 
\ee
and 
\be
\tilde{\Omega}_b= 1 - \frac{\p h_b}{\p N_b} \sum_k \nu_k \frac{\p W(r_{bk},h_b)}{\p h_b}.
\ee
The (canonical) energy variable
\be
\hat{\epsilon}_a = \vec{v}_a \cdot \vec{S}_a + \frac{1+u_a}{\gamma_a}
\ee
is evolved according to
\be
\frac{d \hat{\epsilon}_a}{dt} = - \sum_b \nu_b 
\left(
\frac{P_a \vec{v}_b}{\tilde{\Omega}_a N_a^2}\cdot \nabla_a W_{ab}(h_a) + 
\frac{P_b \vec{v}_a}{\tilde{\Omega}_b N_b^2}\cdot \nabla_a W_{ab}(h_b)
\right).
\label{eq:SR_summary:en_eq}
\ee
The computing frame number density can be found either by summation,
\be
N_a= \sum_b \nu_b W_{ab}(h_a),\label{eq:SR_summary:dens}
\ee
or, alternatively, by integration of
\be
\frac{dN_a}{dt} = \frac{1}{\tilde{\Omega}_a} \sum_b \nu_b \vec{v}_{ab} \cdot \nabla_a W_{ab}(h_a).\label{eq:SR_summary:dN_dt}
\ee
}}

\subsection{General-relativistic SPH on a fixed background metric}
\label{chap:GR}
\label{page_GR_SPH}
\subsubsection{The strategy}
We now generalize the previous approach by assuming that a fixed metric 
is given as a function of 
the coordinates and that corrections to the metric induced by the fluid 
are negligible. Such a situation occurs, for example, during the tidal 
disruption of a star by a supermassive black hole in the center of a galaxy.
Again, we start from the discretized Lagrangian of an ideal fluid and 
use the canonical momentum/energy as a guide for the choice of the numerical
variables. As a matter of course, this new set of equations should reduce 
in the flat-space limit to the special-relativistic equations, see
Eq.~(\ref{eq:SR_summary:mom_eq}) - (\ref{eq:SR_summary:dens}), which, in 
their low-velocity limit, should be equivalent to the non-relativistic, 
standard SPH equations. 

\subsubsection{The Lagrangian}
We start from the general-relativistic Lagrangian of a perfect fluid
 \cite{fock64}
 \be
L_{\rm pf,GR}= - \int T^{\mu\nu} U_\mu U_\nu \; \sqrt{-g}\; dV,\label{eq:fluid_Lag_GR}
\ee
where $\sqrt{-g}\; dV$ is the proper volume element and $g= {\rm det}(g_{\mu
  \nu})$ is the determinant of the metric tensor, $g_{\mu  \nu}$. As before,
the energy momentum tensor is defined by
\be
T^{\mu\nu}= \{n [1 + u(n,s)] + P\}  U^\mu U^\nu + P g^{\mu\nu},
\label{eq:GR:en_mom_tensor} 
\ee
where the fluid quantities are measured in their local rest frame and the
four-velocity is defined as in Eq.~(\ref{eq:SR:4vel}). Obviously,
\be
\frac{d\tau}{dt}=
\left(- g_{\mu\nu} \frac{dx^{\mu}}{dt}
  \frac{dx^{\nu}}{dt}\right)^{\frac{1}{2}}= 
\left(- g_{\mu\nu} v^{\mu} v^{\nu}\right)^{\frac{1}{2}}, \label{eq:GR:dtau_dt}
\ee
where we used $v^\rho= dx^\rho/dt$. Comparing with the special-relativistic 
relation $d\tau/dt = 1/\gamma$, one sees that the quantity 
\be
\Theta \equiv \left(- g_{\mu\nu} v^{\mu} v^{\nu}\right)^{-\frac{1}{2}}
\ee
takes over the role of and reduces in the flat spacetime limit to the 
special-relativistic Lorentz factor $\gamma= 1/\sqrt{1-v^2}$. Similar to 
Eq.~(\ref{eq:SR:3vel}) we can also write
\be
v^{\mu}= \frac{dx^\mu}{dt}= \frac{dx^\mu}{d\tau}\frac{d\tau}{dt}=
\frac{U^\mu}{U^0}, \label{eq:GR:v_mu}
\ee
where we have used $dt/d\tau= U^0= \Theta$, see Eq.~(\ref{eq:GR:dtau_dt}).
Using
the normalization condition of the four-velocity, Eq.~(\ref{eq:SR:U_norm}), the Lagrangian reduces to
\bea
L_{\rm pf,GR}
&=& -\int \left[\left\{n(1+u)+P\right\}U^\mu U^\nu U_\mu U_\nu + P g^{\mu \nu}
  U_\mu U_\nu\right] \; \sqrt{-g}\; dV\nonumber \\ 
&=& -\int n[1+u(n,s)] \; \sqrt{-g}\; dV \label{eq:GR:cont_Lag},
\eea
which is, of course, up to the volume element, the same result as in special
relativity, Eq.(\ref{eq:fluid_Lag_SRT_simp}).\\
\noindent To discretize the Lagrangian, we first have to find a suitable
density variable so that we can apply the usual SPH discretization
prescription. The general-relativistic baryon number conservation
equation reads  
\be
\frac{1}{\sqrt{-g}} \frac{\partial}{\partial x^{\nu}} \left(\sqrt{-g}  U^\nu n
\right)= 0,
\ee 
see e.g. \cite{schutz89}, and this suggests to use the modified number density
\be
N^\ast= \sqrt{-g} \; U^0 \; n= \sqrt{-g} \; \Theta \; n \label{eq:GR:Nstar_n},
\ee
since it can be written with the help of Eq.~(\ref{eq:GR:v_mu}) as
\be
\frac{\partial N^\ast}{\partial t} + \frac{\partial (N^\ast v^i)}{\partial
  x^i}= 0.
\ee
Eq.~(\ref{eq:GR:Nstar_n}) generalizes the earlier relation between the 
computing frame and the local rest frame, Eq.~(\ref{eq:N_vs_n}).
We use $N^\ast$ in the SPH interpolations
\be
f(\vec{r})= \sum_b f_b \frac{\nu_b}{N^\ast_b}
W(|\vec{r}-\vec{r}_b|),\label{eq:GR:SPH_discret} 
\ee
where 
\be
\nu_b= N_b^\ast \Delta V_b = n_b \Theta_b
\sqrt{-g_b} \Delta V_b 
\ee
is the number of baryons contained in the volume $\sqrt{-g_b} \Delta
V_b$. Applying Eq.~(\ref{eq:GR:SPH_discret}) to $N^\ast$ one obtains
a density estimate by summation
\be
N_a^\ast= \sum_b \nu_b W_{ab}(h_a),
\label{eq:GR:num_dens}
\ee
similar to the earlier results, Eqs.~(\ref{eq:Advanced_rho}) and 
(\ref{eq:dens_summ_SR}). We can now use the relation between $N^\ast$ and $n$, 
Eq.~(\ref{eq:GR:Nstar_n}), to write the Lagrangian as
\bea
\hspace*{-0.4cm}L_{\rm pf,GR}&=& -\int \frac{N^\ast}{\sqrt{-g} \; \Theta} \; [1+u(n,s)] \;
\sqrt{-g}\; dV 
= -\int \frac{N^\ast}{\Theta} [1+u(n,s)] dV,
\eea
i.e. the metric tensor contribution $\sqrt{-g}$ drops out, and using 
 $\Delta V_b= \nu_b/N_b^\ast$, the SPH form of the Lagrangian reads
\be
L_{\rm SPH,GR}= - \sum_b \frac{\nu_b}{\Theta_b}
\left[1+u(n_b,s_b)\right].\label{eq:GR:L_SPH} 
\ee
In the flat spacetime limit $\Theta_b \rightarrow \gamma_b$ and we recover the
discretized, special-relativistic Lagrangian, Eq.~(\ref{eq:SR:L_SPH}).

\subsubsection{The momentum equation}
To identify a suitable numerical variable, we use like before, see Eq.~(\ref{eq_Sa}),
the canonical momentum
\bea
\frac{\partial L}{\partial v^i_a}
&=& - \sum_b \nu_b  \left(\frac{\partial }{\partial v^i_a}
  \frac{1}{\Theta_b}\right) [1 + u_b] - \sum_b \frac{\nu_b}{\Theta_b}
\left(\frac{\partial }{\partial v^i_a} u_b\right).
\label{eq:GR:can_mom}
\eea
We thus need
\be
\frac{\partial }{\partial v^i_a} \left(\frac{1}{\Theta_b}\right)=
\frac{\partial }{\partial v^i_a} (-g_{\mu\nu} v^\mu v^\nu)^{1/2}_b
= - \Theta_b \; (g_{i\mu} v^\mu)_a \; \delta_{ab} \label{eq:GR:d_Theta_1dv}
\ee
and the derivative $\partial u_b / \partial v^i_a$, where, like in the 
special-relativistic case, the velocity dependence comes from the Lorentz-factor-like
quantity $\Theta$ relating the densities in the local frame and the computing
frame
\bea
\frac{\partial u_b}{\partial v^i_a} &=& \frac{\partial u_b}{\p n_b} \frac{\p
  n_b}{\p v^i_a}= \frac{P_b}{n_b^2} \frac{\p}{\p v^i_a}\left(
  \frac{N^\ast_b}{\sqrt{-g}_b \; \Theta_b}\right) = \frac{P_b}{n_b^2}
\frac{N^\ast_b}{\sqrt{-g}_b} \frac{\p}{\p
  v^i_a}\left(\frac{1}{\Theta_b}\right)\nonumber\\
&=& - \frac{P_b}{n_b^2} \frac{N^\ast_b}{\sqrt{-g}_b} \Theta_b (g_{i\mu}
v^\mu)_a \; \delta_{ab}.\label{eq:GR:du_dv}
\eea
Here we have used Eqs.~(\ref{eq:first_law_4}), (\ref{eq:GR:Nstar_n}) and  
(\ref{eq:GR:d_Theta_1dv}). By inserting Eqs.~(\ref{eq:GR:d_Theta_1dv}) and 
(\ref{eq:GR:du_dv}) into Eq. (\ref{eq:GR:can_mom}) we find
\bea
\frac{\partial L}{\partial v^i_a} = && 
-\sum_b \nu_b [1+u_b] \; (- \Theta_b \; (g_{i\mu} v^\mu)_a \; \delta_{ab})\
 -\sum_b \frac{\nu_b}{\Theta_b} \left(-\frac{P_b}{n_b^2}
  \frac{N^\ast_b}{\sqrt{-g}_b} \Theta_b \; (g_{i\mu} v^\mu)_a \; \delta_{ab} \right)\nonumber \\
=&& \nu_a \Theta_a \left( 1 + u_a + \frac{P_a}{n_a^2}
  \frac{N^\ast_a}{\sqrt{-g}_a \Theta_a}\right)\; (g_{i\mu}
v^\mu)_a\nonumber \\
=&& \nu_a \Theta_a \left(1 + u_a + \frac{P_a}{n_a} \right) \; (g_{i\mu}
v^\mu)_a, \label{eq:GR:dL_dv}
\eea
where Eq.~(\ref{eq:GR:Nstar_n}) has been used.
This is the $i$-component of the canonical momentum and, like before, we use the
canonical momentum per baryon,
\be
S_{i,a}\equiv \frac{1}{\nu_a} \frac{\partial L}{\partial v^i_a} =\Theta_a
\left(1 + u_a + \frac{P_a}{n_a} \right) \; (g_{i\mu} v^\mu)_a,
\label{eq:GR:mom_var}
\ee
as numerical variable. Again, the term in the first bracket is the specific
enthalpy. In the flat-space limit, 
$\Theta_a \rightarrow \gamma_a$, $g_{\mu\nu}
\rightarrow \eta_{\mu\nu}$ and $g_{i\mu} v^\mu \rightarrow v_i= v^i$, this
expression reduces exactly to the special-relativistic case,
Eq.~(\ref{eq:SR_summary:mom_var}), as it should.\\
The evolution of $S_{i,a}$ is determined by 
the Euler-Lagrange equations, so we need 
\bea
\frac{\p L}{\p x^i_a} &=& -\sum_b \nu_b \left(\frac{\p}{\p x^i_a}
  \frac{1}{\Theta_b}\right) [1+u_b] -\sum_b \frac{\nu_b}{\Theta_b} 
  \left(\frac{\p u_b}{\p x^i_a}\right). \label{eq:GR:dL_dx-1}
\eea
The first derivative can be written as
\bea
\left(\frac{\p}{\p x^i_a} \frac{1}{\Theta_b}\right) &=& \frac{\p}{\p x^i_a}
\left(- g_{\mu\nu} v^{\mu} v^{\nu}\right)^{\frac{1}{2}}_b
= -\frac{\Theta_b}{2} \left(\frac{\p g_{\mu\nu}}{\p x^i_a}  \right)_b
v^\mu_b v^\nu_b \; \delta_{ab},
\eea
and due to Eq.~(\ref{eq:GR:v_mu})
\be
 v^\mu v^\nu = \frac{U^\mu U^\nu}{\Theta^2},\nonumber
\ee
so it becomes
\bea
\left(\frac{\p}{\p x^i_a} \frac{1}{\Theta_b}\right) 
&=& -\left(\frac{U^\mu U^\nu}{2 \Theta}\right)_b \left(\frac{\p g_{\mu\nu}}{\p x^i_a} 
\right)_b  \; \delta_{ab}. \label{eq:GR:d_1_theta_dx}
\eea
The derivative of the internal energy is
\bea
\frac{\p u_b}{\p x^i_a}= && \frac{\p u_b}{\p n_b} \frac{\p n_b}{\p x^i_a}
= \frac{P_b}{n_b^2} \frac{\p}{\p x^i_a}\left(\frac{N^\ast_b}{\sqrt{-g}_b \;
    \Theta_b} \right) \nonumber\\
= && \frac{P_b}{n_b^2 \sqrt{-g}_b \; \Theta_b}\left(\frac{\p N^\ast_b}{\p
    x^i_a} \right) + \frac{P_b N^\ast_b}{n_b^2 \sqrt{-g}_b} \left(\frac{\p}{\p
    x^i_a} \frac{1}{\Theta_b}\right)\nonumber\\
&&  + \frac{P_b N^\ast_b}{n_b^2
  \Theta_b} \left(\frac{\p}{\p x^i_a} \frac{1}{\sqrt{-g}_b}\right),
\label{eq:GR:du_dx}
\eea
where we have used the first law of thermodynamics, Eq.~(\ref{eq:first_law_4}).
By using Eq.~(\ref{eq:k3}), the derivative of the number density becomes
\bea
\frac{\p N^\ast_b}{\p x^i_a}&=& \frac{\p}{\p x^i_a} \left(\sum_k \nu_k W_{bk}
\right) = \sum_k \nu_k \frac{\p W_{bk}}{\p x^i_b} (\delta_{ba} - \delta_{ka}).
\label{eq:GR:dNdx}
\eea
The last remaining derivative is
\bea
\left(\frac{\p}{\p x^i_a} \frac{1}{\sqrt{-g}_b}\right)= 
- \left(\frac{g^{\alpha\beta}}{2 \sqrt{-g}}  \frac{\p g_{\alpha\beta}}{\p x^i_a}
\right)_b \delta_{ab},
\eea
see the box on the derivatives of the metric tensor determinant. 

%
%
\hspace*{0cm}\fbox{
\parbox{14cm}{
\vspace*{0.5cm}
\centerline{\bf Derivatives of the metric tensor determinant $g$}

\vspace*{0.5cm}

We will collect here some formulae related to the quantity $g$ that are needed
in several places. 
The derivative of the metric tensor determinant is generally given by (see
e.g. \cite{schutz89})
\be
\frac{\p g}{\p x^\mu}= g g^{\alpha\beta} \frac{\p g_{\alpha\beta}}{\p x^\mu}.
\ee
Therefore
\be
\frac{\p \sqrt{- g}}{\p x^i}
= \frac{\sqrt{-g}}{2} g^{\alpha\beta} \frac{\p g_{\alpha\beta}}{\p x^i},
\quad \quad 
\frac{\p}{\p x^i} \left( \frac{1}{\sqrt{-g}} \right)= -\frac{g^{\alpha\beta}}{2 \sqrt{-g}}  \frac{\p g_{\alpha\beta}}{\p x^i}\label{eq:GR:dsqrtg1}
\ee
and
\be
\frac{d}{dt} \left(\sqrt{-g}\right) = \frac{\p}{\p x^\mu} \left(\sqrt{-g} \right)
\frac{dx^\mu}{dt} = \frac{\sqrt{-g}}{2} g^{\alpha\beta} \frac{\p
  g_{\alpha\beta}}{\p x^u} v^\mu = \frac{\sqrt{-g}}{2} g^{\alpha\beta} \frac{d g_{\alpha \beta}}{dt}.
\label{eq:GR:dsqrtg}
\ee
Of course, if the metric tensor at the position of particle $b$ is to be
differentiated with respect to a property of particle $a$, the appropriate
Kronecker deltas, $\delta_{ab}$, have to be applied.
}}

With the help of
 Eqs.~(\ref{eq:GR:d_1_theta_dx}), (\ref{eq:GR:du_dx}), (\ref{eq:GR:dNdx}) and (\ref{eq:GR:dsqrtg1})
Eq.~(\ref{eq:GR:dL_dx-1}) becomes
\bea
\frac{\p L}{\p x^i_a} = && +\sum_b \nu_b [1+u_b] \left(\frac{U^\mu U^\nu}{2
      \Theta}\right)_b \left(\frac{\p g_{\mu\nu}}{\p x^i_a}
\right)_b  \; \delta_{ab} \nonumber\\
 && -\sum_b \frac{\nu_b}{\Theta_b} 
\left\{
\frac{P_b}{n_b^2 \sqrt{-g}_b \Theta_b} \sum_k \nu_k \frac{\p W_{bk}}{\p x^i_b} 
(\delta_{ba} - \delta_{ka}) \right.\nonumber\\
 && \hspace*{1.7cm}+ \frac{P_b N^\ast_b}{n^2_b \sqrt{-g}_b} \left[ -\left(\frac{U^\mu
       U^\nu}{2 \Theta} \right)_b  
      \left(\frac{\p g_{\mu\nu}}{\p x^i_a} \right)_b  \;
      \delta_{ab}\right]\nonumber\\
&&\left. \hspace*{1.7cm}+ \frac{P_b N_b^\ast}{n_b^2 \Theta_b}
\left[ - \left(\frac{g^{\mu\nu}}{2\sqrt{-g}} \frac{\p g_{\mu\nu}}{\p x^i_a} \right)_b 
\delta_{ab}\right]
\right\}.
\eea
The terms that involve the derivatives of the kernel represent the 
hydrodynamic part of the equations, the others the action of gravity. 
After eliminating one sum in the hydrodynamic terms via the Kronecker 
symbol, relabeling the summation index from $k$ to $b$ and using kernel property
Eq.~(\ref{eq:k4}) they read
\bea
\left(\frac{\p L}{\p x^i_a}\right)_{\rm h}&=&
-\nu_a \sum_b \nu_b \left( \frac{P_a}{n_a^2 \sqrt{-g}_a \Theta_a^2} + 
\frac{P_b}{n_b^2 \sqrt{-g}_b \Theta_b^2} \right) \frac{\p  W_{ab}}{\p x^i_a}.
\eea
and on using Eq.~(\ref{eq:GR:Nstar_n}) 
\bea
\left(\frac{\p L}{\p x^i_a}\right)_{\rm h}
&=& -\nu_a \sum_b \nu_b \left(\frac{\sqrt{-g}_a P_a}
  {N_a^{\ast 2}} + \frac{\sqrt{-g}_b P_b}{N_b^{\ast 2}} \right) 
  \frac{\p  W_{ab}}{\p x^i_a}.
\eea
The terms that involve derivatives of the metric (``gravity terms'') 
can be written as
\bea
\left(\frac{\p L}{\p x^i_a}\right)_{\rm g}&=& \frac{\nu_a}{2\Theta_a} 
\left[ 
\left(1+u_a + \frac{P_a}{n_a^2}
  \frac{N_a^\ast}{\sqrt{-g}_a \Theta_a}\right) U^\mu U^\nu +
  \frac{P_a}{n_a^2}\frac{N_a^\ast}{\sqrt{-g}_a \Theta_a} g^{\mu\nu}\right]\nonumber\\
 && \left(\frac{\p g_{\mu\nu}}{\p x^i}  \right)_a.
\eea
If we apply once more Eq.~(\ref{eq:GR:Nstar_n}) and remember the form of the 
energy-momentum tensor of the fluid, Eq.~(\ref{eq:GR:en_mom_tensor}),
the above term simply becomes
\bea
\left(\frac{\p L}{\p x^i_a}\right)_{\rm g}&=&\frac{\nu_a \sqrt{-g}_a}{2N_a^\ast} 
\left(T^{\mu\nu} \frac{\p g_{\mu\nu}}{\p x^i} \right)_a.
\eea
Putting all together, the spatial derivative of the Lagrangian reads
\be
\frac{\p L}{\p x^i_a}= \nu_a \left\{-\sum_b \nu_b \left(\frac{\sqrt{-g}_a P_a}{N^{\ast
      2}_a} + \frac{\sqrt{-g}_b P_b}{N^{\ast 2}_b} \right) \frac{\p W_{ab}}{\p
  x^i_a}  
+ \frac{\sqrt{-g}_a}{2 N^\ast_a} \left(T^{\mu\nu} \frac{\p g_{\mu
      \nu}}{\p x^i} \right)_a \right\}.
\label{eq:GR:dL_dx}
\ee 
Note that the first, hydrodynamical term has the symmetry in the particle
indices, $a$ and $b$, that we know from previous forms of SPH equations, while
the gravity term is exclusively evaluated at the position of the particle
under consideration, $a$.\\
Inserting Eqs.~(\ref{eq:GR:dL_dv}), (\ref{eq:GR:mom_var}) and 
(\ref{eq:GR:dL_dx}) into the Euler-Lagrange equations yields the final,
general-relativistic momentum equation
\be
\frac{d S_{i,a}}{dt}= -\sum_b \nu_b \left(\frac{\sqrt{-g}_a P_a}{N^{\ast
      2}_a} + \frac{\sqrt{-g}_b P_b}{N^{\ast 2}_b} \right) \frac{\p W_{ab}}{\p
  x^i_a}  
+ \frac{\sqrt{-g}_a}{2 N^\ast_a} \left(T^{\mu\nu} \frac{\p g_{\mu
      \nu}}{\p x^i} \right)_a.
\label{eq:GR:mom_evol}
\ee
 
\subsubsection{The energy equation}

We use once more guidance from the canonical energy to choose the
numerical energy variable:
\bea
E &\equiv& \sum_a \frac{\p L}{\p \vec{v}_a} \cdot \vec{v}_a - L
= \sum_a \nu_a S_{i,a} v^i_a + \frac{\nu_a}{\Theta_a} \left[1+u_a\right]
\nonumber\\
&=& \sum_a \nu_a  \left[\Theta_a \left(1 + u_a + \frac{P_a}{n_a} \right) \;
  (g_{i\mu} v^\mu v^i)_a + \frac{1+u_a}{\Theta_a} \right]
= \sum_a \nu_a \hat{e}_a.
\eea
Here we have introduced the energy per baryon
\bea
 \hat{e}_a &\equiv& \Theta_a \left(1 + u_a + \frac{P_a}{n_a} \right) \;
  (g_{i\mu} v^\mu v^i)_a + \frac{1+u_a}{\Theta_a}
= S_{i,a} v^i_a + \frac{1+u_a}{\Theta_a}, 
\eea
very similar to the special-relativistic energy variable, 
Eq.~(\ref{eq:SR:epsilon_a}). Its temporal change is given by
\bea
\frac{d \hat{e}_a}{dt} &=& \frac{d S_{i,a}}{dt} v^i_a + S_{i,a} \frac{d
  v^i_a}{dt} + \frac{d}{dt} \left(\frac{1+u_a}{\Theta_a}\right).
\label{eq:GR:dedt1}
\eea
The first time derivative is known from the momentum equation, the second 
can be hoped to cancel out with a term resulting from the last term, as in
the special-relativistic case, see Eq.~(\ref{eq:SR:h1}). For this last term
we need $du_a/dt$ and $d\Theta_a/dt$.
To obtain the time derivative of $u_a$ we use the first law of thermodynamics,
Eq.~(\ref{eq:first_law_4}),
the number density relation between the frames, Eq.~(\ref{eq:GR:Nstar_n}), and 
$d(\sqrt{-g})/dt$, Eq.~(\ref{eq:GR:dsqrtg}). This delivers
\bea
\frac{d u_a}{dt}&=& \frac{P_a}{n_a^2} \frac{d}{dt} 
\left(\frac{N^\ast}{\sqrt{-g} \Theta}\right)_a = \frac{P_a \sqrt{-g}_a \Theta_a}{N^{\ast 2}_a} \frac{d N^{\ast}_a}{dt} + \frac{P_a N^\ast_a}{n_a^2} \frac{d}{dt} \left(\frac{1}{\sqrt{-g} \Theta} \right)_a.
\eea
The last derivative is given by 
\bea
 \frac{d}{dt}\left(\frac{1}{\sqrt{-g} \Theta}\right)_a &=& 
 - \left(\frac{1}{2 \sqrt{-g} \Theta} g^{\alpha \beta} \frac{d g_{\alpha \beta}}{dt}\right)_a - \left(\frac{1}{\sqrt{-g} \Theta^2} \frac{d\Theta}{dt}\right)_a,
\eea
thus the change in $u_a$ becomes
\bea
\frac{d u_a}{dt}
&=& \frac{P_a \sqrt{-g}_a \Theta_a}{N^{\ast 2}_a} \frac{d N^{\ast}_a}{dt}
- \frac{P_a}{2 n_a} g^{\alpha \beta} \frac{d g_{\alpha \beta}}{dt} 
- \frac{P_a}{n_a \Theta_a} \frac{d\Theta_a}{dt}.\label{eq:du_dt}
\eea
It remains to calculate the derivative of the generalized Lorentz-factor
\bea
\frac{d\Theta_a}{dt}&=& \frac{d}{dt} \left(- g_{\alpha \beta} v^\alpha v^\beta \right)_a^{-1/2}=
 \left(\frac{\Theta^3}{2} v^\alpha v^\beta \frac{d g_{\alpha \beta}}{dt}
+ \Theta^3 g_{\alpha \beta} \frac{d v^\alpha}{dt} v^\beta\right)_a.\label{eq:dtheta_dt}
\eea
On using Eqs.~(\ref{eq:du_dt}) and (\ref{eq:dtheta_dt}) we find
\bea
\frac{d}{dt}\left(\frac{1+u_a}{\Theta_a} \right) =& & 
\frac{1}{\Theta_a} \left\{ \frac{P_a \sqrt{-g}_a \Theta_a}{N^{\ast 2}_a} \frac{d N^{\ast}_a}{dt}
- \frac{P_a}{2 n_a} g^{\alpha \beta} \frac{d g_{\alpha \beta}}{dt} 
- \frac{P_a}{n_a \Theta_a} \frac{d\Theta}{dt} \right\} \nonumber\\
& & - \frac{1+u_a}{\Theta_a^2} \frac{d\Theta}{dt}\nonumber\\
&=& \frac{P_a \sqrt{-g}_a}{N^{\ast 2}_a}\frac{d N^{\ast}_a}{dt}
- \Theta_a \left(1 + u_a + \frac{P_a}{n_a} \right) \left(g_{\alpha \beta} \; v^\alpha \; 
\frac{dv^\beta}{dt}\right)_a\nonumber\\
&& - \left(\frac{P_a}{2 n_a \Theta_a} g^{\alpha \beta}_a  + \frac{\Theta_a}{2} \left[1+u_a+\frac{P_a}{n_a}\right] v^\alpha v^\beta
\right)  \left(\frac{dg_{\alpha\beta}}{dt}\right)_a.
\eea
If we replace the velocities via Eq.~(\ref{eq:GR:v_mu}) by four-velocities and use the definition
of the energy momentum tensor, Eq.~(\ref{eq:GR:en_mom_tensor}), the equation simplifies to
\bea
\frac{d}{dt}\left(\frac{1+u_a}{\Theta_a} \right) =& & 
\frac{P_a \sqrt{-g}_a}{N^{\ast 2}_a}\frac{d N^{\ast}_a}{dt} - \Theta_a \left(1+u_a+\frac{P_a}{n_a}\right)
 \left(g_{\alpha \beta} \; v^\alpha \; 
\frac{dv^\beta}{dt}\right)_a\nonumber\\
& & - \left(\frac{\sqrt{-g}T^{\alpha\beta}}{2 N^{\ast}}\frac{dg_{\alpha\beta}}{dt}\right)_a\nonumber\\
=&&\frac{P_a \sqrt{-g}_a}{N^{\ast 2}_a} \frac{d N^{\ast}_a}{dt}
- (S_i)_a \frac{d v^i_a}{dt}
 - \left(\frac{\sqrt{-g}T^{\alpha\beta}}{2 N^{\ast}}\frac{dg_{\alpha\beta}}{dt}\right)_a.
\label{eq:dtemr2_dt}
\eea
Here we have split the contraction in the term containing $dv^\beta/dt$ into a spatial and a temporal
part and we made use of $v^0=1$, see Eq.~(\ref{eq:GR:v_mu}). This equation is very similar the 
special-relativistic one, Eq.~(\ref{eq:SR:h1}), but it also contains a gravity contribution.
If we insert Eq.~(\ref{eq:dtemr2_dt}) into the energy evolution equation 
(\ref{eq:GR:dedt1}), use the SPH prescription for $dN^\ast/dt$ and collect 
terms into ``hydro'' (kernels) and ``gravity'' (metric derivatives) we
find
\bea
\frac{d \hat{e}_a}{dt}
=& &- \sum_b \nu_b \left( \frac{\sqrt{-g}_a P_a \vec{v}_b}{N_a^{\ast 2}} 
+ \frac{\sqrt{-g}_b P_b \vec{v}_a}{N_b^{\ast 2}}\right) \cdot \nabla_a W_{ab}
+ \frac{\sqrt{-g}_a}{2 N^\ast_a} T^{\alpha \beta}_a \left(\frac{\p g_{\alpha
      \beta}}{\p x^i} v^i - \frac{d g_{\alpha \beta}}{dt}\right)_a\nonumber\\
=& &- \sum_b \nu_b \left( \frac{\sqrt{-g}_a P_a \vec{v}_b}{N_a^{\ast 2}} 
+ \frac{\sqrt{-g}_b P_b \vec{v}_a}{N_b^{\ast 2}}\right) \cdot \nabla_a W_{ab}
- \frac{\sqrt{-g}_a}{2 N^\ast_a} \left(T^{\alpha \beta} \frac{\p g_{\alpha \beta}}{\p t}\right)_a
\label{eq:GR:energy_eq}
\eea
This is the final, general-relativistic SPH energy equation.
The equations (\ref{eq:GR:num_dens}), (\ref{eq:GR:mom_evol}), and (\ref{eq:GR:energy_eq}) 
together with an equation of state form the set of general-relativistic SPH equations in a 
fixed background metric.

%
%
\hspace*{0cm}\fbox{
\parbox{14cm}{
\vspace*{0.5cm}
\centerline{\bf Summary of the general-relativistic SPH}
\centerline{\bf equations on a fixed background metric}

\vspace*{0.5cm}

Ignoring derivatives from the smoothing lengths, the momentum
equation reads
\be
\frac{d S_{i,a}}{dt}= -\sum_b \nu_b \left(\frac{\sqrt{-g}_a P_a}{N^{\ast
      2}_a} + \frac{\sqrt{-g}_b P_b}{N^{\ast 2}_b} \right) \frac{\p W_{ab}}{\p
  x^i_a}  
+ \frac{\sqrt{-g}_a}{2 N^\ast_a} \left(T^{\mu\nu} \frac{\p g_{\mu
      \nu}}{\p x^i} \right)_a
\ee
where 
\be
S_{i,a} = \Theta_a
\left(1 + u_a + \frac{P_a}{n_a} \right) \; (g_{i\mu} v^\mu)_a
\ee
is the canonical momentum per baryon and 
\be
\Theta_a= \left(- g_{\mu\nu} v^{\mu} v^{\nu}\right)^{-\frac{1}{2}}_a
\ee
the generalized Lorentz factor.
The energy equation reads
\be
\hspace*{-1cm}\frac{d \hat{\epsilon}_a}{dt}=  -\sum_b \nu_b \left(\frac{\sqrt{-g}_a P_a}{N^{\ast
      2}_a} \vec{v}_b + \frac{\sqrt{-g}_b P_b}{N^{\ast 2}_b} \vec{v}_a\right) 
\cdot \nabla_a W_{ab}
- \frac{\sqrt{-g}_a}{2 N^\ast_a} \left(T^{\mu\nu} \frac{\p g_{\mu
      \nu}}{\p t} \right)_a,
\label{eq:GR_summary:en_eq}
\ee
where 
\be
\hat{\epsilon}_a= S_{i,a} v^i_a + \frac{1+u_a}{\Theta_a}
\label{eq:GR_summary:en_var}
\ee
is the canonical energy per nucleon.
The number density can again be calculated via summation,
\be
N^\ast_a= \sum_b \nu_b W_{ab}(h_a)\label{eq:GR_summary:dens}.
\ee
}}

\section{Summary}
\label{page_summary}
Since the Smooth Particle Hydrodynamics method was suggested in the late 
seventies it has undergone a long sequence of technical improvements. In 
parallel, the method has been applied to a large variety of problems both
inside and outside astrophysics, and consequently a slew of different physical
processes has been included in SPH-based simulations.\\
In this review we have only very briefly touched upon this latter development,
we merely provided pointers to the current literature. Instead 
we have focused on what we hope is a pedagogical introduction leading to an
in-depth understanding of how the method works. All essential equations 
were derived, and in particular, a commonly used (``vanilla ice'')
set of SPH equations that directly discretizes the Lagrangian equations of an
ideal fluid. We have also reviewed concepts such as 
adaptive numerical resolution, the reasoning behind artificial viscosity 
and basic aspects of the ODE integration in the SPH context.\\
The ``vanilla ice'' equation set works well in practice and possesses the 
property of ``hard-wired'' conservation of mass, energy, linear and angular
momentum. The symmetrization in the particle indices is, however, 
somewhat arbitrary and was enforced ad hoc. In Sec. 3, a 
modern form of the SPH equations was derived that improves on this weakness: 
it uses nothing more than the Lagrangian of an ideal fluid, the first 
law of thermodynamics and a prescription how to obtain the density by
a summation over particles. After 
the Lagrangian has been written in its SPH-discrete form, there is no more 
arbitrariness left in the rest of the derivation: the equations follow
stringently from the Euler-Lagrange equations. In addition to curing the
aesthetical shortcoming of the previous approach, the latter also naturally
leads to corrective terms due to the derivatives of the smoothing lengths, 
in SPH usually called ``grad-h-terms''.\\
The elegant variational concept naturally carries over to both the special- 
and general-relativistic (fixed-metric) case. Here, we derived for the 
first time the 
special-relativistic SPH equations that include ``grad-h-terms''. The
numerical variables are chosen as canonical momentum and energy per baryon.
While this has obvious numerical advantages, it comes at the price of
recovering the physical variables from the numerical ones at each time step,
a tribute that is also payed in modern grid-based relativistic approaches, see
e.g. \cite{marti03}. The performance of this new equation set has been briefly
illustrated at the example of a relativistic shock tube test, for a more exhaustive 
set of tests we refer to \cite{rosswog09d}.\\
We conclude this review with a derivation of the general-relativistic
case in which the space-time metric can be considered unperturbed by the
self-gravity of the fluid, a case that is realized, for example, in the
tidal disruption of a star by a supermassive black hole. The resulting equations
are a natural generalization of the special-relativistic case and do not
introduce much of an additional numerical burden. The resulting extra terms
can be analytically calculated from the metric that is assumed to be known.\\
In the future, more physical processes and their corresponding numerical
schemes will certainly find their ways into SPH. On the numerical 
hydrodynamics side the recent past has seen a couple suggestions on 
how to improve on known weaknesses of SPH. One of them was
SPH's unsatisfactory performance on fluid instabilities across  
discontinuities with large density jumps \cite{agertz07}. A modification 
of artificial dissipation terms \cite{price08a} or, alternatively, a more 
general formulation of the SPH equations \cite{read09} have been suggested 
as cures for this problem. 
In terms of artificial dissipation, a tensor artificial viscosity approach
such as the one suggested in \cite{owen04} should definitely be
explored further for its potential and usability in astrophysics. While the 
special-relativistic SPH equations as derived in Sec. \ref{sec:rel_SPH} 
have been carefully tested, the general-relativistic equation set has so far 
not been implemented and applied to astrophysical problems.
This task is left for future investigations.

{\bf Acknowledgments}\\
It is a great pleasure to thank Marcus Br\"uggen, Marius Dan, James Guillochon and Roland Speith for their careful reading of the manuscript.

\label{page_references}
\hyphenation{Post-Script Sprin-ger}

\end{document}